\pdfoutput=1

\newif \ifdraft \draftfalse
\newif \iffull \fulltrue
\newif \ifanonymous \anonymousfalse

\iffull
\documentclass[acmsmall,screen,nonacm]{acmart}
\else
\documentclass[acmsmall]{acmart}
\fi

\setcopyright{rightsretained}
\acmPrice{}
\acmDOI{10.1145/3428209}
\acmYear{2020}
\copyrightyear{2020}
\acmSubmissionID{oopsla20main-p65-p}
\acmJournal{PACMPL}
\acmVolume{4}
\acmNumber{OOPSLA}
\acmArticle{141}
\acmMonth{11}

\newcommand{\BNFALT}{\;\;|\;\;}

\newcommand{\hdr}[2]{%
\flushleft
\textbf{#1} \hfill {#2} \\
\centering
}

\usepackage{enumitem}
\newlist{proofcases}{itemize}{3}
\setlist[proofcases]{leftmargin=2.25cm,style=sameline,parsep=3pt plus 1pt minus 1pt}

\usepackage{bcprules}
\usepackage{xargs}

\newcommandx{\sidebyside}[5][1=.45,2=.45,3=b]{%
\begin{minipage}[#3]{#1\linewidth}
\centering
{#4}
\end{minipage}\hfill
\begin{minipage}[#3]{#2\linewidth}
\centering
{#5}
\end{minipage}%
}

\newcommandx{\threesidebyside}[7][1=.3,2=.3,3=.3,4=b]{%
\begin{minipage}[#4]{#1\linewidth}
\centering
{#5}
\end{minipage}\hfill
\begin{minipage}[#4]{#2\linewidth}
\centering
{#6}
\end{minipage}\hfill
\begin{minipage}[#4]{#3\linewidth}
\centering
{#7}
\end{minipage}%
}

\newcommandx{\foursidebyside}[8][1=.225,2=.225,3=.225,4=.225]{%
\begin{minipage}[b]{#1\linewidth}
\centering
{#5}
\end{minipage}\hfill
\begin{minipage}[b]{#2\linewidth}
\centering
{#6}
\end{minipage}\hfill
\begin{minipage}[b]{#3\linewidth}
\centering
{#7}
\end{minipage}\hfill
\begin{minipage}[b]{#4\linewidth}
\centering
  {#8}
\end{minipage}%
}

\theoremstyle{definition}
\newtheorem{exmp}{Example}

\usepackage{booktabs}   %% For formal tables:
\usepackage{subcaption} %% For complex figures with subfigures/subcaptions
\usepackage{xcolor}
\usepackage{xspace}
\usepackage{listings}
\usepackage{wrapfig}
\usepackage{framed}
\usepackage{multirow}
\usepackage{fnpct}
\usepackage{microtype}
\usepackage[utf8]{inputenc}
\usepackage[T1]{fontenc}

\ifdraft
\fi

\begin{document}

\newcommand{\Name}{Formulog\xspace}

% macros for writing notes
\newcommand{\SHOWCOMMENT}[1]{\ifdraft {$\blacksquare$ [{#1}]}\xspace\fi}
\newcommand{\TODO}[1]{\SHOWCOMMENT{\color{red}{#1}}}
\newcommand{\michael}[1]{\SHOWCOMMENT{MMG: \color{olive}{#1}}}
\newcommand{\aaron}[1]{\SHOWCOMMENT{AB: \color{blue}{#1}}}
\newcommand{\steve}[1]{\SHOWCOMMENT{SC: \color{brickred}{#1}}}
\newcommand{\xxx}{\TODO{XXX}}

\newcommand{\denot}[1]{\ensuremath{[ \! [#1] \! ]}}

%inline code
\newcommand{\ic}[1]{\text{\lstinline;#1;}}
\newcommand{\icfoot}[1]{\text{\lstinline[{basicstyle=\footnotesize\ttfamily}];#1;}}
\newcommand{\iccapt}[1]{\text{\lstinline[{basicstyle=\small\ttfamily}];#1;}}
%backtick
\newcommand{\bk}{\textsf{\`{}}}
\newcommand{\set}[1]{\mathbf{#1}}
\newcommand{\dom}{\ensuremath{\mathrm{dom}}}

\newcommand{\prog}{\ensuremath{\mathrm{prog}}}
\newcommand{\uf}{\ensuremath{\mathit{uf}}}

\newcommand{\bool}{\ensuremath{\mathsf{bool}}}
\newcommand{\strtype}{\ensuremath{\mathsf{string}}}
\newcommand{\bv}[1]{\ensuremath{\mathsf{bv[\ensuremath{#1}]}}}
\newcommand{\fp}[2]{\ensuremath{\mathsf{fp[\ensuremath{#1},\ensuremath{#2}]}}}
\newcommand{\wildcard}{\ensuremath{\mathsf{??}}}

\newcommand{\fun}[4]{\ensuremath{\mathsf{fun } ~ {#1}({#2}) : {#3} = {#4}}}
\newcommand{\horn}{\ensuremath{\mathrel{\mathsf{\mathord{:}\mathord{-}}}}}
\newcommand{\match}[2]{\ensuremath{\mathsf{match} ~ {#1} ~ \mathsf{with} ~ {#2}}}
\newcommand{\qq}[1]{\ensuremath{\bk{#1}\bk}}
\newcommand{\unq}[1]{\ensuremath{\textbf{,}{#1}}}
\newcommand{\letin}[3]{\ensuremath{\mathsf{let} ~ {#1} = {#2} ~ \mathsf{in} ~ {#3}}}
\newcommand{\letphi}[3]{\ensuremath{\mathsf{let}_\phi ~ {#1} = {#2} ~ \mathsf{in} ~ {#3}}}
\newcommand{\ite}[3]{\ensuremath{\mathsf{if} ~ {#1} ~ \mathsf{then} ~ {#2} ~ \mathsf{else} ~ {#3}}}
\newcommand{\forallq}[2]{\ensuremath{\forall {#1}.~{#2}}}

\newcommand{\typeof}{\ensuremath{\mathrm{typeof}}}
\newcommand{\implicit}[1]{{\color{gray} {#1}}}
\newcommand{\makemered}[1]{{\color{red} {#1}}}

%smt and sym types
\newcommand{\mEXP}{\ensuremath{\mathsf{exp}}}
\newcommand{\mSMT}{\ensuremath{\mathsf{smt}}}
\newcommand{\smt}[1]{\ensuremath{{#1} ~ \mathsf{smt}}}
\newcommand{\sym}[1]{\ensuremath{{#1} ~ \mathsf{sym}}}
\newcommand{\MODEL}{\ensuremath{\mathsf{model}}}

\newcommand{\tosmt}{\ensuremath{\mathrm{toSMT}}}
\newcommand{\erase}{\ensuremath{\mathrm{erase}}}

%operational semantics
\newcommand{\world}{\ensuremath{\mathcal{W}}}
\newcommand{\apply}[2]{\ensuremath{\mathrm{apply}(#1, #2)}}
\newcommand{\unify}{\ensuremath{\sim}}
\newcommand{\Unify}{\ensuremath{\sim}}
\newcommand{\length}{\ensuremath{\mathrm{length}}}
\newcommand{\stepsto}{\ensuremath{\rightarrow}}
\newcommand{\StepstoE}{\ensuremath{\Downarrow_e}}
\newcommand{\StepstoVecE}{\ensuremath{\Downarrow_{\vec{e}}}}
\newcommand{\StepstoPhi}{\ensuremath{\Downarrow_\phi}}
\newcommand{\StepstoVecPhi}{\ensuremath{\Downarrow_{\vec{\phi}}}}
\newcommand{\lookup}{\ensuremath{\mathrm{lookup}}}
\newcommand{\isgrnd}{\ensuremath{\mathrm{isGround}}}
\newcommand{\vars}{\ensuremath{\mathrm{vars}}}
\newcommand{\true}{\ensuremath{\mathrm{true}}}
\newcommand{\false}{\ensuremath{\mathrm{false}}}
\newcommand{\cSMT}[1]{\ensuremath{c^\mathsf{SMT}_{\mathsf{#1}}}}
\newcommand{\cCONST}{\ensuremath{\cSMT{const}}}
\newcommand{\cVAR}{\ensuremath{\cSMT{var}}}
\newcommand{\cCTOR}{\ensuremath{\cSMT{ctor}}}
\newcommand{\cUF}{\ensuremath{\cSMT{uf}}}
\newcommand{\cLET}{\ensuremath{\cSMT{let}}}
\newcommand{\cFORALL}{\ensuremath{\cSMT{forall}}}
\newcommand{\grayb}[1]{\implicit{[}{#1}\implicit{]}}
\newcommand{\operation}[1]{\ensuremath{\mathord{\otimes}({#1})}}

%rule-level typing
\newcommand{\sig}{\ensuremath{\mathrm{sig}}}
\newcommand{\inst}{\ensuremath{\mathrm{inst}}}

\lstset{
    basicstyle=\small\ttfamily,
    mathescape=true
}

%% Title information
%\title[Short Title]{Full Title}         %% [Short Title] is optional;
\title{\Name: Datalog for SMT-Based Static Analysis}
                                        %% when present, will be used in
                                        %% header instead of Full Title.
%\titlenote{with title note}             %% \titlenote is optional;
                                        %% can be repeated if necessary;
                                        %% contents suppressed with 'anonymous'
\iffull
\subtitle{Extended Version}                     %% \subtitle is optional
\subtitlenote{This article extends one published in PACMPL~\citep{formulog} with technical appendices.}\fi       %% \subtitlenote is optional;
                                        %% can be repeated if necessary;
                                        %% contents suppressed with 'anonymous'

%% Author information
%% Contents and number of authors suppressed with 'anonymous'.
%% Each author should be introduced by \author, followed by
%% \authornote (optional), \orcid (optional), \affiliation, and
%% \email.
%% An author may have multiple affiliations and/or emails; repeat the
%% appropriate command.
%% Many elements are not rendered, but should be provided for metadata
%% extraction tools.

%% Author with single affiliation.
\author{Aaron Bembenek}
%\authornote{with author1 note}          %% \authornote is optional;
                                        %% can be repeated if necessary
%\orcid{nnnn-nnnn-nnnn-nnnn}             %% \orcid is optional
\affiliation{
%  \position{Position1}
%  \department{Department of Computer Science}              %% \department is recommended
  \institution{Harvard University}            %% \institution is required
%  \streetaddress{Street1 Address1}
%  \city{City1}
%  \state{State1}
%  \postcode{Post-Code1}
  \country{USA}                    %% \country is recommended
}
\email{bembenek@g.harvard.edu}          %% \email is recommended

%% Author with two affiliations and emails.
\author{Michael Greenberg}
\authornote{Work done while on sabbatical at Harvard University.}          %% \authornote is optional;
                                        %% can be repeated if necessary
%\orcid{nnnn-nnnn-nnnn-nnnn}             %% \orcid is optional
\affiliation{
%  \position{Position2a}
%  \department{Department of Computer Science}             %% \department is recommended
  \institution{Pomona College}           %% \institution is required
%  \streetaddress{Street2a Address2a}
%  \city{City2a}
%  \state{State2a}
%  \postcode{Post-Code2a}
  \country{USA}                   %% \country is recommended
}
\email{michael@cs.pomona.edu}         %% \email is recommended

\author{Stephen Chong}
\affiliation{
%  \position{Position2b}
%  \department{Department2b}             %% \department is recommended
  \institution{Harvard University}           %% \institution is required
%  \streetaddress{Street3b Address2b}
%  \city{City2b}
%  \state{State2b}
%  \postcode{Post-Code2b}
  \country{USA}                   %% \country is recommended
}
\email{chong@seas.harvard.edu}         %% \email is recommended

%% Abstract
%% Note: \begin{abstract}...\end{abstract} environment must come
%% before \maketitle command
\begin{abstract}
	Satisfiability modulo theories (SMT) solving has become a critical part of
  many static analyses, including symbolic execution, refinement type checking,
  and model checking.
We propose \Name, a domain-specific language that makes it possible to write a
  range of SMT-based static analyses in a way that is both close to their
  formal specifications and amenable to high-level optimizations and efficient
  evaluation.

\Name extends the logic programming language Datalog with a first-order
  functional language and mechanisms for representing and reasoning about SMT
  formulas; a novel type system supports the construction of expressive
  formulas, while ensuring that neither normal evaluation nor SMT solving goes
  wrong.
Our case studies demonstrate that a range of SMT-based analyses can naturally
  and concisely be encoded in \Name, and that --- thanks to this encoding ---
  high-level Datalog-style optimizations can be automatically and
  advantageously applied to these analyses.
\end{abstract}

%% 2012 ACM Computing Classification System (CSS) concepts
%% Generate at 'http://dl.acm.org/ccs/ccs.cfm'.
\begin{CCSXML}
<ccs2012>
   <concept>
       <concept_id>10011007.10010940.10010992.10010998.10011000</concept_id>
       <concept_desc>Software and its engineering~Automated static analysis</concept_desc>
       <concept_significance>500</concept_significance>
       </concept>
   <concept>
       <concept_id>10011007.10011006.10011050.10011017</concept_id>
       <concept_desc>Software and its engineering~Domain specific languages</concept_desc>
       <concept_significance>500</concept_significance>
       </concept>
   <concept>
       <concept_id>10011007.10011006.10011008.10011009.10011015</concept_id>
       <concept_desc>Software and its engineering~Constraint and logic languages</concept_desc>
       <concept_significance>500</concept_significance>
       </concept>
 </ccs2012>
\end{CCSXML}

\ccsdesc[500]{Software and its engineering~Automated static analysis}
\ccsdesc[500]{Software and its engineering~Domain specific languages}
\ccsdesc[500]{Software and its engineering~Constraint and logic languages}
%% End of generated code

%% Keywords
%% comma separated list
\keywords{Datalog, SMT solving}  %% \keywords are mandatory in final camera-ready submission

%% \maketitle
%% Note: \maketitle command must come after title commands, author
%% commands, abstract environment, Computing Classification System
%% environment and commands, and keywords command.
\maketitle

\section{Introduction}

Satisfiability modulo theories (SMT) solving provides a way to reason logically
about common program constructs such as arrays and bit vectors, and as such has
become a key component of many static analyses. 
For example, symbolic execution tools use SMT solving to prune infeasible
execution paths~\citep{klee,symexecsurvey}; type checkers use it to prove
subtyping relations between refinement types~\citep{dminor,Rondon08liquid}; and
model checkers use it to abstract program
states~\citep{lazyabsinterpolants,cimatti2012software}.
This paper presents \Name, a domain-specific language for writing SMT-based
static analyses.
\Name makes it possible to concisely encode a range of SMT-based static
analyses in a way that is close to their formal specifications.
Furthermore, \Name is designed so that analyses implemented in it are amenable
to efficient evaluation and powerful, high-level optimizations, including
parallelization and automatic transformation of exhaustive analyses into
goal-directed ones.

% \Name is based on Datalog, a logic programming language used to implement
% static analyses ranging from points-to analyses for object-oriented
% languages~\citep{bddbddbPointer,doop}, to
% decompilers~\citep{gigahorse,datalogdisassembly}, to security analyses for
% Java~\citep{bddbddbSecurity,souffle1}, JavaScript~\citep{gatekeeper}, and
% Ethereum smart contracts~\citep{madmax,securify}.
\Name is based on Datalog, a logic programming language used to implement
static analyses ranging from points-to analyses~\citep{bddbddbPointer,doop} to
decompilers~\citep{gigahorse,datalogdisassembly} to security
analyses~\citep{bddbddbSecurity,souffle1,gatekeeper,madmax,securify}.
Embodying the principle of separating the logic of a computation from the
control necessary to perform that computation~\citep{kowalski}, Datalog frees
analysis designers from low-level implementation details and enables them to
program at the level of specifications (such as formal inference rules).
This leads to concise implementations~\citep{bddbddbUsing} that can be easier
to reason about and improve at the algorithmic level compared to
analyses in more traditional languages~\citep{fastandeasy}.
Datalog-based analyses can be fast and scalable, even outperforming the
non-Datalog state-of-the-art~\citep{doop}.
Indeed, Datalog's high-level nature makes it amenable to high-level
optimizations, such as parallelization~\citep{souffle2} and synthesis of
goal-directed analyses from exhaustive ones~\citep{reps}.

However, despite the appeal of Datalog for static analysis and the importance
of SMT solving in static analysis, until now there has not been a focused study
of how to effectively extend the benefits of Datalog to SMT-based analyses;
our work bridges this gap.

\smallskip

\Name augments Datalog with an interface to an external SMT solver and a
first-order fragment of the functional language ML.
It provides a library of constructors for building terms that are interpreted
as logical formulas when applied to special SMT operators; in the backend,
these operators are implemented by calls to an external SMT solver.
A \Name program is essentially a set of ML-style function definitions and
Datalog-style rules; both pieces can refer to each other and invoke the SMT
operators.
As in Datalog, the goal of \Name evaluation is to compute all possible
inferences with respect to the rules, which correspond to logical implications.
Unlike Datalog, rule evaluation might involve both ML evaluation and calls to
an SMT solver.

The way this design combines Datalog, ML, and SMT solving gives \Name some
desirable properties.
First, \Name programs can use SMT solving the way it is used in SMT-based
analyses.
This results from the choice to represent SMT formulas as ML terms, and
contrasts with the approach of most prior work combining logic programming and
constraint solving (where, e.g., checking for formula validity is hard).
Second, the combination of Datalog-style rules and ML-style functions mirrors
the combination of inference rules and helper functions commonly used in
analysis specifications, making it easier to translate formal analysis
specifications into executable code.
This close correspondence between specification and implementation means
that specification-level reasoning is still applicable to analysis
implementations (and vice versa: unexpected behavior in \Name programs has
revealed bugs in specifications).
Third, because \Name is based on Datalog, analyses written in it can be
effectively optimized and evaluated via powerful Datalog algorithms, making
them competitive with analyses written in more mature languages.

It takes care to fit Datalog, ML, and SMT solving together in a way that truly
achieves these properties.
Along these lines, part of our technical contribution is a novel bimodal type
system that treats terms appearing in SMT formulas more liberally than terms
appearing outside of formulas, making it possible to construct expressive
logical formulas, while still ensuring that neither concrete (i.e., Datalog/ML)
evaluation nor SMT solving goes wrong.
\smallskip

To test the practicality of \Name, we implemented a fully featured, prototype
\Name runtime and wrote three substantial SMT-based analyses in \Name: a type
checker for a refinement type system, a bottom-up points-to analysis for JVM
bytecode, and a bounded symbolic evaluator for a subset of LLVM bitcode.
Our implementations for the first two case studies are almost direct
translations of previously published formal
specifications~\citep{dminor,bottomupPointsto}; indeed, \Name allowed us to
program close enough to the specifications to uncover bugs in both of them.
Despite encoding complex analysis logic, each of our analyses is
concise (no more than 1.5K LOC).
Furthermore, our \Name-based implementations have acceptable performance, even
when compared against reference implementations running on more mature language
platforms.
In some cases, we actually achieve substantial speedups over the reference
implementations.

These performance results are possible only because \Name's design allows
our runtime to automatically and effectively apply high-level optimizations to
\Name programs.
Our third case study makes this point emphatically.
Due to automatic parallelization, our symbolic evaluator achieves a speedup
of 8$\times$ over %a recent version of
the symbolic execution tool
KLEE~\citep{klee}.
Moreover, this speedup increases to 12$\times$ when we use the magic set
transformation~\citep{magicsets1,magicsets2} to automatically transform our
exhaustive symbolic evaluator into a goal-directed one that explores only paths
potentially leading to assertion failures.
That Datalog can speed up analyses like points-to analysis is well established
~\citep{bddbddbPointer,doop}; that it can automatically scale symbolic
evaluation is a novel result.
\smallskip

In sum, this paper makes the following contributions:
\begin{itemize}
  \item the design of \Name (Section~\ref{sec:language}), a domain-specific
    language for writing SMT-based static analyses that judiciously combines
    Datalog, a fragment of ML, and SMT solving;
  \item a lightweight bimodal type system (Section~\ref{sec:typesystem}) that
    mediates the interface between concrete evaluation and SMT solving,
    enabling the construction of expressive formulas while preventing many
    kinds of runtime errors in both concrete evaluation and SMT solving;
  \item a fully featured prototype and three substantial case studies
    (Section~\ref{sec:examples}), showing that the design of \Name can be the
    basis of a practical tool for writing SMT-based analyses; and
  \item an evaluation of \Name's design in light of these case studies
    (Section~\ref{sec:eval}), demonstrating how careful design decisions make
    \Name an effective medium for encoding a range of SMT-based analyses in a
    way that is both close to their formal specifications and amenable to
    efficient evaluation and high-level optimizations.
\end{itemize}

\section{Background}\label{sec:background}

\begin{figure}[t!]
  \begin{minipage}{0.85\linewidth}
    \sidebyside[0.5][0.45][t]{
    \[\begin{array}{lrcl}
      \text{Programs}     & \prog &::=& H^* \\
      \text{Horn clauses} & H     &::=& p(e^*) \horn P^* \\
      \text{Premises}     & P     &::=& A \BNFALT !A \\
      \text{Atoms}        & A     &::=& p(e^*) \BNFALT e = e \\
      \text{Expressions}  & e     &::=& X \BNFALT c 
    \end{array}\]
  }{
    \[\begin{array}{lrcl}
      \text{Variables} & X &\in& \mathrm{Var} \\
      \text{Constructors} & c     &\in& \mathrm{CtorVar} \\
      \text{Predicates} & p &\in& \mathrm{PredVar}
    \end{array}\]
  }
  \end{minipage}
  \caption{A Datalog program is a collection of Horn clauses that represent rules for making inferences.}
  \label{fig:datalog}
\end{figure}

The starting point for \Name is Datalog with stratified negation
(Figure~\ref{fig:datalog})~\citep{gallaire1978logic,apt1988towards,przymusinski1988declarative,vangelder1989negation,recursivequery}.
A Datalog program is a collection of Horn clauses, where a \emph{clause} $H$
consists of a head predicate $p(e^*)$ and a sequence of body premises $P$.
Each \emph{premise} $P$ is either a positive atom $A$ or a negated atom $!A$.
An \emph{atom} $A$ has one of two forms: It is either a predicate symbol
applied to a list of expressions, or the special equality predicate $e = e$.
An \emph{expression} $e$ is a variable $X$ or a nullary constructor $c$, i.e.,
an uninterpreted constant. 
Each predicate symbol $p$ is associated with an extensional database (EDB)
relation or an intensional database (IDB) relation.
An \emph{EDB relation} is tabulated explicitly through \emph{facts} (clauses with
empty bodies), whereas an \emph{IDB relation} is computed through \emph{rules}
(clauses with non-empty bodies).
A rule should be read as a universally quantified logical implication, with the
conjunction of the body premises implying the predicate in the head.
Datalog evaluation amounts to computing every possible inference with respect
to these implications; the restriction to stratified negation (a relation
cannot be defined, either directly or indirectly, by its complement) ensures
that this can be done via a sequence of fixed point computations. 

Datalog has proven to be a natural and effective way to encode a range of
static
analyses~\citep{bddbddbPointer,doop,gigahorse,datalogdisassembly,bddbddbSecurity,souffle1,gatekeeper,madmax,securify}.
EDB relations are used to represent the program under analysis; for example,
EDB relations might encode a control flow graph (CFG) of the input program.
The logic of the analysis is encoded using rules that define IDB relations;
these rules are fixed and do not depend on the program under analysis (which is
already captured by the EDB relations).
The Datalog program will compute the contents of the IDB relations, which can
be thought of as the analysis results.

That being said, standard Datalog is a very restricted language and there are
many other analyses that cannot easily be encoded in it, if at all.
Recent variants extend Datalog for analyses that operate over interesting
lattices~\citep{flix,inca}.
Following in this spirit, \Name proposes a way to support analyses that need
access to SMT solving.

\section{Language design}\label{sec:language}

The design of \Name is driven by three main desiderata.
First, it should be possible to implement SMT-based static analyses in a form
close to their formal specifications.
Second, it should be easy to use logical terms the way that they are commonly
used in many analyses.
For example, analyses often need to create formulas about entities such as
arrays and machine integers, test those formulas for satisfiability or
validity, and generate models of them.
Third, \Name programs should still be amenable to powerful Datalog
optimizations and evaluable using scalable Datalog algorithms.

Section~\ref{sec:eval} demonstrates how the design of \Name largely meets these
desiderata.
Here, we give a warm-up example of \Name, provide an overview of its language
features, discuss how these features support logical formulas, and conclude
with its operational semantics.

\subsection{\Name by Example}

To give the flavor of \Name-based analyses, this section presents a bounded
symbolic evaluator for CFGs of a simple imperative language
(Figures~\ref{fig:ex_repr} and~\ref{fig:ex_logic}).
A symbolic evaluator~\citep{king} interprets a program in which some values are
unknown.
When the evaluator reaches a condition that depends on one of these symbolic
values, it forks into two processes, one in which the condition is assumed to
be true and one in which it is assumed to be false.
At this point, it can avoid exploring an impossible path by checking whether
the condition along that branch is consistent with the conditions encountered
so far during execution (which are known collectively as the ``path
condition'').
Our symbolic evaluator uses fuel to bound the depth of its execution.

We use algebraic data types to represent the input language to the evaluator
(lines 1-9).
Values are formulas representing 32-bit vectors (i.e., terms of type \ic{i32
smt}), and operands are either values or variables.
Our input language has binary operations, conditional jumps, and fail
instructions (indicating that control flow has reached, e.g., an assertion
failure).
The program-to-analyze is given by two EDB (i.e., \ic{input}) relations (lines 11-12).
The relation \ic{node_has_inst} maps a CFG node to the corresponding
instruction, and the relation \ic{node_has_succ} relates it to its fall-through
successor.

The state of the symbolic evaluator (line 14) is a record with a store mapping
variables to values, and a path condition; an initial state (line 16) consists
of an empty map and the \ic{true} path condition.%
\footnote{We omit the definitions for maps; we use association lists with the
standard operations \icfoot{empty_map}, \icfoot{get}, and \icfoot{put}.}
ML-style functions are used to update and query the state.
The function \ic{update_store} (lines 18-19) updates a store binding, while the
function \ic{update_path_cond} (lines 21-24) adds another conjunct to the path
condition, returning \ic{none} in the case that the resulting path condition is
unsatisfiable.
The built-in operator \ic{is_sat} queries an external SMT solver for the
satisfiability of its argument (an SMT proposition).
The function \ic{operand_value} (lines 26-34) looks up the value of an operand in the
state, returning a pair of the value and a (possibly) new state.
In the case that the operand is a value or a mapped variable, the relevant
value is returned with the input state.
In the case that the operand is a variable that is not in the store, the
function returns a fresh symbolic bit vector (\ic{`#\{st\}[i32]`}) with an
updated state mapping the variable to that value.
%As explained in Section~\ref{sec:repformulas},
% This value is a
% bit-vector-valued SMT variable that is guaranteed to be fresh with respect to
% any SMT variables already present in the state.

\begin{figure}[t]
  \begin{lstlisting}[numbers=left,xleftmargin=2em,framexleftmargin=1.5em]
type val = i32 smt
type var = string
type operand = o_val(val) | o_var(var)
type binop = b_add | b_mul | b_eq | b_lt
type node = i32
type inst =
  | i_binop(var, binop, operand, operand)
  | i_jnz(operand, node) (* jump if the operand is not zero *)
  | i_fail

input node_has_inst(node, inst)
input node_has_succ(node, node)

type state = { store: (var, val) map; path_cond: bool smt; }

fun initial_state : state = { store=empty_map; path_cond=`true`; }

fun update_store(x: var, v: val, st: state) : state =
  { st with store=put(x, v, store(st)) }

fun update_path_cond(x: bool smt, st: state) : state option =
  let y = path_cond(st) in
  let z = `x /\ y` in
  if is_sat(z) then some({ st with path_cond=z }) else none

fun operand_value(o: operand, st: state) : val * state =
  match o with
  | o_val(v) => (v, st)
  | o_var(x) =>
    match get(x, store(st)) with
    | some(v) => (v, st)
    | none => let v = `#{st}[i32]` in (v, update_store(x, v, st))
    end
  end
\end{lstlisting}
  \caption{A combination of types and input relations represent the program
  under evaluation; ML-style functions are defined for manipulating the complex
  types that represent evaluator state.}
  \label{fig:ex_repr}
\end{figure}

\begin{figure}[hbtp]
\begin{lstlisting}[numbers=left,xleftmargin=2em,framexleftmargin=1.5em,firstnumber=36]
output reached(node, state, i32 option)
output failed(node, state)

fun decr(n: i32) : i32 option = if n > 0 then some(n - 1) else none

fun do_binop(b: binop, op1: operand, op2: operand, st: state) :
    val * state =
  let (v1, st1) = operand_value(op1, st) in
  let (v2, st2) = operand_value(op2, st1) in
  let fun b2i(x: bool smt) : i32 smt = `#if x then 1 else 0` in
  let v = match b with
          | b_add => `bv_add(v1, v2)`
          | b_mul => `bv_mul(v1, v2)`
          | b_eq => b2i(`v1 #= v2`)
          | b_lt => b2i(`bv_slt(v1, v2)`)
          end in
  (v, st2)

reached(0, initial_state, some(10)). (* start with 10 units of fuel *)

reached(Next, St2, decr(N)) :-
  reached(Curr, St, some(N)),
  node_has_inst(Curr, i_binop(Def, B, Op1, Op2)),
  node_has_succ(Curr, Next),
  St2 =
    let (v, st1) = do_binop(B, Op1, Op2, St) in
    update_store(Def, v, st1).

reached(Dst, St2, decr(N)) :-
  reached(Curr, St, some(N)),
  node_has_inst(Curr, i_jnz(Op, Dst)),
  some(St2) =
    let (v, st1) = operand_value(Op, St) in
    update_path_cond(`~(v #= 0)`, st1).

reached(Next, St2, decr(N)) :-
  reached(Curr, St, some(N)),
  node_has_inst(Curr, i_jnz(Op, _)),
  node_has_succ(Curr, Next),
  some(St2) =
    let (v, st1) = operand_value(Op, St) in
    update_path_cond(`v #= 0`, st1).

failed(Node, St) :-
  reached(Node, St, _),
  node_has_inst(Node, i_fail).
\end{lstlisting}
  \caption{Horn clauses and ML-style helper functions define the logic of the
  symbolic evaluator.}
  \label{fig:ex_logic}
\end{figure}

The symbolic evaluator itself is defined through two IDB (i.e., \ic{output})
relations (lines 36-37).
The relation \ic{reached} consists of tuples $(node, st, fuel)$ that indicate
that the symbolic evaluator has reached a node $node$ with state $st$ and the
amount of fuel $fuel$.
The relation \ic{failed} consists of pairs $(node, st)$ that indicate that the
evaluator has reached a failure node $node$ with the state $st$.

Before defining these relations, we define some helper functions.
The function \ic{decr} (line 39) decrements an integer if it is greater than
zero, and else returns \ic{none}; it is used to decrement the amount of fuel.
The function \ic{do_binop} (lines 41-52) is used to perform a binary operation
on two operands.
It looks up the value of those operands in the given state, and then returns a
bit-vector-valued SMT formula representing the binary operation applied to
those values.
It also returns a state, since the resolution of the operands might have
resulted in an updated state (if one of the operands is an unmapped variable).
The locally scoped function \ic{b2i} converts an SMT proposition to a bit-vector-valued SMT formula by building an if-then-else SMT expression (via the
$\ic{\#if} \cdot \ic{then} \cdot \ic{else}$ constructor) that is \ic{1} if the
proposition is true and \ic{0} otherwise.

Four rules define the \ic{reached} relation.
The first one (line 54) states the base case: node 0 (the start of the CFG) is
reachable with the initial state and 10 units of fuel.
The remaining recursive rules match each possible step of execution and have a
shared form:
They check whether execution has reached a particular type of instruction with
a non-zero amount of fuel, do whatever operation is required for that
instruction, and then, if successful, step to the appropriate successor
instruction with one less unit of fuel (computed via \ic{decr}). 
For example, the second rule (lines 56-62) handles a binary operation: the operation is
performed symbolically (via the function \ic{do_binop}), the store is updated
with the resulting value, and evaluation steps to the fall-through successor
node \ic{Next}.

The third and fourth rules define what happens when evaluation reaches a
conditional jump.
The first of these (lines 64-69) handles the case where the jump condition
succeeds (i.e., when the operand in the jump can be nonzero) in which case the
evaluator steps to the jump destination \ic{Dst} with an updated path condition
constraining the operand to be nonzero.
The second of these (lines 71-77) handles the case where the jump condition
fails (i.e, the operand can be zero).
Note that these two cases are not mutually exclusive; the fact that these two
rules can ``fire'' at the same time means that the symbolic evaluator can
explore both branches in parallel.

One final rule (lines 79-81) defines the \ic{failed} relation, and states that
evaluation has uncovered a failure if it has reached a node with a fail
instruction.

The symbolic evaluator can correctly determine that this program is safe:
\begin{lstlisting}
if (x < y) { x++; assert(x <= y); }
\end{lstlisting}
It can also determine that this program is not (because the bit vector \ic{y} can wrap around):
\begin{lstlisting}
if (x < y) { x++; y++; assert(x <= y); }
\end{lstlisting}
While seemingly simple, this toy symbolic evaluator captures the essence of the
more developed symbolic evaluator we describe as a case study
(Section~\ref{sec:symexec}).

\subsection{Overview}\label{sec:features}

\begin{figure}[t]
  \[ \begin{array}{lrcl}
    \multicolumn{4}{l}{\textbf{Types}} \\ \hline
    \text{Types}        & \tau   &::=& t \BNFALT \smt{t} \BNFALT \sym{t} \BNFALT \mathsf{model} \\
    \text{Pre-types}    & t      &::=& B \BNFALT D ~ \tau^* \BNFALT \alpha \\
    \text{Base types}   & B      &::=& \bool \BNFALT \strtype \BNFALT \bv{k}_{k \in {\mathbb{N}^+}} \BNFALT
                                     \dots\\[1em]
    
    \multicolumn{4}{l}{\textbf{Terms}} \\ \hline

    \text{Programs}     & \prog &::=& H^*~T^*~F^*~Z^* \\
    \text{Horn clauses} & H     &::=& p(e^*) \horn P^* \\
    \text{Premises}     & P     &::=& A \BNFALT \mathord{!}A \\
    \text{Atoms}        & A     &::=& p(e^*) \BNFALT e=e \\
    \text{Type definitions} & T &::=& \mathsf{type}~\alpha^*~D = \grayb{c(\tau^*)}^*\\
    \text{Functions}    & F     &::=& \mathsf{fun}~f(\grayb{X : \tau}^*) : \tau = e \\
    \text{SMT declarations} & Z &::=& \mathsf{uninterpreted~fun}~c(\grayb{\smt{t}}^*) : \smt{t} \BNFALT\\
                            &   &   & \mathsf{uninterpreted~sort}~\alpha^*~D \\[0.25em]
    \text{Expressions}  & e     &::=& X \BNFALT c(e^*) \BNFALT k \BNFALT f(e^*) \BNFALT
                                      \match{e}{\grayb{c(X^*) \rightarrow e}^*} \BNFALT \\
                        &       &   & \letin{X}{e}{e} \BNFALT \ite{e}{e}{e} \BNFALT
                                      \operation{e^*} \BNFALT \qq{\phi} \BNFALT p(w^*) \\
    \text{Constants}    & k     &::=& \mathsf{true} \BNFALT \mathsf{false} \BNFALT \mathsf{0} \BNFALT
                                      \mathsf{1} \BNFALT \dots \\
    \text{SMT formulas} & \phi  &::=& \unq{e} \BNFALT \cFORALL(\phi, \phi) \BNFALT \cLET(\phi, \phi, \phi) \BNFALT
                                      \cCTOR[c](\phi^*) \BNFALT \dots \\
    \text{Wildcard}     & w     &::=& \wildcard \BNFALT e \\[0.25em]
    \text{Values}       & v \in \mathrm{Val}     &::=& k \BNFALT c(v^*) \\[1em]

    \multicolumn{4}{l}{\textbf{Namespaces}} \\ \hline
    
  \end{array} \]
	
	\sidebyside{
		\(\begin{array}{lrcl}
    \text{Data type names} & D &\in& \mathrm{ADTVar} \\
    \text{Type variables} & \alpha &\in& \mathrm{TVar} \\
    \text{Constructors} & c     &\in& \mathrm{CtorVar} \\
		\end{array}\)
	}{
		\(\begin{array}{lrcl}
    \text{Variables} & X &\in& \mathrm{Var} \\
    \text{Predicates} & p &\in& \mathrm{PredVar} \\
    \text{Functions} & f &\in& \mathrm{FunVar}
		\end{array}\)
	}

  \caption{\Name extends the abstract syntax of Datalog with type definitions,
  functions, SMT declarations, and a richer language of expressions.}
  \label{fig:grammar}
\end{figure}

\Name extends Datalog with a fragment of first-order ML and a language of SMT
formulas (Figure~\ref{fig:grammar}).
Accordingly, a program consists of Horn clauses, type and function definitions,
and SMT declarations.
The Horn clause fragment is the same as in Datalog, except with a richer
variety of expressions $e$ that can occur as arguments to predicates.

\paragraph{Type definitions}
\Name users can define ML-style algebraic data types, which can be
polymorphic and mutually recursive.
An algebraic data type definition consists of a list of type variables $\alpha$, a type
name $D$, and a list of constructors $c$ with their argument types $\tau$.
Section~\ref{sec:typesystem} explains \Name's type system in more detail; we
provide a brief sketch now.
Algebraic data types $D~\tau^*$, base types $B$, and type variables $\alpha$
are treated as {\em pre-types}; intuitively, a pre-type $t$ is the type of a
concrete (non-formula) term.
In addition to pre-types, there are types that represent SMT-relevant terms: a
$t$-valued SMT formula has type $\smt{t}$, a $t$-valued SMT variable has type
$\sym{t}$, and an SMT model --- a finite map from formula variables to concrete
terms --- has type $\MODEL$.
The \Name type system distinguishes the first three types where it is
computationally relevant (i.e., during concrete evaluation, where
confusing a $t$-valued formula for a concrete $t$ term might lead to a
computation getting stuck), and collapses them where it is not (i.e.,
during SMT evaluation, where there is no meaningful distinction
between a $t$-valued formula and a concrete $t$ value).
It also prevents SMT models, which are not representable as SMT expressions,
from flowing into SMT formulas.

\paragraph{Functions}
\Name supports ML-style function definitions, although functions are limited to
being first-order and are not first-class values.
They can be polymorphic and mutually recursive.

\paragraph{SMT declarations}
\Name users can declare uninterpreted functions and polymorphic uninterpreted
sorts.
An uninterpreted function amounts to a special constructor for building a purely symbolic
term of type $\smt{t}$ (for some pre-type $t$).
An uninterpreted sort amounts to a special symbolic pre-type $t$, where $t$ is
not inhabited by any value, but $\sym{t}$ and $\smt{t}$ are.

\paragraph{Expressions and formulas}
Expressions $e$ occur as function bodies and as predicate arguments in Horn
clauses.
Although Datalog traditionally limits ground terms to nullary constructors, we
admit $n$-ary constructors.
While this comes with the cost of possibly-diverging programs --- adding
$n$-ary constructors makes Datalog Turing-complete~\citep{recursivequery} ---
many recent Datalog variants allow complex terms, including
Souffl\'e~\citep{souffle2}, LogicBlox~\citep{logicblox}, and Flix
~\citep{flix}.
For us, complex terms provide a natural way to reify logical formulas, and they
also can be used to create data structures that make it easier to encode
certain analyses.

Additional \Name expressions include standard ML fare like constants (booleans,
strings, machine integers, and floats), function calls, and match, let, and
if-then-else expressions.
The expression $\operation{e^*}$ represents the application of a primitive
operator to a sequence of subexpressions.
These cover both basic arithmetic operations (e.g., addition) and SMT-specific
operations (e.g., checking for satisfiability, generating $\MODEL$s; see
Section~\ref{sec:useformulas}).

The expression $\qq{\phi}$ is a quasi-quoted SMT formula, where the language of
formulas $\phi$ consists of unquoted expressions $\unq{e}$ and formula
constructors of the form $\cSMT{c'}$ applied to SMT formulas.
Some of these constructors directly reflect SMT formula constructs; for
example, the constructor $\cFORALL$  builds a universally quantified formula,
and the constructor $\cLET$ builds an SMT let formula.
Formula constructors can appear only in formulas, and non-formula constructors
cannot appear directly in formulas.
We embed algebraic data type constructors in formulas using a family of
\emph{formula constructors}. Each formula constructor $\cCTOR[c]$ lifts the user-defined algebraic data type constructor $c$ to SMT.
Quotes are used to delineate formulas and trigger a different type checking
mode, in which the types $t$, $\smt{t}$, and $\sym{t}$ are conflated (with some
restrictions, as explained in Section~\ref{sec:typesystem}).
The unquote operator $\textbf{,}$ escapes from this type checking mode and
makes it possible to inject a non-formula expression into a formula.
Section~\ref{sec:formulas} discusses formulas in more detail.

We have already seen how the Datalog fragment of \Name can include expressions
from the ML fragment; the final expression $p(w^*)$ ties the loop by
providing a way for the ML fragment to reference the Datalog fragment.
The expression $p(w^*)$ acts like a function call that queries the
contents of the relation $p$.
Its exact behavior depends on its arguments, which are either expressions or
the special wildcard term \wildcard.
If its arguments contain no wildcards, then $p(e^*)$ returns a boolean
indicating whether the tuple identified by its arguments is in the $p$
relation.
If it has $k > 0$ wildcards, it returns a list of $k$-tuples: For each tuple
$v^*$ in the relation corresponding to $p$, there is a corresponding $k$-tuple
in this list that is $v^*$ projected to the wildcard positions; if there are
$n$ matching tuples in $p$, then the list is of length $n$.
In other words, given complete arguments, a predicate is really just a
predicate; given partial arguments with wildcards, a predicate is the multiset
consisting of matching tuples after they have been appropriately projected.%
\footnote{Multisets can arise if the ``anonymous'' variable \icfoot{_} is used
to project out unwanted columns. For example, given that exactly \icfoot{p(1,
2)} and \icfoot{p(3, 2)} hold, the expression \icfoot{p(_, ??)} would evaluate
to a multiset represented by the list \icfoot{[2, 2]}.}

\paragraph{Remarks}
Extending Datalog with our fragment of ML is not foundational, as
it can relatively easily be translated to Datalog rules (this would not
necessarily be the case for a higher-order fragment of ML).
However, despite the fact that the ML fragment could be treated as just
syntactic sugar,
it has a significant positive impact on the usability of \Name, as we argue in
Section~\ref{sec:eval}.

The concrete syntax of formulas in our prototype (and in the examples we give
in this paper) differs from the abstract syntax given here. % in two notable ways.
We differentiate between ML variables (initial lowercase) and Datalog variables (initial caps).
Algebraic data type constructors are allowed to appear directly in
formulas, and are implicitly lifted to the appropriate formula constructor (so
data type constructor $c$ is automatically lifted to $\cCTOR[c]$).
We do not support an explicit unquote operator; instead, we implicitly
unquote variables, constants, and invocations of nullary functions. %\michael{Function calls? Do we need an exta premise?}
We support additional features (records, locally scoped functions, etc.) that
can be easily compiled to the abstract syntax.

\subsection{Logical Formulas}\label{sec:formulas}

\Name uses data types and operators to support constructing and reasoning about
logical formulas.
\Name provides a library of data types that define logical terms.
Most of the time during evaluation, these terms are unremarkable and treated
just like any other ground term.
However, these terms are interpreted as logical formulas when they are used as
arguments to built-in operators that make calls to an external SMT solver.
In our current prototype, it is possible to create logical terms in first-order
logic extended with (fragments of) the SMT-LIB theories of uninterpreted
functions, integers, bit vectors, floating point numbers, arrays, and algebraic
data types~\citep{smtlib}, as well as the theory of strings shared by the SMT
solvers Z3~\citep{z3} and CVC4~\citep{cvc4}.

\subsubsection{Representing Formulas}\label{sec:repformulas}

\begin{figure*}[t!]
\begin{center}
$\begin{array}{lrcl}
  \text{Negation} & \ic{\~} & : & \ic{bool smt} \rightarrow \ic{bool smt}\\
  \text{Conjunction} & \ic{/\\} &
    : & (\ic{bool smt},~\ic{bool smt}) \rightarrow \ic{bool smt}\\
  \text{Implication} & \ic{==>} &
    : & (\ic{bool smt},~\ic{bool smt}) \rightarrow \ic{bool smt}\\
  \text{Equality} & \ic{smt_eq[$t$]} &
    : & (t~\ic{smt},~t~\ic{smt}) \rightarrow \ic{bool smt}\\
  \text{SMT variable} & \ic{smt_var[$t^\prime,t$]} &
    : & t^\prime \rightarrow t~\ic{sym} \\
  \text{Bit vector constant} & \ic{bv_const[$k$]} &
    : & \ic{bv[32]} \rightarrow \ic{bv[$k$] smt} \\
  \text{Bit vector addition} & \ic{bv_add} &
    : & (\ic{bv[$k$] smt},~\ic{bv[$k$] smt}) \rightarrow \ic{bv[$k$] smt} \\
  \end{array}$
\end{center}
    \caption{Logical formulas are created in \Name via built-in constructors, such
    as the ones shown here.}
    \label{fig:formulas}
\end{figure*}

Users create logical terms through constants and formula constructors.
For example, to represent the formula $False \implies True$, one would use the
term \ic{`false ==> true`}, where \ic{false} and \ic{true} are the standard
boolean values and \ic{==>} is the infix constructor for implication.

Our current prototype offers around 70 constructors for creating logical terms
ranging from symbolic string concatenation to logical quantifiers; others could
be added in the future.
Figure~\ref{fig:formulas} shows a sample of these constructors and their types.
Some constructors require explicit indices, either to guarantee that type
information is available at runtime when the formula is serialized to SMT-LIB,
or to make sure that the type of the arguments can be determined by the type of
the constructed term (which makes type checking easier).
For example, \ic{bv_const[$k$]} creates a symbolic $k$-bit-vector value from a
concrete 32-bit vector; at runtime, it is necessary to know the width $k$ so
that we can serialize it correctly.
The constructor \ic{smt_eq[$t$]}  denotes the
equality of two terms of type $t$ \ic{smt} (alternatively stated using the infix notation \ic{\#=}); here, the index makes sure that the type checker knows what types
the arguments should have.
A programmer typically does not need to provide these indices explicitly, as
they can often be inferred (our prototype does this).

\Name distinguishes between logic programming variables and formula variables.
A formula variable is a ground term that, when interpreted logically,
represents a symbolic value.
A term \ic{smt_var[$t^\prime,t$]($v$)} --- typically abbreviated as \ic{\#\{$v$\}[$t$]} ---
is a formula variable of type $t$ \ic{sym} identified
by a value $v$ of type $t^\prime$.
Intuitively, $v$ is the ``name'' of the variable.
The term \ic{\#\{$v$\}[$t$]} is guaranteed not to occur in $v$, which means
that the variable it represents is fresh with respect to the set of formula
variables in $v$; this makes it easy to deterministically construct a new
variable that is fresh with respect to an environment, a trick we use often in
our case studies.
For example, if \ic{X} is bound to a list of boolean formula variables, the
formula variable \ic{\#\{X\}[bool]} will not unify with any term in \ic{X}.
The shorthand \ic{\#$id$[$t$]} is equivalent to \ic{\#\{"$id$"\}[$t$]},
where $id$ is a syntactically valid identifier.

Importantly, because formula variables are ground terms, we can derive facts
containing formula variables without violating Datalog's range restriction,
which requires that every derived fact is variable-free.
This restriction enables efficient evaluation by simplifying table lookups, one
of the fundamental operations in Datalog evaluation.

\subsubsection{Using Formulas}\label{sec:useformulas}

\begin{figure*}[t!]
  \centering
  $\begin{array}{lrcl}
        \text{Satisfiability} & \ic{is_sat} &
            : & \ic{bool smt} \rightarrow \ic{bool}\\
                              & \ic{is_sat_opt} &
            : & (\ic{bool smt list},~\ic{bv[32] option}) \rightarrow \ic{bool option}\\
        \text{Validity} & \ic{is_valid} &
            : & \ic{bool smt} \rightarrow \ic{bool}\\
        \text{Model generation} & \ic{get_model} &
            : & (\ic{bool smt list},~\ic{bv[32] option}) \rightarrow \ic{model option}\\
        \text{Model inspection} & \ic{query_model} &
            : & (\ic{'a sym},~\ic{model}) \rightarrow \ic{'a option}
  \end{array}$
    \caption{\Name provides built-in operators for reasoning about logical terms.}
    \label{fig:logicfuncs}
\end{figure*}

Built-in operators provide a way to reason about logical terms as formulas
(Figure~\ref{fig:logicfuncs}).
When an operator in the SMT interface is invoked, its formula argument is
serialized into the SMT-LIB format and a call is made to an external SMT
solver.
These operators are assumed to act deterministically during a single \Name run;
an implementation can achieve this in the presence of a non-deterministic SMT
solver by memoizing operations.

For example, to test the validity of the principle of explosion (any
proposition follows from false premises), one could make the call
\ic{is_valid(`false ==> #x[bool]`)}.
Like other operators, the SMT interface operators can be invoked from the
bodies of rules, as here:
\begin{lstlisting}
ok :-  #x[bool] != #y[bool],
       is_sat(`#x[bool] #= #y[bool]`) = true,
       is_sat(`~(#x[bool] #= #y[bool])`) = true.
\end{lstlisting}
This rule derives the fact \ic{ok}: The term \ic{\#x[bool]} is not unifiable
with the term \ic{\#y[bool]}, since they are different formulas, representing different SMT variables. But these terms both may and may not be equal
when interpreted as formula variables via the operator \ic{is_sat}.
Within an invocation of \ic{is\_sat}, constraints are formed between
\ic{\#x[bool]} and \ic{\#y[bool]} --- in the first case they must be equal, and
in the second case they must not be --- but these constraints do not leak into
the larger context.
This is an intentional design decision and differs from the approach taken by
paradigms like constraint logic programming (see Section~\ref{sec:eval}).

\Name provides two sets of operators for testing the satisfiability and logical
validity of propositions.
In general, an SMT solver can return three possible answers to such a query:
``yes,'' ``no,'' and ``unknown.''
The operators \ic{is_sat} and \ic{is_valid} return booleans.
In the case that the backend SMT solver is not be able to determine whether a
formula $\phi$ is satisfiable, these operators fail (as explained in
Section~\ref{sec:typesystem}).
The operator \ic{is_sat_opt($\phi^*$, timeout)} provides more fine-grained
control: it takes a list of propositions (interpreted as conjuncts) and an
optional timeout, and returns an optional boolean, with \ic{none} corresponding
to ``unknown.''
While we suspect that the simpler versions will be sufficient for most
applications, this more complex version does allow applications to explicitly
handle the ``unknown'' case if need be (e.g., pruning paths in symbolic
execution).

The operator \ic{get_model} takes a list of propositions and an optional timeout; it
returns a model for the conjunction of the propositions if the SMT solver is able to find one in
time, and \ic{none} otherwise.
The values of formula variables in this model can be inspected using
\ic{query_model}, which returns \ic{none} if the variable does not occur free
in the formula or if a concrete value for it is not representable in \Name (for
example, \Name does not have a type for a concrete 13-bit vector).
The values of symbolic expressions can be indirectly extracted through formula
variables: Before finding the model, add the equality
$\ic{`}x~\ic{\#=}~e\ic{`}$ to the formula, where $x$ is a fresh formula
variable and $e$ is an expression; in the extracted satisfying model, $x$ will be assigned
the value of $e$ in that model. 

\subsubsection{Custom Types in Formulas}

\Name's algebraic data types can be reflected in SMT formulas via SMT-LIB's
support for algebraic data types.
Thus, \Name permits arbitrary term constructors to be used within logical
formulas.
For example, we can define a type \ic{foo} with a single nullary constructor
\ic{bar} and then write formulas involving \ic{foo}-valued terms:
\begin{lstlisting}
type foo = | bar
ok :- is_valid(`#x[foo] #= bar`) = true.
\end{lstlisting}
This program would derive the fact \ic{ok}: Since there is only one way to
construct a \ic{foo} --- through the constructor \ic{bar} --- any symbolic
value of type \ic{foo} must be the term \ic{bar}.

For each algebraic data type, we automatically generate two kinds of
constructors that make it easier to write formulas involving terms of that
type.
The first kind is a constructor tester.
For each constructor $c$ of a type $t$, \Name provides a constructor
\ic{\#is_$c$} of type $t$ \ic{smt} $\rightarrow$ \ic{bool smt}.
The proposition \ic{\#is_$c$($e$)} holds if the outermost constructor of $e$ is
$c$.
The second kind is an argument getter.
If $c$ is a constructor for type $t$ with $n$ arguments of types $t_i$ for $1
\le 1 \le n$, \Name generates $n$ argument getters of the form \ic{\#$c$_$i$},
where \ic{\#$c$_$i$} has the type $t$ \ic{smt} $\rightarrow t_i$ \ic{smt}.
When interpreted as a formula, the term \ic{\#$c$_$i$($e$)} represents the
value of the $i^{\text{th}}$ argument of $e$.
For example, we can state that a symbolic list of booleans is non-empty and its
first argument is \ic{true}:
$$\ic{`\#is_cons(\#x[bool list]) /\\ \#cons_1(\#x[bool list])`}$$
We could use the operator \ic{get\_model} to find a model of this satisfiable
formula; in this model, \ic{\#x[bool list]} might be assigned the concrete
value \ic{cons(true, nil)}.

\subsection{Operational Semantics}\label{sec:opsem_main}

\begin{figure}[t]
  \hdr{Namespaces and constructs}{}
  \sidebyside{
    \[\begin{array}{lrcl}
      \text{World} & \world &\in& \mathrm{PredVar} \rightarrow \mathcal{P}
                                  (\mathrm{Val}^*) \\
      \text{Substitution} & \theta     &\in& \mathrm{Var} \rightharpoonup \mathrm{Val} \\
    \end{array} \]
  }{
    \[\begin{array}{lrcl}
      \text{Error} & \bot &\in& \mathrm{Err} \\
      \text{$u$-term} & u &::=& X \BNFALT k \BNFALT c(\vec{u_i}) \\
    \end{array} \]
  }

  \hdr{Clause semantics}{ \quad \fbox{$\vec{F}; \world \vdash
    H \stepsto \world_\bot$} }
  
  \infrule[Clause]{
     |\vec{P_i}| = n \quad
     \theta_0 = \cdot \quad
     \forall i \in [0,n),~\theta_i \vdash P_{i} \stepsto \theta_{i+1}
  }{
    \vec{F}; \world \vdash p(\vec{X_j}) \horn \vec{P_i} \stepsto \world[p \mapsto \world(p) \cup \{\theta_n(\vec{X_j})\}]
  }

  \hdr{Premise semantics}{
    \quad \fbox{$\implicit{\vec{F}; {}} \world; \theta \vdash P \stepsto \theta_\bot$}
  }
  
  ~ \\[0.25em]

  \sidebyside
  {
    \infrule[PosAtom]{
      \vec{v} \in \world(p) \andalso
      \theta \vdash \vec{X} \Unify \vec{v} : \theta'_\bot
    }{
      \world; \theta \vdash p(\vec{X}) \stepsto \theta'_\bot
    }
  }{
    \infrule[EqCtor]{
      \theta \vdash Y \unify c(\vec{X}) : \theta'_\bot
    }{
      \world; \theta \vdash {Y = c(\vec{X})} \stepsto \theta'_\bot
    }
  }

  \hdr{Expression semantics}{
    \quad \fbox{$\implicit{\vec{F}; {}} \world; \theta \vdash e \StepstoE v_\bot$}
    \quad \fbox{$\implicit{\vec{F}; {}} \world; \theta \vdash \vec{e} \StepstoVecE \vec{v}_\bot$}
  }

  ~ \\[0.25em]

    \sidebyside[.4][.5]
    {
      \infrule[$\StepstoE$-Op]{
        \world; \theta \vdash \vec{e} \StepstoVecE \vec{v}
        \andalso \denot{\otimes}(\vec{v}) = v
      }{
        \world; \theta \vdash \operation{\vec{e}} \StepstoE v
      }
    }
    {
      \infrule[$\StepstoE$-Quote]{
        \world; \theta \vdash \phi \StepstoPhi v_\bot
      }{
        \world; \theta \vdash \qq{\phi} \StepstoE v_\bot
      }
    }

  \hdr{Formula semantics}{
    \quad \fbox{$\implicit{\vec{F}; {}} \world; \theta \vdash \phi \StepstoPhi v_\bot$}
    \quad \fbox{$\implicit{\vec{F}; {}} \world; \theta \vdash \vec{\phi} \StepstoVecPhi \vec{v}_\bot$}
  }

  ~ \\[0.25em]

  \sidebyside
  {
    \infrule[$\StepstoPhi$-Ctor]{
      \world; \theta \vdash \vec{\phi} \StepstoVecPhi \vec{v}
    }{
      \world; \theta \vdash \cSMT{c'}(\vec{\phi}) \StepstoPhi \cSMT{c'}(\vec{v})
    }
  }
  {
    \infrule[$\StepstoPhi$-Unquote]{
      \world; \theta \vdash e \StepstoE v
    }{
      \world; \theta \vdash \unq{e} \StepstoPhi \tosmt(v)
    }
  }

  \hdr{SMT conversion}{
    \quad \fbox{$\tosmt(v) = v$}
  }

  \sidebyside{
  \[ 
   \begin{array}{@{}rcl@{}}
     \tosmt(\cLET(v_1, v_2, v_3)) &=& \cLET(v_1, v_2, v_3) \\
     \tosmt(\cFORALL(v_1, v_2)) &=& \cFORALL(v_1, v_2) \\
    \end{array} 
  \]
  }{
  \[
   \begin{array}{@{}rcl@{}}
     \tosmt(c(\vec{v_i})) &=& \cCTOR[c](\overrightarrow{\tosmt(v_i)}) \\
     &\dots&
    \end{array} 
  \]
  }

  \caption{A fragment of \Name's operational semantics\iffull{} (see Appendix~\ref{sec:opsem} for full formalization)\fi.}
  \label{fig:opsem1}
\end{figure}

This section presents \Name's operational semantics, making reference
to a selection of the formal rules (Figures~\ref{fig:opsem1}).%
\footnote{%
\Name can also be given a model-theoretic semantics: because the ML
features can be desugared into Datalog rules, the model theory of \Name is
essentially that of stratified Datalog.
\iffull Appendix~\ref{sec:modeltheory} sketches this out further.
\else
The extended version of this paper~\citep{formulog_ext} sketches this out further,
and includes the full formalism for the operational semantics.\fi}%
\footnote{In the boxed rule schemata, implicit parameters are in
\implicit{gray}; we conserve space by stating the rules without threading
implicit parameters through, which are unchanging.
We write $\vec{x_i}$ for some metavariable $x$ to mean a possibly empty sequence
of $x$s indexed by $i$, and write $S_\bot$ for some set $S$ to mean the set $S +
\mathrm{Err}$.}
\Name imposes the standard stratification requirements upon programs: no
recursive dependencies involving negation or aggregation between relations.
As a stratifiable program can be evaluated one stratum at a time, we focus on
the evaluation of a single stratum.

A stratum is evaluated by repeatedly evaluating its Horn clauses until no new
inferences can be made.
The semantics of a Horn clause $H$ is defined through the judgment $\vec{F};
\world \vdash H \stepsto \world_\bot$, where a world \world\ is a map from
predicate symbols to sets of tuples (i.e., those that have been derived so
far).
A Horn clause takes a world to either a new world or the error value $\bot$.
Going wrong can result for two reasons: either because a variable is unbound at
a point where it needs to be bound, or because an operator is applied to a
value outside of its domain.
It is important to distinguish between a rule going wrong and a rule failing to
complete because two terms fail to unify:
The first is an undesirable error (ruled out by our type system), whereas the
second is expected behavior.

A rule is evaluated by evaluating its premises one-by-one, using a
left-to-right order (\rn{Clause}).
The judgment $\vec{F}; \world; \theta \vdash P \stepsto \theta_\bot$ defines
the semantics of a premise, which takes a world and a substitution $\theta$ (a
partial function from variables to values) and returns a new substitution or
an error.
The substitution produced by one premise is used as the input to the next one.
A successful inference extends the input world with a (potentially novel) tuple
$\theta_n(\vec{X_j})$, i.e., the result of element-wise applying the
substitution produced by the rightmost premise to the variables in the head of
the rule.
Clause evaluation goes wrong if the evaluation of one of the premises goes
wrong.

Without loss of generality, we assume that premises occur in a limited form:
predicates are applied to only variables, written $p(\vec{X_i})$, and
equality predicates bind variables, as in $Y = e$.
(Our prototype similarly desugars premises.)
An atom $p(\vec{X})$ is evaluated by non-deterministically choosing a tuple
$\vec{v}$ from the tuples in $\world(p)$, and then pairwise unifying its
elements with the variables $\vec{X}$ (\rn{PosAtom}).
The premise $Y = c(\vec{X})$ unifies its two terms (\rn{EqCtor}).
The judgment $\theta \vdash u_1 \unify u_2 : \theta_\bot$ defines the
unification of terms $u_1$ and $u_2$ under the substitution $\theta$; it
results in an error if $u_1$ and $u_2$ both contain unbound variables, and a
new substitution if they are otherwise unifiable.

%The semantics for many \Name expressions are standard.
Most expressions have standard semantics.
An operator produces a value if its arguments are evaluated
to values in its domain (\rn{$\StepstoE$-Op}); it goes wrong if the argument
values are outside its domain, e.g., if a string and number are added together.
A quoted formula \qq{\phi} evaluates to whatever $\phi$ evaluates to
(\rn{$\StepstoE$-Quote}).
Formula $\cSMT{c'}(\vec{\phi})$ evaluates to formula $\cSMT{c'}(\vec{v})$ if
arguments $\vec{\phi}$ evaluate to values $\vec{v}$ (\rn{$\StepstoPhi$-Ctor}).
If the expression $e$ evaluates to the value $v$, then the formula \unq{e}
evaluates to the term $\tosmt(v)$ (\rn{$\StepstoPhi$-Unquote}), where the
function $\tosmt$ lifts a term to its formula version.

\section{Type system}\label{sec:typesystem}

\Name's type system is designed to meet three desiderata.
The first desideratum is that concrete evaluation should never go wrong, which
might happen if an operator is applied to an
operand outside its domain or a variable is unbound at a point when it needs to
be evaluated.
The second desideratum is that SMT solving should never go wrong, which might
happen if a term that does not represent a well-sorted formula under the
SMT-LIB standard reaches the external SMT solver (e.g., a formula representing
the addition of a 16-bit vector and 32-bit vector).
The third desideratum is that the type system should make it easy to construct
expressive logical formulas, including formulas that involve terms drawn from
user-defined types.

There is some tension between the first and third of these desiderata.
The first one requires that we differentiate between, for example, a concrete
bit-vector value and a symbolic bit-vector value (e.g., a bit-vector-valued
formula) since an operator that is expecting a concrete bit vector might get
stuck if its argument is a symbolic bit vector.
For instance, we want to rule out this program:
\begin{exmp}[A bad program we would want to reject]
\label{example:not_ok}
\begin{lstlisting}

type foo = | bar(bv[32])
fun f(x: foo) : bv[32] = match x with bar(y) => y + y end
not_ok :-  X = #x[bv[32]],
           f(bar(X)) = 42.
\end{lstlisting}
\end{exmp}
\noindent This program gets stuck evaluating \ic{f(bar(X))}, since \ic{y} is
bound to a symbolic value in \ic{f} but the ML fragment's addition operator
needs concrete arguments.
On the other hand, we are able to construct more expressive formulas if we can
occasionally conflate concrete and symbolic expressions:
\begin{exmp}[A good program we would want to accept]
\label{example:ok}
\begin{lstlisting}

ok :-  X = #x[bv[32]],
       is_sat(`bar(X) #= bar(5)`) = true.
\end{lstlisting}
\end{exmp}
\noindent This rule asks whether there exists a symbolic bit vector $x$ such
that $\ic{bar($x$)}$ equals \ic{bar(5)}, where \ic{bar} is the constructor
defined above.
This reasonable formula is not well-typed under a type system that uniformly
distinguishes between concrete and symbolic values, since the constructor
\ic{bar} expects a concrete bit vector argument but instead receives the
symbolic one $x$.

\begin{figure}[t]

  \hdr{Contexts}{}
  \vspace{-10pt}
  \[ \begin{array}{lrcl}
    \text{Data type declarations} & \Delta &::=& \cdot \BNFALT \Delta, D:\forall \vec{\alpha_i}. ~ \{ \overrightarrow{c_j : \vec{\tau_k}} \} \\
    \text{Program declarations}  & \Phi   &::=&
      \cdot \BNFALT
      \Phi, f:\forall \vec{\alpha}, \vec{\tau} \rightarrow \tau \BNFALT
      \Phi, p \subseteq \vec{\tau} \\
    \text{Variable contexts}    & \Gamma &::=&
      \cdot \BNFALT
      \Gamma, x:\tau \BNFALT
      \Gamma, \alpha \\
  \end{array} \]

  \hdr{Clause typing}{
    \quad \fbox{$\implicit{\Delta; \Phi}  \vdash H$}
  }
  
  ~ \\[0.25em]
  
  \infrule[$H$-Clause]{
     \cdot \vdash P_0 \rhd \Gamma_1 \andalso \dots \andalso
     \Gamma_j \vdash P_j \rhd \Gamma_{j+1} \andalso \dots \andalso
     \Gamma_n \vdash P_n \rhd \Gamma' \\
     p \subseteq \vec{\tau_i} \in \Phi \andalso
     \Gamma' \vdash \vec{X_i}, \vec{\tau_i} \rhd \Gamma'
  }{
    \vdash p(\vec{X_i}) \horn \vec{P_j}
  }

  ~ \\[0.25em]

  \hdr{Variable binding and typing}{
    \quad \fbox{$\Gamma \vdash x,\tau \rhd \Gamma$}
    \quad \fbox{$\Gamma \vdash \vec{x},\vec{\tau} \rhd \Gamma$}
  }

  ~ \\[0.25em]

  \sidebyside
   {
    \infrule[$X\tau$-Bind]{
      X \not \in \dom(\Gamma)
    }{
      \Gamma \vdash X, \tau \rhd \Gamma, X \mathord{:} \tau
    }
  }{
    \infrule[$X\tau$-Check]{
      \Gamma(X) = \tau
    }{
      \Gamma \vdash X, \tau \rhd \Gamma
    }
  }

  ~ \\[0.25em]

  \hdr{Premise typing}{
    \quad \fbox{$\implicit{\Delta; \Phi; {}} \Gamma \vdash P \rhd \Gamma$}
  }
  
  ~ \\[0.25em]

  \sidebyside[.44][.56]
  {
    \infrule[$P$-PosAtom]{
      p \subseteq \vec{\tau_i} \in \Phi \andalso
      \Gamma \vdash \vec{X_i}, \vec{\tau_i} \rhd \Gamma'
    }{
      \Gamma \vdash p(\vec{X_i}) \rhd \Gamma'
    }
  }{
    \infrule[$P$-Eq-FB]{
      \Gamma \vdash e : \tau \andalso
      \Gamma \vdash Y, \tau \rhd \Gamma'
    }{
      \Gamma \vdash {Y = e} \rhd \Gamma'
    }
  }

  ~ \\[0.25em]

  \hdr{Function and expression well
    formedness}{ \quad \fbox{$\implicit{\Delta; \Phi} \vdash
    F$} \quad \fbox{$\implicit{\Delta; \Phi; {}} \Gamma \vdash e
    : \tau$} }

  ~ \\[0.25em]

  \sidebyside{
    {\infrule[$e$-Op]
       {\typeof(\otimes) = \vec{\tau_i} \rightarrow \tau \andalso
        \Gamma \vdash e_i : \tau_i}
       {\Gamma \vdash \operation{\vec{e_i}} : \tau}}
  }{
  \infrule[$e$-Quote]
   {\Gamma \vdash \phi : \tau}
   {\Gamma \vdash \qq{\phi} : \tau}
  }

  ~ \\[0.25em]

  \hdr{SMT constructors and formula well formedness}{
    \fbox{$\implicit{\Delta; \Phi; {}} \Gamma \vdash \cSMT{...} : \vec{\tau_i} \rightarrow \tau$}
    \quad
    \fbox{$\implicit{\Delta; \Phi; {}} \Gamma \vdash \phi : \tau$}
  }

  ~ \\[0.25em]

  \threesidebyside[.4][.25][.25]
  {
    \infrule[$\phi$-Ctor]
    {\Gamma \vdash \cSMT{c} : \vec{\tau_i} \rightarrow \tau \andalso
     \Gamma \vdash \phi_i : \tau_i
    }
    {\Gamma \vdash \cSMT{c}(\vec{\phi_i}) : \tau}
  }
  {
    \infrule[$\phi$-Unquote]
    {\Gamma \vdash e : \tau}
    {\Gamma \vdash \unq{e} : \tosmt(\tau)}
  }
  {
    \infrule[$\phi$-Promote]
    {\Gamma \vdash \phi : \sym{t}}
    {\Gamma \vdash \phi : \smt{t}}
  }

  ~ \\[0.25em]

  \hdr{SMT representations}{
    \quad \fbox{$\erase(\tau) = t$}
    \quad \fbox{$\tosmt(\tau) = \tau$}}

  \threesidebyside[.3][.27][.35][t]
  {\[\begin{array}{rcl}
      \erase(B) &=& B \\
      \erase(D ~ \vec{\tau_i}) &=& D ~ \overrightarrow{\erase(\tau_i)} \\
    \end{array}\]}
  {\[\begin{array}{rcl}
      \erase(\smt{t}) &=& \erase(t) \\
      \erase(\sym{t}) &=& \erase(t) \\
    \end{array}\]}
  {\[\begin{array}{rcl}
      \tosmt(t) &=& \smt{\erase(t)} \\
      \tosmt(\smt{t}) &=& \smt{\erase(t)} \\
      \tosmt(\sym{t}) &=& \sym{\erase(t)} \\
    \end{array}\]}

  \caption{A fragment of \Name's type system\iffull{} (see Appendix~\ref{sec:types} for full formalization)\fi.}
  \label{fig:typesystem}
\end{figure}

\Name resolves the tension between these desiderata through a bimodal type
system that acts differently inside and outside formulas (which are demarcated
by quotations).
In essence, the \Name type system differentiates between the pre-type $t$, the
SMT formula type $\smt{t}$, and the SMT variable type $\sym{t}$ outside of
formulas, but typically conflates them within formulas.%
\footnote{It does not conflate them in binding positions where formula
variables are required, such as in quantifiers.}
This bimodal approach disallows Example~\ref{example:not_ok} (since outside a
formula, a term of type $\sym{\bv{32}}$ cannot be used where a term of type
$\bv{32}$ is expected), while permitting Example~\ref{example:ok} (since within
a formula, a term of type $\sym{\bv{32}}$ can be used anywhere a term of type
$\bv{32}$ is expected).

Intuitively, this bimodal approach is safe because it distinguishes between
concrete and symbolic values during concrete evaluation --- where conflating them
might lead to going wrong --- and conflates them only during SMT evaluation,
where the distinction is not meaningful.
We have formalized the \Name type system and proven it sound with respect to
the operational semantics of \Name.
We present only a small subset of it here (Figure~\ref{fig:typesystem})\iffull; the
full system is in Appendix~\ref{sec:types}.\else%
.\footnote{See the extended version of this paper~\citep{formulog_ext} for the full formalism and proofs.}\fi

The rule defining a well-typed Horn clause (\rn{$H$-Clause}) depends on two
notable judgments.
The premise typing judgment $\Gamma \vdash P \rhd \Gamma'$ takes a variable
typing context $\Gamma$ and a premise $P$ and produces a new variable typing
context $\Gamma'$.
The variable binding and typing judgment $\Gamma \vdash x,\tau \rhd \Gamma'$
holds if either $X$ is not in $\Gamma$, in which case $\Gamma'$ extends
$\Gamma$ with $X$ mapped to $\tau$ (\rn{$X\tau$-Bind}), or $X$ is mapped to
$\tau$ by $\Gamma$, in which case $\Gamma=\Gamma'$ (\rn{$X\tau$-Check}).
As can be seen from rule \rn{$H$-clause}, the type checking of Horn clauses is
flow-sensitive and proceeds left-to-right across the clause, with the
``output'' context of checking premise $P_i$ used as the ``input'' context for
checking premise $P_{i+1}$.
This left-to-right type checking mirrors the left-to-right evaluation strategy
\Name uses; this is important for ensuring that variables are bound at the
correct points.%
\footnote{The fact that the operational semantics and type system assume a
certain order of evaluation does not prohibit a \Name runtime from reordering
premises within rules (for example, when applying database-style query planning
optimizations); it just needs to check that the new order is also well typed.
This type of rewriting does not affect the result of running the
rule provided that all subexpressions terminate (an assumption we make).
}
The second line of premises in rule \rn{$H$-clause} ensures that every variable
in the head of the rule is bound at the type specified by the head relation's
signature.

A positive atom is well typed if each of its variable arguments has the type
given to that argument by the relation's signature (\rn{$P$-PosAtom}).
A premise of the form $Y = e$ is typed according to a few different rules
depending on which side of the equation is ground with respect to the input
context $\Gamma$.
The key is that our type system only types premises of the form $Y = e$ when
unification is guaranteed to not go wrong at runtime.

The typing rules for most expressions are standard.
An operation is well-typed if its arguments match its type signature
(\rn{$\phi$-Ctor}).
A quoted formula $\qq{\phi}$ types at whatever $\phi$ types at
(\rn{$e$-Quote}).
The formula constructor $\cSMT{c}$ is well typed if the types of its arguments
match its type signature (\rn{$\phi$-Ctor}); in the case of a constructor for
an algebraic data type that has been lifted to a formula constructor, that
signature will require the constructed term and all of its arguments to have
types of the form $\smt{t}$.
If an expression types at $\tau$, then the formula $\unq{e}$ types at
$\tosmt(\tau)$ (\rn{$\phi$-Unquote}).
The helper function $\tosmt$ lifts a type to a formula type; for example, it
lifts $\bool$ to $\smt{\bool}$.
The typing rules for formulas also include a rule promoting from $\sym{t}$ to
$\smt{t}$, reflecting the fact that, within a formula, a $t$-valued formula
variable can be used anywhere a $t$-valued formula can be.%
\footnote{The opposite is not true, since some formula constructors (i.e.,
quantifiers and let expressions) bind formula variables.}

Type soundness with respect to the semantics of \Name comes from safety and
preservation:
\begin{theorem}[Safety]
\label{thm:main-typesafe-program-safe}
  If $\Delta; \Phi \vdash \vec{F_i} ~ \vec{H_j}$ and $\Delta; \Phi \models \world$ then
  for all $H \in \vec{H_j}, ~ \neg (\vec{F_i}; \world \vdash H \stepsto \bot)$.
\end{theorem}%
\begin{theorem}[Preservation]
\label{thm:main-typesafe-program-pres}
  If $\Delta; \Phi \vdash \vec{F_i} ~ \vec{H_j}$ and $\Delta; \Phi \models \world$ and
  $\vec{F_i}; \world \vdash H \stepsto \world'$ for some $H \in \vec{H_j}$ then
  $\Delta; \Phi \models \world'$.
\end{theorem}%
Safety (Theorem~\ref{thm:main-typesafe-program-safe}) guarantees that a Horn
clause from a well-typed program, evaluated on a well-typed world (i.e., one
where all the tuples have the right types), cannot step to error.
Thus, safety means that an operator is never applied to an operand outside its
domain, and a variable is never unbound when it needs to be bound. 
Preservation (Theorem~\ref{thm:main-typesafe-program-safe}) guarantees that if
a Horn clause, from a well-typed program, is evaluated on a well-typed world
and results in a new world, then that new world is also well-typed.
Taken together, these theorems imply that a well-typed \Name program does not
go wrong during concrete evaluation\iffull{} (see Appendix~\ref{sec:metatheory-safety}
for proofs)\fi.

The type system is sound with respect to the semantics of SMT-LIB because the
types of the formula constructors provided by \Name are consistent with the
types given by the SMT-LIB standard. The \Name type system guarantees that, at
runtime, terms (including formulas) are well-typed, and the type system
prevents terms that are not representable in SMT (such as those of type \MODEL)
from flowing into SMT formulas.
We distinguish between SMT-compatible types and non-SMT types formally
by indexing the type well formedness judgment with a \emph{mode}, which is
either $\mSMT$ (for those types that can be sent to the solver) or $\mEXP$ (for
those types that cannot).
It is fair to think of these modes as kinds with a subkinding
relationship: types of kind $\mSMT$ can safely be treated as general types of kind
$\mEXP$, but not the other way round\iffull{} (Lemma~\ref{lem:lift-msmt})\fi.

\paragraph{Assumptions}
An actual implementation of \Name, such as our prototype, has to contend with a
few sources of going wrong that are not captured in our formal model.
First, our model assumes that patterns in match clauses are exhaustive; this
is just for simplicity, and could be statically checked using standard
algorithms.
Second, our model assumes that operators are total with respect to terms with
the correct type.
There are three places where this assumption might break:
1) division or remainder by zero;
2) the operators \ic{is\_sat} or \ic{is\_valid} may induce an ``unknown''
response from the external SMT solver; and
3) the SMT solver may reject patterns used in trigger-based quantifier
instantiation that it considers to be ill formed (for example, if the pattern
contains a binding operation).
The first case is standard for many languages; the second can be avoided if the
programmer uses the option-returning SMT operator \ic{is\_sat\_opt}.
The last case would be hard to check statically; however, an implementation could
dynamically check patterns before making a call to the SMT solver, dropping
invalid patterns and issuing a warning to the user.
Our prototype uses ``hard exceptions'' by default, aborting the program.
We also support a ``soft exception'' mode, which treats all these cases
analogously to unification failures, halting execution on the current path but
allowing execution on other paths to continue.

\section{Implementation and case studies}\label{sec:examples}

In this section we briefly describe our prototype implementation of \Name, and
then discuss three analyses we have built as case studies: refinement type
checking, bottom-up points-to analysis, and bounded symbolic evaluation.

\subsection{Prototype}\label{sec:prototype}

Our prototype runtime ($\sim$17.5K LOC Java) works in five stages: parsing,
type checking, rewriting (for query specialization), validation, and evaluation.
Stratification and the range restriction are checked during the validation
phase.
Our parallel implementation of semi-naive evaluation~\citep{bancilhon1986naive}
uses a work-stealing thread pool; worker threads dispatch SMT queries to external solvers (Z3 by default~\citep{z3}).
Our prototype is feature complete, but not very optimized.%
\footnote{Our prototype is available at
\url{https://github.com/HarvardPL/formulog}.}

Unless otherwise noted, we ran experiments on an Ubuntu Server 16.04 LTS
machine with a 3.1 GHz Intel Xeon Platinum 8175 processor (24 physical CPUs,
each hyperthreaded) and 192 GiB of memory.
We configured our \Name runtime to use up to 40 threads and up to 40 Z3
instances (v4.8.7); all comparison systems were set to use the same version of
Z3 (with one exception, noted later).
For each result, we report the median of three trials.%
\footnote{For each case study, we use a tool to translate the input programs
into \Name facts.  We do not include these times, which are typically quite
short. Extracting libraries can take a few minutes, but this needs to be done
only once per library.}
Times are given as minutes:seconds.

\subsection{Refinement Type Checking}\label{sec:dminor}

We have implemented a type checker in \Name for Dminor, a first-order
functional programming language for data processing that combines refinement
types with dynamic type tests~\citep{dminor}.
This type system can, e.g., prove that
$$\ic{x in Int ? x : (x ? 1 : 0)}$$
type checks as \ic{Int} in a context in which \ic{x} has the union type
\ic{(Int|Bool)}.
Proving this entails encoding types and expressions as logical formulas and
invoking an SMT solver over these formulas.
We built a type checker for Dminor by almost directly translating the formal
inference rules used to describe the bidirectional Dminor type system.
In fact, we programmed so closely to the formalism that debugging an infinite
loop in our implementation helped us, along with the Dminor authors, uncover a
subtle typo in the formal presentation!
Our Dminor type checker is 1.2K lines of \Name.
The implementation of Bierman et al. is 3.2K lines of F$^\sharp$ and 400 lines
of SMT-LIB; we estimate that the functionality we implemented accounts for over
two thousand of these lines.%
\footnote{The reference implementation is closed source; the authors have
kindly provided us with line counts for each file.}

The encoding of Dminor types and expressions is complex, requiring
uninterpreted sorts, uninterpreted functions, universally quantified axioms,
and arrays (among other features).
The fact that we were able to code this relatively concisely speaks to the
expressiveness of \Name's formula language.
For example, Figure~\ref{fig:accum} shows an axiom describing the denotation of
the base case of a Dminor accumulate expression, which is essentially a fold
over a multiset.
Here, the type \ic{closure} is an uninterpreted sort, \ic{enc\_val} is an
algebraic data type that represents an encoded Dminor value, and \ic{accum} and
\ic{v\_zero} are uninterpreted functions, where the latter represents an empty
multiset.

\begin{figure}[!t]
    \begin{lstlisting}
fun accum_nil_axiom : bool smt =
  let (f, i) = (#func[closure], #init[enc_val]) in
  `forall f, i : accum(f, v_zero, i). accum(f, v_zero, i) #= i`
    \end{lstlisting}
		\caption{This axiom encodes the denotation of a Dminor accumulate
  expression over an empty multiset. The term in the formula between \ic{:} 
  and \ic{.} is a quantifier pattern~\citep{simplify}.}
    \label{fig:accum}
\end{figure}

We defined a set of mutually-recursive functions that encode expressions,
environments, and types.
For example, the type encoding function (fragment, Figure~\ref{fig:encode_type})
takes a type $\tau$ and an (encoded) Dminor value $v$, and returns two
propositions.
The first is true when $v$ has type $\tau$.
The second is a conjunction of axioms: new axioms are created to describe the
denotation of the bodies of accumulate expressions as they are encountered when
encoding expressions.
The first case in the figure encodes the fact that any value has type \ic{Any}.
The second one says that a value has type \ic{Bool} if it is constructed using
the constructor \ic{ev\_bool}; the constructor \ic{#is\_ev\_bool} is an
automatically-generated constructor tester.
The third case handles multiset types.
It creates a fresh encoded value \ic{x}, uses \ic{x} to recursively create a
proposition representing the encoding of the type \ic{s} of items in the
multiset, and then returns a proposition requiring the value to be a ``good''
collection (defined using the uninterpreted function \ic{good\_c}) and every
item in the multiset to have type \ic{s} (where \ic{mem} is another
uninterpreted function).

\begin{figure}[!t]
    \begin{lstlisting}
fun encode_type(t: typ, v: enc_val smt) : bool smt * bool smt =
  match t with
  | t_any => (`true`, `true`)
  | t_bool => (`#is_ev_bool(v)`, `true`)
  | t_coll(s) =>
    let x = #{(s, v)}[enc_val] in
    let (phi, ax) = encode_type(s, `x`) in
    (`good_c(v) /\ forall x : mem(x, v). mem(x, v) ==> phi`, ax)
    \end{lstlisting}
    \caption{This function (fragment) constructs a formula capturing the
    logical denotation of a Dminor type.}
    \label{fig:encode_type}
\end{figure}

Although we use ML-style functions to define the logical denotation of
expressions, environments, and types, we use logic programming rules to define
the bidirectional type checker, which allows us to write rules that are very
similar to the inference rules given in the paper.
Figure~\ref{fig:subtype} gives the one rule defining the subtype relation:
\ic{T} is a subtype of \ic{T1} in environment \ic{Env} if \ic{T1} is
well formed and the denotation of \ic{T}, given our axioms and the denotation
of \ic{Env}, implies the denotation of \ic{T1}.
This rule is an almost exact translation of the inference rule given in the
paper.

\begin{figure}[!t]
    \begin{lstlisting}
subtype(Env, T, T1) :-
  type_wf(Env, T1),
  encode_env(Env) = Phi_env,
  X = `#{(Env, T, T1)}[enc_val]`,
  encode_type(T, X) = (Phi_t, Axioms1),
  encode_type(T1, X) = (Phi_t1, Axioms2),
  Premises = [Phi_t, Phi_env, Axioms2, Axioms1, axiomatization],
  is_sat_opt(`~Phi_t1` :: Premises, z3_timeout) = some(false).
    \end{lstlisting}
    \caption{This rule defines Dminor's semantic subtyping relation. It uses
    the operator \ic{is\_sat\_opt} instead of \ic{is\_valid} because its SMT
    queries can sometimes result in ``unknown.''}
    \label{fig:subtype}
\end{figure}

Finally, the type checker needs to ensure that any expressions that occur
in refinements are pure (i.e., terminate and are deterministic).
We have written a termination checker based on the size-change
principle~\cite{sizechange}.
Our implementation is another good example of the synergy between ML-style
functions and Datalog rules, as we use the former to define the
composition of two size-change graphs and use the latter to find the fixed
point of composing size-change graphs.

We tested our type checker on six of the sample programs included in the Dminor
documentation (the other three examples make use of a feature --- the ability
to generate an instance of a type --- that we did not implement, although it
should be possible to do so; to the best of our knowledge, these are the only
publicly available Dminor programs).
We combined these examples into a single aggregate program of $\sim$150 LOC.
The reference implementation type checked this program in 1.5 seconds using an
optimization that tries syntactic subtyping before semantic subtyping; with
this optimization disabled, it took 3.6 seconds.%
\footnote{Here we used a machine with Microsoft Windows Server 2019 and the
same hardware specs as our Ubuntu machine.}
Our implementation completed in 4.7 seconds; it did not use this optimization
(which is not detailed in the paper), but did use a newer version of Z3.
Thanks to parallelization, our implementation automatically scaled to larger
programs: On a synthetic program consisting of ten copies of the original
aggregate program, it completed in 19.8 seconds (2.0 seconds per program copy);
on a synthetic program consisting of 100 copies, it completed in 153.6 seconds
(1.5 seconds per program copy).
In contrast, the reference implementation did not scale: even with the
syntactic-subtyping optimization enabled, it took 68 seconds on the ten-copy
program and over 100 minutes on the 100-copy program.

\subsection{Bottom-up Points-to Analysis}
\label{sec:pointsto}

We have implemented the bottom-up context-sensitive points-to analysis for Java
proposed by \citet{bottomupPointsto}.
A points-to analysis computes a static approximation of the objects that stack
variables and heap locations can point to at runtime.
A bottom-up points-to analysis does this through constructing method summaries
that describe the effect of a method on the heap; it is bottom-up in the sense
that summaries are propagated up the call graph, from callees to callers.

In Feng et al.'s algorithm, a method summary is an abstract heap that maps
abstract locations to heap objects, where an abstract location might be a stack
variable, an explicitly allocated heap object, or an argument-derived heap
location.
Edges in the abstract heap are labeled with logical formulas that describe
the conditions under which the edges hold;
when a method summary is instantiated at a call site, a constraint solver can
be used to filter out edges with unsatisfiable labels.

Feng et al.'s tool based on this algorithm, Scuba, is $\sim$15K lines of Java,
builds on the Chord analysis framework~\citep{chord}, and uses Z3 to discharge
constraints.
As for many realistic static analysis tools, there is a gap between what is
implemented in Scuba and the formal specification of the analysis.
This is partly because Scuba is written in Java: Object-oriented programming
does not naturally capture inference rules, the form of the specification.
In contrast, our \Name implementation, which is $\sim$1.5K LOC, closely mirrors
the inference rules.
For example, we can directly state how a points-to edge is instantiated at a
call site (Figure~\ref{fig:ptsto}), one step of summary instantiation, a
complex process defined through half a dozen mutually recursive relations that
need to be computed as a fixed point.
The Java code for encoding this logic is more complex and further from the
formal specification.
Programming close to the specification also helps check the specifications' correctness: while implementing in \Name one of the judgments specified by
Feng et al., we discovered an inconsistency between the judgment's definition
and its type signature.

Scuba employs a range of sophisticated heuristics that are essential to making
the algorithm perform in practice, as they tune precision to achieve scalability.
Some go far beyond the algorithm described in the paper and are interesting in
their own right.
Our implementation uses some heuristics based on the ones in Scuba.
The fact that we were able to implement useful heuristics --- a necessity for a
realistic static analysis tool --- argues for the practicality of \Name.
Moreover, we were able to do so such that our code still closely
reflects the core algorithm specified in the paper.

\begin{figure}[!t]
    \begin{lstlisting}
instantiate_ptsto(C, O1, Phi1, O2, widen(C, Phi_all)) :-
  instantiate_loc(C, heap(O1), heap(O2), Phi2),
  instantiate_constraint(C, Phi1, Phi3),
  Phi_all = conjoin(Phi2, Phi3).
    \end{lstlisting}
    \caption{
      This rule describes how a points-to edge to object \iccapt{O1} labeled
      with constraint \iccapt{Phi1} is instantiated at a call site \iccapt{C}:
      if at \iccapt{C} a heap location \iccapt{heap(O1)} can be instantiated to a
      heap location \iccapt{heap(O2)} under constraint \iccapt{Phi2}, and the
      original constraint on the edge \iccapt{Phi1} can be instantiated to a
      constraint \iccapt{Phi3}, then the points-to edge to \iccapt{O1} labeled
      with \iccapt{Phi1} instantiates to a points-to edge to \iccapt{O2}
      labeled with \iccapt{widen(C, Phi\_all)}, where \iccapt{Phi\_all} is the
      conjunction of \iccapt{Phi2} and \iccapt{Phi3} and \iccapt{widen} is a
      function that widens constraints in mutually-recursive functions (one of
      the heuristics we borrowed from Scuba).
    }
    \label{fig:ptsto}
\end{figure}

\begin{table}[!t]
	\caption{In the median, our implementation of a bottom-up points-to analysis
  for Java was 6.7$\times$ slower than Scuba, the reference implementation (times in mm:ss);
	however, the two tools use different heuristics and thus compute very
	different things, as indicated by the discrepancy in the number of points-to
	edges computed in the summary for \iccapt{main} (which also captures the
	effect on the heap of methods invoked transitively from it).}
\begin{center}
  \begin{tabular}{ l|ll|lll }
    & \multicolumn{2}{c|}{Scuba} & \multicolumn{2}{c}{\Name} \\
    Benchmark & Time & \# \ic{main} edges & Time & \# \ic{main} edges \\
    \hline
    antlr & 1:11 & 3,313 & 12:16 & 112,415 \\
    avrora & 1:05 & 714 & 7:40 & 127,535 \\
    hedc & 0:57 & 867 & 5:04 & 2,962 \\
    hsqldb & 0:51 & 780 & 4:53 & 7,039 \\
    luindex & 1:40 & 3,395 & T/O & - \\
    polyglot & 0:55 & 117 & 4:52 & 4,245 \\
    sunflow & 3:48 & 7,456 & T/O & - \\
    toba-s & 0:58 & 521 & 4:57 & 12,284 \\
    weblech & 1:10 & 1,262 & 17:58 & 6,785 \\
    xalan & 0:54 & 183 & 5:40 & 55,722 \\
  \end{tabular}
\end{center}
  \label{tab:ptsto}
\end{table}

We ran both tools on the benchmarks used in the evaluation by Feng
et al., which represent a selection from the pjbench suite plus the benchmark
polyglot.%
\footnote{The pjbench suite is available at
\url{https://bitbucket.org/psl-lab/pjbench/src/master/}.}
These experiments include library code and use a context-sensitivity of two
call sites; reflection is ignored, as are many native methods.
Given an hour timeout, our implementation completed on eight of the ten
benchmarks, with times ranging from five to 18 minutes (Table~\ref{tab:ptsto}).
In the median, we were 6.7$\times$ slower than Scuba.
However, a performance comparison between the tools should be taken with a
grain of salt: Since they use different heuristics, they compute very different
things.

In sum, we were able to implement the algorithm in a way that is still very
close to its specification and achieve decent performance on many realistic
benchmarks while implementing only a small selection of heuristics.
Other heuristics might have helped our version complete on the two benchmarks
it timed out on.
Making the algorithm practical is a significant engineering challenge: Even
with its sophisticated heuristics, Scuba does not complete on all benchmarks in
pjbench.%
\footnote{For example, we found it timed out on batik, chart, fop, lusearch,
and pmd (as did our implementation).}

Moreover, our implementation could be used as a platform for exploring
potential optimizations to Scuba.
First, because it is automatically parallelized (with a user-chosen number of
worker threads), it could be used to evaluate how well the underlying points-to
algorithm parallelizes before going through all the trouble of parallelizing
Scuba, which uses mutable state in a complex way.
Second, thanks to the magic set transformation, we have automatically derived a
goal-directed version of the analysis that computes only the summaries
necessary for constructing user-requested summaries.
The points-to algorithm resulting from this transformation could be used as a
road map for implementing a demand-driven version of Scuba, which Feng et al.
describe as future work.

\subsection{Bounded Symbolic Evaluation}
\label{sec:symexec}

We have written a symbolic evaluator ($\sim$1K LOC) for a fragment of LLVM
bitcode~\citep{llvm} corresponding to a simple imperative language with
integer arrays and symbolic integers (a symbolic integer represents a set of
integer values that might occur at runtime).
It implements a form of bounded symbolic execution~\citep{king}, exploring all
feasible program paths up to a given length, evaluating concretely whenever
possible, and aggressively pruning infeasible paths.

Our implementation uses a different logic rule to define each of the possible
cases during evaluation, and uses ML functions to manipulate and reason about
complex terms representing evaluator state.
For example, one rule defines when an assertion fails
(Figure~\ref{fig:failed_assume}).
This rule says that the path \ic{Path} ends in a failure with evaluator state
\ic{St} if: (1) there is an assert instruction \ic{Instr} with argument \ic{X},
(2) following \ic{Path} has led the evaluator to that instruction with state
\ic{St}, (3) \ic{X} could have the (possibly symbolic) integer value \ic{V} in state \ic{St},
and (4) \ic{V} may be zero.
The function \ic{may\_be\_zero(V, St)} returns true if and only if \ic{V} may
be zero given \ic{St}.
We represent symbolic values as SMT formulas, so when \ic{V} is symbolic, this
function invokes the SMT solver.

We have evaluated our symbolic evaluator on ten benchmarks based on five
template programs.
The first template (shuffle-$N$) non-deterministically shuffles an array of
size $N$ and asserts that the resulting array represents the same set as the
input array.
The second template (sort-$N$; Figure~\ref{fig:sort})
splits into two branches, sorts an array using selection sort in both branches,
and asserts that the resulting array is sorted in the second branch.
The third template completes a partially filled-in 4$\times$4 grid of integers,
such that there is a path from 1 to 16 where each integer follows its
predecessor and only horizontal and vertical movements are used;
the benchmark numbrix-sat runs this program on a satisfiable instance, while
the benchmark numbrix-unsat runs it on an unsatisfiable one.
The fourth template (prioqueue-$N$) tests the equivalence of two
implementations of a priority queue (one based on a heap, the other on an
unsorted array) by pushing the same $N$ symbolic integers on them and verifying
that they have the same behavior during a sequence of operations. 
The fifth template (interp-$N$) runs an interpreter for a simple bytecode
language for $N$ steps; the input bytecode is represented by an array of
symbolic integers that can be interpreted as commands for binary operations,
register loads and stores, and conditional jumps.

\begin{figure}[!t]
  \begin{minipage}[b]{0.45\textwidth}
  \begin{lstlisting}
failed_assert(Path, St) :-
  assert_instruction(Instr, X),
  stepped(Instr, St, _, Path),
  has_value(X, St, v_int(V)),
  may_be_zero(V, St) = true.\end{lstlisting}
	\caption{This rule states that the symbolic evaluator has reached a failing
assertion when the argument of the assert instruction may be zero.}
  \label{fig:failed_assume}
  \end{minipage}\hspace{2em}%
  \begin{minipage}[b]{0.45\textwidth}
  \begin{lstlisting}
a := array of $N$ symbolic ints;
b := symbolic int;
if (b) { sort a; }
else { sort a; assert a sorted; }
\end{lstlisting}
	\caption{This pseudocode sketches a C program that creates an array,
branches, sorts it in each branch, but asserts that the result is sorted
only in one branch.}
  \label{fig:sort}
  \end{minipage}
\end{figure}

\begin{table}
  \caption{We report absolute times (mm:ss) for KLEE, CBMC, and a \Name-based symbolic
  evaluation tool on ten benchmark programs; for the latter, we also report
  speedups ($\uparrow$) and slowdowns ($\downarrow$) relative to KLEE.}
\begin{center}
  \begin{tabular}{lllll}
    Benchmark & \# paths & KLEE & CBMC & \Name \\
    \hline
    shuffle-4 & 125 & 0:06 & 0:01 & 0:02 ($\uparrow3.0\times$) \\
    shuffle-5 & 1,296 & 1:56 & 0:01 & 0:07 ($\uparrow16.6\times$)\\
    sort-6 & 2,718 & 2:29 & 0:24 & 0:18 ($\uparrow8.3\times$)\\
    sort-7 & 22,070 & 27:13 & 2:46 & 3:16 ($\uparrow8.3\times$)\\
    numbrix-sat & 1 & 0:15 & 0:01 & 1:10 ($\downarrow4.7\times$)\\
    numbrix-unsat & 1 & 0:15 & 0:01 & 0:59 ($\downarrow3.9\times$)\\
    prioqueue-5 & 1,132 & 0:43 & 6:45 & 0:16 ($\uparrow2.7\times$)\\
    prioqueue-6 & 4,409 & 3:24 & T/O & 1:10 ($\uparrow2.9\times$)\\
    interp-5 & 994 & 0:55 & 0:05 & 0:39 ($\uparrow1.4\times$)\\
    interp-6 & 3,433 & 3:19 & 0:12 & T/O ($\downarrow\infty\times$)\\
  \end{tabular}
\end{center}
	\label{tab:symexec}
\end{table}

We compared our times on these benchmarks against the symbolic execution tool
KLEE (v2.1) \citep{klee} and the bounded model checker CBMC
(v5.11)~\citep{cbmc} (Table~\ref{tab:symexec}); we used a timeout of 30
minutes.
These should not be taken as apples-to-apples comparisons: KLEE operates over
all of LLVM bitcode and CBMC operates over C source code, whereas we handle
just a fragment of LLVM bitcode; CBMC implements bounded model checking and not
symbolic execution, with the result that it generates many fewer (but
presumably more complex) SMT queries; and all three tools might translate
program constructs into SMT formulas in different ways, leading to different
external solver performance.
Nonetheless, these comparisons provide some context for our evaluation numbers.

In general, our tool achieved speedups over KLEE, but did not quite match the
performance of CBMC.
It performed relatively poorly for benchmarks with a single path
(numbrix-sat and numbrix-unsat), but on most other programs we were able to
achieve substantial speedups ($1.4\times$-$16.6\times$) over KLEE and perform
within striking distance of CBMC.
This was at least partly due to the fact that our analysis is automatically
parallelized, whereas KLEE and CBMC are single threaded.
The interp-$N$ benchmarks caused trouble for our tool: Our trials for interp-5
had an unusually high degree of variance (with two trials taking less than 40
seconds, and one trial taking $\sim$18 minutes), and our tool timed out on
interp-6.
We suspect that this might be because, on this benchmark, our tool generates
SMT queries involving the theory of arrays, and our particular naive encoding
might be leading to slowdowns with the external SMT solver.%
\footnote{The shuffle-$N$ benchmarks are the only other ones during which our
tool generates SMT queries with array constructs.}

Additionally, our tool can be run in a goal-directed mode:
If we only want to check that no assertion fails, we can add the query
\ic{failed\_assert(\_Path, \_St)}, triggering the \Name runtime to rewrite our
evaluator to explore only paths that could potentially lead to a failed
assertion.
For the sorting benchmarks (Figure~\ref{fig:sort}), this means the symbolic
evaluator can ignore the first branch of the program.
This leads to significant performance gains, as we completed sort-6 in 13
seconds and sort-7 in 2:18, representing increased speedups of 11.5$\times$ and
11.8$\times$, respectively, over KLEE.
We ran CBMC in a similar directed mode (it can use program
slicing~\citep{slicing} to ignore parts of the program irrelevant to
assertions); it was slightly slower than our \Name implementation, completing
sort-6 in 28 seconds and sort-7 in 2:25.
This suggests the potential of \Name's automatic optimizations, which help make
it competitive with hand-optimized systems.

\section{Design evaluation}\label{sec:eval}

\begin{table}
  \caption{
    \Name analyses can be concise and close to the formal specifications. This
    table gives the number of rules, non-nullary functions, and line counts for our
    case studies; in parentheses, we give the number of rules and functions
    that correspond to the formal specifications (the rest handle other parts of
    the analyses, e.g., the termination checker in Dminor, and the
    context-insensitive points-to analysis used by the bottom-up points-to
    analysis).
  For comparison, we provide the number of rules and functions used in the formal
  specifications, as well as the line count of the reference implementations.
  Our symbolic evaluator is not based on a particular specification; we omit a
  LOC comparison with the reference implementations, KLEE and CBMC, as they
  handle much larger input languages and it would be difficult to isolate the
  parts of their codebases that correspond to the language our analysis
  supports.
  }
\begin{center}
  \begin{tabular}{l|lll||ll|l}
    & \multicolumn{3}{l||}{\Name impl.} & \multicolumn{2}{l|}{Specification} & Reference impl. \\
    Analysis & \# rules & \# funcs & LOC & \# rules & \# funcs & LOC \\
    \hline
    Dminor type checker & 78 (50) & 61 (43) & 1.2K & 34 & 15 & $\sim$2K lines F$^\sharp$ \& SMT-LIB \\
    Bottom-up points-to & 203 (47) & 49 (28) & 1.5K & 19 & 8 & 15K lines Java\\
    Symbolic evaluator & 51 & 37 & 1K & & & \\
  \end{tabular}
\end{center}
	\label{tab:summary}
\end{table}

In this section, we evaluate the design of \Name with respect to our case
studies.
We argue that \Name is an effective and usable tool for writing SMT-based
analyses.

\paragraph{\Name makes it possible to write SMT-based analyses in a way that is
close to their mathematical specification, leading to concise encodings
{\normalfont (Table~\ref{tab:summary})}}
Our implementations of the Dminor type checker (Section~\ref{sec:dminor}) and
the bottom-up points-to analysis (Section~\ref{sec:pointsto}) directly mirror
their published formal specifications; our third case study
(Section~\ref{sec:symexec}), which was not based on any particular
formalization, would itself be the basis of a reasonable specification of
symbolic evaluation.
\Name provides language features that are a good match for the way that
SMT-based analyses are specified: algebraic data types naturally encode BNF
grammars (a common feature in analysis specifications); Horn clauses match
judgments; ML functions fit helper functions; and the reification of formulas
as terms captures the way that formulas are treated in analysis specifications.

As a corollary, analyses written in \Name can be concise.
Despite encoding quite complex logic, each of our case studies is less than
1.5K lines of code.
In the case of the points-to analysis, this is 10$\times$ smaller than the
reference implementation (which also uses functionality defined externally in
Chord).
This is partly because Scuba implements heuristics that we do not and Java is a
verbose language; however, we suggest that much of the difference is because
\Name is a better fit for encoding the logic of the analysis than an
imperative, object-oriented language like Java.
The relative concision of \Name matches the results reported by previous work
on Datalog-based static analysis, which found that Datalog-based analyses can
be orders of magnitude more concise than counterparts written in more
traditional languages~\citep{bddbddbUsing}.
The ML fragment of \Name also helps it be concise, since ML expressions ---
through supporting sequenced, nested, and scoped computation --- can encode
logic that would be more verbose to write in Datalog.

We have shown that three diverse case studies can be naturally encoded in
\Name, suggesting that its design is a good match for a range of SMT-based
analyses.
However, not all analysis logic can be easily encoded in \Name.
There is currently no way to join facts, a useful operation for abstract
interpretation-based analyses \citep{cousot1977abstract}.
The restriction to stratified negation is sometimes too severe:
For example, one Dminor rule for the type synthesis relation \ic{synth} is not
directly expressible in \Name, because it is defined in terms of the negation
of the type well formedness relation, which is in turn defined using the
relation \ic{synth}.%
\footnote{To get around this, our implementation uses a less precise rule that
drops the negated premise.}
Finally, given its lack of mutable state, \Name is probably not a good fit for
analyses that can most naturally be specified in an imperative manner, such as
lazy abstraction model checking~\citep{lazyabs}.

Because \Name is designed to be compatible with Datalog, we can expand the type
of analysis logic it supports by taking advantage of research on Datalog
extensions. 
For instance, lattice-based recursive aggregation~\citep{flix,inca} would make
it possible to join facts, and local
stratification~\citep{przymusinski1988declarative} would support the Dminor
logic we previously cited.

\paragraph{\Name provides a rich and flexible language of formulas that supports
the type of logic-based reasoning found in SMT-based analyses}
The formula fragment of \Name makes it possible to use formulas the way they
need to be used by static analyses.
A good example of this is the decision to reify logical formulas as terms, a
departure from the approach taken by constraint logic
programming~\citep{clp,clpsurvey} and constrained Horn clause (CHC)
solving~\citep{hfc,gurfinkel2015seahorn,hoder2012generalized,bjorner2015horn},
the two major previous paradigms for combining logic programming and constraint
solving.
In these systems, constraints are represented as predicates, not terms, and an
inference is made if the constraints in the body of a rule are satisfiable.
This approach makes sense in the context of programming with {\em constraints};
however, it seems overly restrictive in the context of programming with {\em
formulas}, which do not necessarily have to be used directly as constraints.
For example, analyses like our Dminor type checker need to check the validity
of a formula, which is the unsatisfiability of its negation.
Checking validity does not easily fit in constraint-based paradigms, since
constraint programming is built around satisfiability.
Similarly, we might want to write an analysis that uses Craig
interpolants~\citep{craig}.
One could imagine extending \Name's SMT interface to include an operator
\ic{interpolate} that takes two formulas and returns a third (optional)
formula, the interpolant; it is not clear how to do this in one of the
constraint-based paradigms.

Our treatment of formula variables through the constructor $\#\{e\}[t]$
provides further evidence.
This mechanism makes it easy to identify a formula variable with an
object-level construct (e.g., a variable in the input program) by choosing for
$e$ the expression representing that construct.
It also makes it easy to create a variable that is guaranteed to be fresh
relative to a set of constructs (e.g., fresh with respect to an environment),
an extremely useful operation.
This is done by choosing for $e$ a tuple of the constructs that the
variable needs to be fresh with respect to.
We use this trick in both the Dminor type checker and the symbolic evaluator.
Crucially, this freshness mechanism is deterministic, which means that we can
safely rewrite \Name programs and parallelize them.
The logic programming language Calypso~\citep{calypso,saturn} provides a
similar mechanism, except that it requires that all the variables in a formula
are identified by terms with the same type; this severely limits its usability
and is too restrictive for our case studies.

Our case studies exercise a range of the SMT-LIB standard and demonstrate the
richness of our formula language.
The case studies variously use algebraic data types and uninterpreted functions
(the Dminor type checker and the bottom-up points-to analysis); bit vectors and
arrays (the Dminor type checker and the symbolic evaluator); and integers,
uninterpreted sorts, and quantifiers (the Dminor type checker).
It is easy to extend \Name with additional theories (by adding new
constructors) and different types of logical reasoning (by adding new
operators, like \ic{interpolate}).
As \Name so loosely couples Datalog evaluation and constraint solving, it
is easy to swap in new solver backends without major changes to the \Name
runtime; our prototype currently supports Z3~\citep{z3}, CVC4~\citep{cvc4}, and
Yices~\citep{yices}.

\paragraph{The design of \Name makes it possible to advantageously apply
Datalog-style optimizations to SMT-based analyses, with the result that \Name
programs can compete with analyses written in more mature languages}
All of our case study implementations benefit from automatic parallelization:
this scales our Dminor type checker and symbolic evaluator over the reference
implementations, and helps our bottom-up points-to analysis be reasonably
performant. 
The points-to analysis and symbolic evaluator also demonstrate the potential of
the magic set transformation, as we have used it to derive demand-driven versions
of these SMT-based analyses.
While these types of optimizations could be added by hand to the reference
implementations we compare against, the point is that the design of \Name means
that \Name-based analyses get these optimizations for free, without the
explicit effort of the analysis designer.
Moreover, because of \Name's close affinity to Datalog, a \Name runtime can
be augmented with additional Datalog-style optimizations.
For instance, a \Name runtime could use an incremental Datalog evaluation
algorithm, which efficiently evaluates Datalog programs while facts are added
or retracted from EDB relations~\citep{gupta1993maintaining,inca}.
This would be helpful for using SMT-based analyses in situations where the code
under analysis changes, such as in IDEs or rapidly evolving codebases.

It speaks to the design of \Name that the high-level optimizations it
enables can, in many cases, make up for the naivety of our prototype
runtime. 
Nonetheless, we are optimistic that significantly better performance can be
achieved with a sophisticated backend.
As we have designed \Name to be close to Datalog, we can take advantage of many
of the optimizations that have helped Datalog systems scale.
For example, since we maintain the range restriction (which entails that
every derived fact is ground), we can use concurrent data structures
specialized for Datalog evaluation~\citep{soufflebtree}; since \Name can be
evaluated using standard semi-naive evaluation, we can compile \Name programs
to C++ following Souffl{\'e}'s strategy~\citep{souffle1}.

\paragraph{The ML fragment is an integral part of \Name and has a
substantial impact on its usability.}
As discussed in Section~\ref{sec:features}, the first-order fragment of ML we
use can be translated in a pretty straightforward way to Datalog rules, and
hence can be thought of as syntactic sugar.
Despite this, the ML fragment is an integral part of the \Name programming
experience.
First, it improves the ergonomics of \Name, by making it more natural to
manipulate complex terms.
In particular, pattern matching and let expressions provide a structured way to
reflect on complex terms and sequence computation on them; this same effect is
not always as easy to achieve in Datalog rules.
Second, it helps \Name achieve its design goal of allowing SMT-based analyses
to be implemented in a style close to their specification, since formal
specifications often involve functions in addition to inference rules.
Third, it improves the performance of \Name, as there is more overhead involved
with evaluating Datalog rules than evaluating an ML expression.
A substantial amount of our case study code is in the ML fragment:
The ratio of functions to rules is 1:4 for the bottom-up points-to analysis and
3:4 for the two other case studies (Table~\ref{tab:summary}).
Typically, the case studies use Horn clauses to define the overall structure of
the analysis, and ML functions for structuring lower-level control flow,
mirroring the use of judgments and helper functions in analysis specifications.

The limitation to first-order ML has several advantages.
From a theoretical perspective, it means that there is an easy translation from
it to Datalog rules, which allows us to give the standard Herbrand model-based
semantics to \Name programs.
From a practical perspective, it ensures that we never have to unify functions,
which would require higher-order unification.
The specifications of our case studies did not make heavy use of higher-order
functions, so they were not much missed.
However, a future version of \Name could allow a limited use of higher-order
functions (for example, those programs that can be compiled to the first-order
fragment).

\section{Related work}\label{sec:related}

\paragraph{Datalog-based frameworks and domain-specific languages for static
analysis}
A variety of static analysis frameworks have been developed based on
more-or-less standard Datalog, such as bddbddb~\citep{bddbddbUsing},
Chord~\citep{chord}, Doop~\citep{doop}, QL~\citep{ql}, and
Souffl\'e~\citep{souffle2}.
Recent work has explored synthesizing Datalog-based
analyses~\citep{datalogsynthesis,provenancesynthesis}.
Flix~\citep{flix} and IncA~\citep{inca} extend Datalog for analyses that
operate over lattices besides the powerset lattice.
IncA supports incremental evaluation, while Flix (like \Name) includes
algebraic data types and a pure functional language.
Dataflow analysis is used as a case study for Datafun, a language combining
Datalog and higher-order functional programming~\citep{datafun}.
It might be possible to encode something like \Name in Datafun; however,
although it has recently been shown that Datafun can be evaluated using
semi-naive evaluation~\citep{datafunseminaive}, it is not clear to what extent
other Datalog optimizations can be applied to Datafun programs.
By combining Datalog with functional programming, \Name, Flix, and Datafun are
related to functional logic programming~\citep{flp}.
The functional fragment of \Name is less expressive than what is typically
found in such languages, as \Name functions are not first-class values and not
higher-order.

\paragraph{Logic programming with constraints and formulas}

The two dominant prior paradigms for combining logic programming and constraint
solving are constraint logic programming (CLP)~\citep{clp,clpsurvey} and constrained Horn clause
(CHC) solving~\citep{hfc,gurfinkel2015seahorn,hoder2012generalized,bjorner2015horn}.
As discussed in Section~\ref{sec:eval}, these systems typically encode
constraints as predicates, not terms, and thus support programming with
constraints as things to be satisfied, rather than programming with formulas, which can be manipulated in more interesting ways (e.g., validity checking).
In the context of static analysis, these systems have been used primarily for
model checking, where a model of the input system is encoded using Horn
clauses~\citep{bjorner2015horn,clpmodelchecking,flanagan2004automatic,fribourg1996symbolic,hfc}.
The rules depend on the program being analyzed, and the solutions
to these rules reveal properties of the model; e.g.,
SeaHorn~\citep{gurfinkel2015seahorn} checks programs by solving a CHC
representation of their verification conditions.
This differs than the approach taken in this paper, where the rules encode an
analysis independent of the input program.
The Datalog mode of $\mu$Z~\citep{hoder2011muz} can be thought of as a
bottom-up CLP system with special support for abstract interpretation.

A few existing logic programming systems support programming with formulas (vs
constraints); we would argue that none do so with the same richness and
flexibility as \Name.
\citet{lpsat} extend Prolog with an interface to a SAT solver.
SICStus Prolog~\citep{sicstus}, with its CLP extensions, has been used to write model
checkers~\citep{clpmodelchecking,fribourg1996symbolic,armc,hfc}; these
implementations typically rely on Prolog's non-logical features, like
\ic{assert}, making it harder to apply high-level optimizations like
parallelization.
Calypso~\citep{calypso,saturn} is a Datalog variant that interfaces with
external constraint solvers and has specialized support for bottom-up analyses.
Calypso has been used with SAT and integer constraint solvers; in theory, it
could be connected to an SMT solver.
However, \Name offers several advantages over Calypso for SMT-based analyses.
First, \Name's approach to constructing formulas (via complex terms) and
manipulating them (via its ML fragment) scales to the complex and
heterogeneous formulas that arise in the SMT context, whereas Calypso's
approach to formulas (opaque terms, constructed via predicates) would be
cumbersome in this setting.
Second, \Name's type system supports the construction of expressive (and safe)
formulas involving user-defined terms such as algebraic data types and
uninterpreted functions.
Third, the ML fragment of \Name goes a long way towards making it practical for
SMT-based analyses, by closing the gap between specification and
implementation, and improving ergonomics and performance.

The logic programming language $\lambda$Prolog provides a natural way to
represent logical formulas using a form of higher-order abstract syntax based
on $\lambda$-terms and higher-order unification~\citep{hoas,lambdaprolog}.
Although this representation simplifies some aspects of using formulas, moving
to a higher-order setting would complicate \Name, widen the gap between \Name
and other Datalog variants, and potentially be an impediment to building a
performant and scalable \Name implementation.
Answer set programming (ASP) uses specialized solvers to find a {\em stable
model} (if it exists) of a set of Horn clauses~\citep{stablemodel,aspsurvey}.
Common extensions support constraints on the shape of the stable model that
will be found.
ASP enables concise encoding of classic NP-complete constraint problems such as
graph $k$-coloring, but it is not as obviously applicable to static analysis
problems.

\paragraph{Type system engineering}

PLT Redex~\cite{Felleisen:2009:SEP:1795772} and Spoofax
\cite{Kats:2010:SLW:1869459.1869497} support exploratory type system
engineering.
PLT Redex supports a notion of judgment modeled explicitly on inference rules.
Spoofax's type engineering framework, Statix, uses a logic programming syntax
to specify type systems, with a custom solver for resolving the binding
information in scope graphs simultaneously with solving typing
constraints~\cite{vanAntwerpen:2018:ST:3288538.3276484}.
Both of these systems use custom approaches to finding typing derivations;
neither supports SMT queries, but Statix's custom solver can resolve constraint
systems that might not always terminate in \Name.

\paragraph{Solver-aided languages}

Scala${^\text{Z3}}$~\citep{scalaz3} supports mixed computations combining
normal Scala evaluation and Z3 solving; we avoid this level of integration.
Smten~\citep{smten} is a solver-aided language that supports both concrete and
symbolic evaluation; Rosette~\citep{rosette} is a framework for creating
solver-aided languages that have this property.

\section{Conclusion}\label{sec:conclusion}

\Name is a domain-specific language for writing SMT-based static analyses that
judiciously combines Datalog, ML, and SMT solving (via an external SMT solver).
As demonstrated by our case studies, it makes it possible to concisely
implement a range of SMT-based analyses --- refinement type checking, bottom-up
points-to analysis, and symbolic evaluation --- in a way close to their formal
specifications, while also making it possible to automatically and
advantageously apply high-level optimizations to these analyses like
parallelization and goal-directed rewriting.

%% Acknowledgments
\begin{acks}                            %% acks environment is optional
                                        %% contents suppressed with 'anonymous'
  %% Commands \grantsponsor{<sponsorID>}{<name>}{<url>} and
  %% \grantnum[<url>]{<sponsorID>}{<number>} should be used to
  %% acknowledge financial support and will be used by metadata
  %% extraction tools.
  % This material is based upon work supported by the
  % \grantsponsor{GS100000001}{National Science
  %   Foundation}{http://dx.doi.org/10.13039/100000001} under Grant
  % No.~\grantnum{GS100000001}{nnnnnnn} and Grant
  % No.~\grantnum{GS100000001}{mmmmmmm}.  Any opinions, findings, and
  % conclusions or recommendations expressed in this material are those
  % of the author and do not necessarily reflect the views of the
  % National Science Foundation.
  This material is based upon work supported by the \grantsponsor{0}{Defense
  Advanced Research Projects Agency (DARPA)}{https://www.darpa.mil/} under
  Contract No.~\grantnum{0}{FA8750-19-C-0004}.  Any opinions, findings and
  conclusions or recommendations expressed in this material are those of the
  author(s) and do not necessarily reflect the views of the Defense Advanced
  Research Projects Agency (DARPA).
  We thank Arlen Cox, Scott Moore, the Harvard PL group, and anonymous
  reviewers for thoughtful feedback on earlier drafts.

\end{acks}

%% Bibliography
\bibliography{main}

\iffull
%% Appendix
\clearpage
\onecolumn
\appendix

\section{\Name's formal model}

\begin{figure}[t]
  \[ \begin{array}{lrcl}
    \multicolumn{4}{l}{\textbf{Types}} \\ \hline
    \text{Types}        & \tau   &::=& t \BNFALT \smt{t} \BNFALT \sym{t} \BNFALT \MODEL \\
    \text{Pre-types}    & t      &::=& B \BNFALT D ~ \vec{\tau} \BNFALT \alpha \\
    \text{Base types}   & B      &::=& \bool \BNFALT \bv{k}_{k \in {\mathbb{N}^+}} \BNFALT \dots \\[1em]
    
    \multicolumn{4}{l}{\textbf{Contexts}} \\ \hline
    \text{Data type declarations} & \Delta &::=& \cdot \BNFALT \Delta, D:\forall \vec{\alpha_i}. ~ \{ \overrightarrow{c_j : \vec{\tau_k}} \} \\
    \text{Program declarations}  & \Phi   &::=&
      \cdot \BNFALT
      \Phi, f:\forall \vec{\alpha}, \vec{\tau} \rightarrow \tau \BNFALT
      \Phi, \uf:\vec{t} \rightarrow t \BNFALT
      \Phi, p \subseteq \vec{\tau} \\
    \text{Variable contexts}    & \Gamma &::=&
      \cdot \BNFALT
      \Gamma, x:\tau \BNFALT
      \Gamma, \alpha \\[1em]
    
    \multicolumn{4}{l}{\textbf{Terms}} \\ \hline

    \text{Programs}     & \prog &::=& \vec{F_i} ~ \vec{H_j} \\
    \text{Functions}    & F     &::=& \fun{f}{\vec{X_i} : \vec{\tau_i}}{\tau}{e} \\
    \text{Horn clauses} & H     &::=& p(\vec{X_i}) \horn \vec{P_j} \\
    \text{Premises}     & P     &::=& A \BNFALT \mathord{!}A \\
    \text{Atoms}        & A     &::=& p(\vec{e_i}) \BNFALT X=e \\
    \text{Expressions}  & e     &::=& k \BNFALT
                                      X \BNFALT
                                      c(\vec{e_i}) \BNFALT
                                      f(\vec{e_i}) \BNFALT
                                      p(\vec{e_i}) \BNFALT
                                      \operation{\vec{e_i}} \BNFALT
                                      \qq{\phi}
                                      \\[0.25em] &&&
                                      \letin{X}{e_1}{e_2} \BNFALT
                                      \ite{e_1}{e_2}{e_3} \BNFALT
                                      \\[0.25em] &&&
                                      \match{e}{\overrightarrow{c_i(\vec{X_j}) \rightarrow e_i}}
                                      \\[0.25em]
    \text{SMT formulas} & \phi  &::=& \cVAR[x,t]() \BNFALT
                                      \cCONST[k]() \BNFALT
                                      \cLET(\phi_1, \phi_2, \phi_3) \BNFALT
                                      \\[0.25em] &&&
                                      \cCTOR[c](\vec{\phi_i}) \BNFALT
                                      \cFORALL(\phi_1, \phi_2) \BNFALT
                                      \cUF[\uf](\vec{\phi_i}) \BNFALT
                                      \unq{e} \\[0.25em]
    \text{Constants}    & k     &::=& \mathsf{true} \BNFALT \mathsf{false} \BNFALT \mathsf{0} \BNFALT \mathsf{1} \BNFALT \dots \\[1em]

    \multicolumn{4}{l}{\textbf{Namespaces}} \\ \hline
    \text{Type modes} & m &::=& \mEXP \BNFALT \mSMT \\
    \text{Data type names} & D &\in& \mathrm{ADTVar} \\
    \text{Type variables} & \alpha &\in& \mathrm{TVar} \\
    \text{Constructors} & c     &\in& \mathrm{CtorVar} \\
    \text{\Name variables} & X &\in& \mathrm{Var} \\
    \text{SMT variables} & x &\in& \mathrm{SMTVar} \\
    \text{Predicates} & p &\in& \mathrm{PredVar} \\
    \text{Functions} & f &\in& \mathrm{FunVar} \\
    \text{Uninterpreted functions} & \uf &\in& \mathrm{UninterpFunVar} \\
    
  \end{array} \]

  \caption{Syntax of \Name's formal model}
  \label{fig:syntax}
\end{figure}

We define a `middleweight' formal model of \Name, designing a type
system (Section~\ref{sec:types}) and an operational semantics
(Section~\ref{sec:opsem}), relating the two in a proof of type safety
(Section~\ref{sec:metatheory}).

Our model characterizes \Name as a two-level system
(Figure~\ref{fig:syntax}), comprising Datalog-esque Horn clauses $H$
and first-order functions $F$; Horn clause ``rules'' are made up of
premises $P$, where each premise is a series of (possibly negated)
atoms $A$. Each atom $A$ either references a Datalog predicate or
binds a variable to an expression $e$. Expressions themselves have two
mutually recursive modes: ordinary functional computation $e$ and
quoted SMT terms $\qq{\phi}$, which can include unquoted expressions
$\unq e$.

The Datalog fragment of \Name is fairly standard syntactically, up to
the addition of the atomic form $X = e$.
We constrain premises to a sort of administrative normal form:
predicate references apply only to variables, written $p(\vec{X_i})$, and
expression constraints bind variables, as in $Y = e$.
Our implementation can handle compound premises like $p(e_1, e_2)$;
our formal model would require rewriting such a premise to three
premises: $p(X, Y)$, $X = e_1$, and $Y = e_2$ (for some fresh $X$ and
$Y$).

The functional programming fragment fully annotates the types on its
functions $F$; variable names in both fragments are written in capital
letters. (Our implementation merely demands that the first letter be
capitalized.)
SMT variables are written using lowercase letters and annotated with their
type, as in $\cVAR[x, t]()$.
As described in Section~\ref{sec:formulas}, our implementation allows any value
to be used as the name of an SMT variable; here, without loss of generality, we
treat SMT variables as being drawn from a distinct universe.
Code in the functional fragment can treat Datalog relations as
predicates, i.e., $p(\vec{v_i})$ returns $\true$ when $\vec{v_i} \in
p$.
In our implementation, some elements of $\vec{v_i}$ can be the
wildcard \texttt{??}, turning a Datalog predicate into a list.
For example, if $p \subseteq \bool \times \bv{32}$, then: $p(\true,
42)$ yields a $\bool$; $p(\texttt{??}, 42)$ returns a list of $\bool$s
$b$ such that $p(b, 42)$ holds; $p(\true, \texttt{??})$ returns a list
of $\bv{32}$s $n$ such that $p(\true, n)$ holds;
$p(\texttt{??}, \texttt{??})$ returns a list of
$\bool \times \bv{32}$, i.e., the relation $p$. We don't include this
behavior in our formal model.

The set of available base types $B$ must include $\bool$ at a minimum;
any other SMT-embeddable base types are acceptable, e.g., $k$-width
bit vectors for a statically known $k$.

As a matter of notation, we write $\vec{e_i}$ for a metavariable $e$
to mean a possibly empty sequence of $e$s, indexed by $i$. When more
than one variable shares the same index, we mean that those sequences
must be of the same length (e.g., in the $\mathsf{match}$ syntax, each
branch of a match is a triple of a constructor $c$, a vector of
variable names for $c$'s arguments, and a corresponding single
expression).

\section{\Name's type system}
\label{sec:types}

\begin{figure}[t]
  \hdr{Type and typing context well formedness}{
    \quad \fbox{$\implicit{\Delta} \vdash \Gamma$}
    \quad \fbox{$\implicit{\Delta; {}} \Gamma \vdash_m \tau$}
  }

  \threesidebyside
  {
  \infrule[$\Gamma$-Empty]{}{\vdash \cdot}
  }
  {
  \infrule[$\Gamma$-Var]{\vdash \Gamma \andalso \Gamma \vdash_\mEXP \tau}{\vdash \Gamma,x:\tau}
  }
  {
  \infrule[$\Gamma$-TVar]{\vdash \Gamma}{\vdash \Gamma,\alpha}
  }

  ~ \\[0.5em]

  \threesidebyside[.25][.25][.49]
  {
  \infrule[$t$-Base]{}{\Gamma \vdash_m B}
  }
  {
  \infrule[$t$-TVar]{\alpha \in \Gamma}{\Gamma \vdash_\mEXP \alpha}
  }
  {
  \infrule[$t$-ADT]{
    \Delta(D) = \forall \vec{\alpha_i}, \{ \dots\} \andalso
    \Gamma \vdash_m \tau_i}{\Gamma \vdash_m D ~ \vec{\tau_i}}
  }

  \threesidebyside
  {
  \infrule[$\tau$-SMT]{\Gamma \vdash_\mSMT t}{\Gamma \vdash_m \smt{t}}
  }
  {
  \infrule[$\tau$-Sym]{\Gamma \vdash_\mSMT t}{\Gamma \vdash_m \sym{t}}
  }
  {
  \infrule[$\tau$-Model]{}{\Gamma \vdash_\mEXP \MODEL}
  }

  ~ \\[0.5em]

  \hdr{Data type and program signature well formedness}{
    \quad \fbox{$\vdash \Delta$}
    \quad \fbox{$\vdash \Phi$}
  }

  \[ \begin{array}{rcl}
    \vdash \Delta &\Leftrightarrow& \forall D:\forall \vec{\alpha}. ~ \{ c_1 : \vec{\tau_1}, \dots, c_n : \vec{\tau_n} \} \in \Delta \forall i, \\
    && \text{  (1) } \forall D' \in \dom(\Delta), ~ c_i \in \Delta(D') \Rightarrow D = D' \\
    && \text{  (2) } \vec{\alpha} \vdash_\mEXP \tau_i \\
    && \text{  (3) } \forall \beta \in \vec{\alpha}, \beta \in \vec{\tau_i}
  \end{array}
  \]

  \sidebyside[.28][.71]
  {\infrule[$\Phi$-empty]{}{\vdash \cdot}}
  {\infrule[$\Phi$-Fun]
     {\vdash \Phi \andalso
      \forall \beta \in \vec{\alpha_i}, ~ \beta \in \vec{\tau_j},\tau \andalso
      \vec{\alpha_i} \vdash_\mEXP \tau_j \andalso
      \vec{\alpha_i} \vdash_\mEXP \tau
     }
     {\vdash \Phi, f:\forall \vec{\alpha_i}, \vec{\tau_j} \rightarrow \tau}}
  
  ~ \\[0.5em]

  \sidebyside[.3][.69]
  {\infrule[$\Phi$-Rel]
     {\vdash \Phi \andalso
      \cdot \vdash_\mEXP \tau_i
     }
     {\vdash \Phi, p \subseteq \vec{\tau_i}}}
  {\infrule[$\Phi$-UFun]
     {\vdash \Phi \andalso
      \cdot \vdash_\mSMT t_i \andalso
      \cdot \vdash_\mSMT t \andalso
     }
     {\vdash \Phi, \uf:\vec{t_i} \rightarrow t}}
  
  ~ \\[0.5em]

  \hdr{Program and function typing}{
    \quad \fbox{$\Delta; \Phi \vdash \prog$}
  }

  \infrule[$\prog$-WF]
    {\vdash \Delta \andalso \vdash \Phi \andalso
     \Delta; \Phi \vdash F_i \andalso
     \Delta; \Phi \vdash H_j
    }
    {\Delta; \Phi \vdash \vec{F_i} ~ \vec{H_j}}

  \caption{Type, context, and definition well formedness; top-level program typing}
  \label{fig:types}
\end{figure}

We begin by presenting type checking rules for \Name
(Figures~\ref{fig:types}, ~\ref{fig:types-rules},
~\ref{fig:types-expressions}, and~\ref{fig:types-formulas}).
Our \emph{implementation} of \Name not only performs type checking,
but can also perform type inference, e.g., automatically finding type
variable substitutions.

Our types are broken into two levels: types $\tau$ and pre-types
$t$. Every pre-type $t$ can be directly considered as a type, but
there are two additional types: $\smt{t}$, the type of SMT formulas
yielding $t$, and $\sym{t}$, the type of SMT variables of type $t$.
We factor the syntax in this way to prevent anomalies like
$\smt{\smt{\bool}}$, which would mean SMT formulas that yield SMT
formulas that yield booleans.
It is \emph{not} the case, however, that every pre-type $t$ is
necessarily representable as an SMT type, because data types may
contain SMT formulas as arguments; we discuss how we categorize
SMT-representable types shortly.

Before we begin, some further notational clarification.
Rules are named by their primary subjects followed by a hyphen and a
descriptive name.
Whenever we use indices in rules, we will always map (stating a single
premise in terms of the index, e.g., \rn{$\prog$-WF}) or fold (stating
first, indexed, and last, e.g., \rn{$\vec{X}\vec{\tau}$-All}) over the
sequence. We omit the indices when selecting an element of a sequence
or set (as in, e.g., \rn{$e$-Match} in
Figure~\ref{fig:types-expressions}).

All of our typing rules are in terms of a fixed set of data type
declarations $\Delta$ and program declarations $\Phi$
(Figure~\ref{fig:syntax}). Data type declarations $\Delta$ map data type
names $D$ to some number of type arguments $\vec{\alpha_i}$ and a set
of constructors $\vec{c_j}$, each of which takes some number of
arguments of type $\vec{\tau_k}$; each $c_j$ can have a different
number of arguments.
Program declarations $\Phi$ collect the signatures of first-order
polymorphic functions
$f:\forall \vec{\alpha}, \vec{\tau} \rightarrow \tau$, uninterpreted
functions for use in the SMT solver $\uf:\vec{t} \rightarrow t$, and
relations $p \subseteq \vec{\tau_i}$.

The highest level typing rule is \rn{\prog-WF}
(Figure~\ref{fig:types}), which ensures that the declarations are well
formed and each part of the program is well formed.

The context and type well formedness rules (Figure~\ref{fig:types})
are mostly straightforward,
type well formedness being the most interesting. Each type can be
found to be well formed in either SMT mode $\mSMT$---i.e., it can be
exported to the SMT solver--or in expression mode $\mEXP$, meaning it
cannot be. There is a sub-moding relationship: well formed types at
$\mSMT$ are also well formed at $\mEXP$, but not necessarily
vice-versa: for example, there is no way to export an SMT formula or
variable as the \emph{object} of another SMT formula, only as a
constituent.
We assume that all \Name constants are SMT representable, i.e.,
$\cdot \vdash_\mSMT \typeof(k)$ for all constants $k$.
Data type declarations are polymorphic, but we disallow phantom type
variables. Data types can freely mutually recurse.
Uninterpreted functions must be in terms of pre-types, and those
pre-types must be closed and SMT representable ($\mSMT$); 
functions and relations can use any well formed types ($\mEXP$).
Functions can be polymorphic but we disallow phantom type variables;
relations have monomorphic types.
Disallowing phantom types in constructors and functions and keeping
relations monomorphic ensure that these forms are ``reverse
determinate'', i.e., the types of their arguments uniquely determine their types.

Since the declaration environments $\Delta$ and $\Phi$ are statically
determined for an entire program, we typically leave
them \emph{implicit}.
Implicit parameters are in \implicit{gray} in the boxed rule schemata
in the figures. In proofs we will treat these parameters explicitly,
but we conserve space by stating the rules without threading implicit
parameters through.
For example, the data type declarations $\Delta$ are necessary to
ensure that \rn{$t$-ADT} only allows us to name data types that have
actually been defined. Rather than threading $\Delta$ through every
rule for context and type well formedness, we write
$\implicit{\Delta}$ in the rule schemata.

\begin{figure}[t]
  \hdr{Variable binding and typing}{
    \quad \fbox{$\Gamma \vdash x,\tau \rhd \Gamma$}
    \quad \fbox{$\Gamma \vdash \vec{x},\vec{\tau} \rhd \Gamma$}
  }
  
  ~ \\[0.5em]
 
  \sidebyside
   {
    \infrule[$X\tau$-Bind]{
      X \not \in \dom(\Gamma)
    }{
      \Gamma \vdash X, \tau \rhd \Gamma, X \mathord{:} \tau
    }
  }{
    \infrule[$X\tau$-Check]{
      \Gamma(X) = \tau
    }{
      \Gamma \vdash X, \tau \rhd \Gamma
    }
  }

  ~ \\[0.5em]

  {
    \infrule[$\vec{X}\vec{\tau}$-All]{
      \Gamma   \vdash X_0, \tau_0 \rhd \Gamma_1 \andalso \dots \andalso
      \Gamma_i \vdash X_i, \tau_i \rhd \Gamma_{i+1} \andalso \dots \andalso
      \Gamma_n \vdash X_n, \tau_n \rhd \Gamma'
    }{
      \Gamma \vdash \vec{X_i}, \vec{\tau_i} \rhd \Gamma'
    }
  }

  ~ \\[0.5em]

  \hdr{Premise typing}{
    \quad \fbox{$\implicit{\Delta; \Phi; {}} \Gamma \vdash P \rhd \Gamma$}
  }
  
  ~ \\[0.5em]

  \sidebyside[.44][.56]
  {
    \infrule[$P$-PosAtom]{
      p \subseteq \vec{\tau_i} \in \Phi \andalso
      \Gamma \vdash \vec{X_i}, \vec{\tau_i} \rhd \Gamma'
    }{
      \Gamma \vdash p(\vec{X_i}) \rhd \Gamma'
    }
  }{
    \infrule[$P$-NegAtom]{
      p \subseteq \vec{\tau_i} \in \Phi \andalso
      \Gamma \vdash \vec{X_i}, \vec{\tau_i} \rhd \Gamma
    }{
      \Gamma \vdash \mathord{!}p(\vec{X_i}) \rhd \Gamma
    }
  }

  \infrule[$P$-EqCtor-BF]{
    \vec{X_i} \not \subseteq \Gamma \andalso
    \Delta(D) = \forall \vec{\alpha_j}, \{ \dots, c : \vec{\tau_i}, \dots \} \andalso
    \Gamma \vdash Y, D ~ \vec{\tau_j'} \rhd \Gamma \andalso
    \Gamma \vdash \vec{X_i}, \vec{\tau_i}[\vec{\tau_j'}/\vec{\alpha_j}] \rhd \Gamma'
  }{
    \Gamma \vdash {Y = c(\vec{X_i})} \rhd \Gamma'
  }

  % INVARIANT: $\Gamma \vdash \cSMT{c'} : ...$ only yields SMT types
  \infrule[$P$-EqSMT-BF]{
    \vec{X_i} \not \subseteq \Gamma \andalso
    \Gamma \vdash \cSMT{c'} : \vec{\tau_i} \rightarrow \tau \andalso
    \Gamma \vdash Y, \tau \rhd \Gamma \andalso
    \Gamma \vdash \vec{X_i}, \vec{\tau_i} \rhd \Gamma'
  }{
    \Gamma \vdash {Y = \qq{\cSMT{c'}(\vec{\unq{X_i}})}} \rhd \Gamma'
  }

  \sidebyside{
  \infrule[$P$-Eq-FB]{
    \Gamma \vdash e : \tau \andalso
    \Gamma \vdash Y, \tau \rhd \Gamma'
  }{
    \Gamma \vdash {Y = e} \rhd \Gamma'
  }
  }{
  \infrule[$P$-NegEq]{
    \Gamma \vdash e : \tau \andalso
    \Gamma \vdash Y, \tau \rhd \Gamma
  }{
    \Gamma \vdash {!(Y = e)} \rhd \Gamma
  }
  }

  ~ \\[0.5em]

  \hdr{Clause typing}{
    \quad \fbox{$\implicit{\Delta; \Phi}  \vdash H$}
  }
  
  ~ \\[0.5em]
  
  \infrule[$H$-Clause]{
     \cdot \vdash P_0 \rhd \Gamma_1 \andalso \dots \andalso
     \Gamma_j \vdash P_j \rhd \Gamma_{j+1} \andalso \dots \andalso
     \Gamma_n \vdash P_n \rhd \Gamma' \\
     p \subseteq \vec{\tau_i} \in \Phi \andalso
     \Gamma' \vdash \vec{X_i}, \vec{\tau_i} \rhd \Gamma'
  }{
    \vdash p(\vec{X_i}) \horn \vec{P_j}
  }

  \caption{Typing rules: Horn clauses (rules)}
  \label{fig:types-rules}
\end{figure}

The type checking of the Datalog fragment of \Name
(Figure~\ref{fig:types-rules}) must encode two Datalog invariants in
addition to conventional typing constraints: the range restriction,
i.e., every variable in the head of a rule appears somewhere in a
premise; and appropriate binding, i.e., it is possible to interpret a
Horn clause in such a way that all of the variables will be bound at
the end.
Our formal rules ensure that the program has correct binding structure
for a left-to-right evaluation of each Horn clause. An implementation
could determine whether or not an ordering would work and could reorder
programs into an appropriate order automatically.
Our formal model does not enforce that the dependencies between relations are
appropriately stratified, though doing so would be easy:
the relation-and-function call graph should not have any ``negative'' edge in a
cycle, where a negative edge is created whenever there is a negated predicate
in a rule body or a predicate is invoked as a function.

Concretely, \rn{$H$-Clause} ensures that (a) a left-to-right binding
order produces some appropriate final context $\Gamma'$ (via the
premise typing judgment), (b) the range restriction is satisfied,
($\vec{X_i} \subseteq \Gamma'$) by making sure that (c) every variable
is well typed and bound
($\Gamma' \vdash \vec{X_i},\vec{\tau_i} \rhd \Gamma'$---having the
same $\Gamma'$ means no new bindings were introduce when checking the
head variables).

Premise typing $\Gamma \vdash P \rhd \Gamma$ and variable binding and typing
$\Gamma \vdash x,\tau \rhd \Gamma$ work together to generate appropriate types
for each premise.
Positive references to relations are well formed in binding $\Gamma'$ according
to \rn{$P$-PosAtom} when (a) the use is well typed
($p \subseteq \vec{\tau_i} \in \Phi$) and (b) the variables used in
the premise yield the binding $\Gamma'$.
Negative references to relations $\mathord{!}p(\vec{X_i})$
additionally require that all of the $X_i$ be already bound, i.e., the
resulting $\Gamma$ is the same as the starting one.
We split expression equality constraints $Y = e$ into three main cases:
\begin{enumerate}
\item $Y$ is bound and $e$ is a constructor $c(\vec{X_i})$ where all of $\vec{X_i}$ are unbound (\rn{$P$-EqCtor-BF}).
\item $Y$ is bound and $e$ is a quoted SMT constructor $\qq{\cSMT{c'}(\vec{\unq{X_i}})}$ where all of $\vec{X_i}$ are unbound (\rn{$P$-EqSMT-BF}).
\item $Y$ is possibly unbound and $e$ has no unbound variables  (\rn{$P$-Eq-FB}).
\end{enumerate}
It is critical that we avoid the case where both $Y$ and some of the
$X_i$ are unbound, in which case we would need to perform true
unification (or even higher-order unification, depending on our
treatment of functional programs).
In the case where $Y$ is bound and the expression has no unbound
variables, only the \rn{$P$-Eq-FB} case could apply.
There is a fourth, irrelevant case: \rn{$P$-NegEq}. No binding can
possibly occur there, so the constraint is imply checked by running
$e$ and making sure it isn't equal to $Y$.

The binding rules come in three forms: \rn{$X\tau$-Bind} for adding a
new binding, \rn{$X\tau$-Check} for ensuring that an already bound
variable is matched at appropriate type, and a vectorized
form \rn{$\vec{X}\vec{\tau}$-All} for folding over a sequence of such
bindings.
Note that the resulting bindings are the same, i.e., $\Gamma \vdash
X,\tau \rhd \Gamma$, if and only if $X \in \dom(\Gamma)$; the same
holds for vectors of variables and types, as well
(Lemma~\ref{lem:bind-contain}).

\begin{figure}[t]

  \hdr{Function and expression well
    formedness}{ \quad \fbox{$\implicit{\Delta; \Phi} \vdash
    F$} \quad \fbox{$\implicit{\Delta; \Phi; {}} \Gamma \vdash e
    : \tau$} }

  \infrule[$F$-WF]
    {f : \forall \vec{\alpha_j}, \vec{\tau_i} \rightarrow \tau \in \Phi \andalso
     \vec{\alpha_j}, \overrightarrow{X_i:\tau_i} \vdash e : \tau
    }
    {\vdash \fun{f}{\vec{X_i} : \vec{\tau_i}}{\tau}{e}}

  \threesidebyside[.25][.3][.45]
  {\infrule[$e$-Var]
     {\vdash \Gamma \andalso \Gamma(X) = \tau}
     {\Gamma \vdash X : \tau}}
  {\infrule[$e$-Const]
     {}
     {\Gamma \vdash k : \typeof(k)}}
  {\infrule[$e$-Let]
     {\Gamma \vdash e_1 : \tau_1 \andalso
      \Gamma, X:\tau_1 \vdash e_2 : \tau_2}
     {\Gamma \vdash \letin{X}{e_1}{e_2} : \tau_2}}

  ~ \\[0.5em]

  \sidebyside[.74][.25]
  {\infrule[$e$-Ctor]
     {\Delta(D) = \forall \vec{\alpha_j}, \{ \dots, c : \vec{\tau_i}, \dots \} \andalso
      \Gamma \vdash_\mEXP \tau_j' \andalso
      \Gamma \vdash e_i : \tau_i[\tau_j'/\alpha_j]}
     {\Gamma \vdash c(\vec{e_i}) : D ~ \vec{\tau_j'}}}
  {\infrule[$e$-Quote]
     {\Gamma \vdash \phi : \tau}
     {\Gamma \vdash \qq{\phi} : \tau}}

  ~ \\[0.5em]

  \sidebyside[.34][.65]
  {\infrule[$e$-Rel]
     {p \subseteq \vec{\tau_i} \in \Phi \andalso
      \Gamma \vdash e_i : \tau_i}
     {\Gamma \vdash p(\vec{e_i}) : \bool}}
  {\infrule[$e$-Fun]
     {f : \forall \vec{\alpha_j}, \vec{\tau_i} \rightarrow \tau \in \Phi \andalso
      \Gamma \vdash_\mEXP \tau_j' \andalso
      \Gamma \vdash e_i : \tau_i[\tau_j'/\alpha_j]}
     {\Gamma \vdash f(\vec{e_i}) : \tau[\tau_j'/\alpha_j]}}

  {\infrule[$e$-Op]
     {\typeof(\mathord{\otimes}) = \forall \vec{\alpha_j}, \vec{\tau_i} \rightarrow \tau \andalso
      \Gamma \vdash_\mEXP \tau_j' \andalso
      \Gamma \vdash e_i : \tau_i[\tau_j'/\alpha_j]}
     {\Gamma \vdash \operation{\vec{e_i}} : \tau}}

  {\infrule[$e$-If]
     {\Gamma \vdash e_1 : \bool \andalso
      \Gamma \vdash e_2 : \tau \andalso
      \Gamma \vdash e_3 : \tau}
     {\Gamma \vdash \ite{e_1}{e_2}{e_3} : \tau}}

  {\infrule[$e$-Match]
     {\Gamma \vdash e : D ~ \vec{\tau_j} \andalso
      \Delta(D) = \forall \vec{\alpha_j}, \{ \dots, c_i : \vec{\tau_k}, \dots \} \andalso
      \Gamma,\overrightarrow{X_k:\tau_k[\tau_j/\alpha_j]} \vdash e_i : \tau}
     {\Gamma \vdash \match{e}{\overrightarrow{c_i(\vec{X_k}) \rightarrow e_i}} : \tau}}

  \caption{Typing rules: expressions; implicit parameters are in \implicit{gray}}
  \label{fig:types-expressions}
\end{figure}

We split the rules for expressions $e$ and formulas $\phi$ in two
parts (Figures~\ref{fig:types-expressions}
and~\ref{fig:types-formulas}, respectively).
Expression typing is conventional for functional languages. 
We adopt a declarative style for type substitutions
(\rn{$e$-Ctor}, \rn{$e$-Fun}, \rn{$e$-Op}, \rn{$e$-Match}). Our actual
implementation uses Hindley--Damas--Milner type
inference~\citep{hindley,damas} to find the correct types to use.
As to \Name-specific features, we ensure type well formedness is in
$\mEXP$-mode; the $\qq{\phi}$ expression switches from expression mode
to formula mode.
Relations $p \subseteq \vec{\tau_i} \in \Phi$ are treated as if they
are functions of type $\vec{\tau_i} \rightarrow \bool$. Since Datalog
predicates can occur in functional terms, we must use the control-flow
graph of the program when analyzing for stratification. Consider the
following program:
\begin{lstlisting}
fun f(X : bv[32]) : bv[32] = if p(X) then ... else X
p(Y) :- q(Y, Y).
q(A, B) :- r(A), B = f(A).
r(42).
\end{lstlisting}
Here the relation \ic{q} calls the function \ic{f}, which in turn relies on the
negation of the relation \ic{p} (since the behavior of \ic{f} is conditioned on
the contents of \ic{p}).
As \ic{p} is defined in terms of \ic{q}, this
leads to a circularity we want to avoid: It is possible to derive \ic{q(42,
42)}, but this derivation nonsensically relies on \ic{q(42, 42)} being false
(since it requires that \ic{p(42)} is false).

\begin{figure}[pt]
  \hdr{SMT constructors and formula well formedness}{
    \fbox{$\implicit{\Delta; \Phi; {}} \Gamma \vdash \cSMT{...} : \vec{\tau_i} \rightarrow \tau$}
    \quad
    \fbox{$\implicit{\Delta; \Phi; {}} \Gamma \vdash \phi : \tau$}
  }

  ~ \\[0.5em]
  
  \sidebyside[.4][.58]
  {\infrule[$c$-SMT-Var]
    {\Gamma \vdash_\mSMT t}
    {\Gamma \vdash \cVAR[x,t] : \cdot \rightarrow \sym{t}}}
  {\infrule[$c$-SMT-Const]
    {}
    {\Gamma \vdash \cCONST[k] : \cdot \rightarrow \smt{\typeof(k)}}}

%  \sidebyside[.54][.45]

  {\infrule[$c$-SMT-Let]
    {\Gamma \vdash_\mSMT t_1 \andalso
     \Gamma \vdash_\mSMT t_2}
    {\Gamma \vdash \cLET : \sym{t_1} \times \smt{t_1} \times \smt{t_2} \rightarrow \smt{t_2}}}
    
  {\infrule[$c$-SMT-Ctor]
    {\Delta(D) = \forall \vec{\alpha_j}, \{ \dots, c : \vec{\tau_i}, \dots \} \andalso
     \Gamma \vdash_\mSMT \tau_i[t_j'/\alpha_j] \andalso
     \Gamma \vdash_\mSMT t_j' \\
}
    {\Gamma \vdash \cCTOR[c] : \overrightarrow{\tosmt(\tau_i[t_j'/\alpha_j])} \rightarrow \smt{(D ~ \overrightarrow{t_j'})}}
  }

  ~ \\[0.25em]

  {\infrule[$c$-SMT-Forall]
    {\Gamma \vdash_\mSMT t_1}
    {\Gamma \vdash \cFORALL : \sym{t_1} \times \smt{\bool} \rightarrow \smt{\bool}}}
  {\infrule[$c$-SMT-UFun]
    {\uf : \vec{t_i} \rightarrow t \in \Phi}
    {\Gamma \vdash \cUF[\uf] : \overrightarrow{\smt{t_i}} \rightarrow \smt{t}}}

  ~ \\[0.75em]

  \threesidebyside[.25][.25][.4]
  {\infrule[$\phi$-Promote]
     {\Gamma \vdash \phi : \sym{t}}
     {\Gamma \vdash \phi : \smt{t}}}
  {\infrule[$\phi$-Unquote]
     {\Gamma \vdash e : \tau \andalso \Gamma \vdash_\mSMT \tau}
     {\Gamma \vdash \unq{e} : \tosmt(\tau)}}
  {\infrule[$\phi$-Ctor]
    {\Gamma \vdash \cSMT{c} : \vec{\tau_i} \rightarrow \tau \andalso
     \Gamma \vdash \phi_i : \tau_i'      
    }
    {\Gamma \vdash \cSMT{c}(\vec{\phi_i}) : \tau}}

  ~ \\[0.5em]

  \hdr{Conversion to SMT types}{
    \quad \fbox{$\erase(\tau) = t$}
    \quad \fbox{$\tosmt(\tau) = \tau$}}

  \sidebyside[.45][.45][t]
  {\[\begin{array}{rcl}
      \erase(B) &=& B \\
      \erase(D ~ \vec{\tau_i}) &=& D ~ \overrightarrow{\erase(\tau_i)} \\
      \erase(\smt{t}) &=& \erase(t) \\
      \erase(\sym{t}) &=& \erase(t) \\
    \end{array}\]}
  {\[\begin{array}{rcl}
      \tosmt(t) &=& \smt{\erase(t)} \\
      \tosmt(\smt{t}) &=& \smt{\erase(t)} \\
      \tosmt(\sym{t}) &=& \sym{\erase(t)} \\
    \end{array}\]}

  \caption{Typing rules: SMT constructors and formulas; conversion to SMT types}
  \label{fig:types-formulas}
\end{figure}

While \Name's expressions \emph{compute} values, the \Name's
formulas \emph{construct} ASTs, to be shipped off to an SMT solver.
Our formal account here uniformly uses SMT constructors to model the
SMT syntax, but our implementation offers special-purpose syntax.
For example, we write $\cVAR[x,\bv{32}]$ in our formalism to name a 32-bit
integer variable $x$, while in our implementation one might write
\ic{\#x[bv[32]]}.

Every \rn{$\phi$-$\dots$} typing rule generates a value with an SMT
type, i.e., either $\sym{t}$ or $\smt{t}$ for SMT types $t$, i.e.,
$\Gamma \vdash_\mSMT t$ (Lemma~\ref{lem:regularity}). The
\rn{$c$-SMT-*} rules yield $\sym{t}$ and $\smt{t}$.
SMT variables $\cVAR[x,t]$ are written in lowercase to emphasize
their distinction from expression variables $X$; these SMT variables
will be used as names in the formulas sent to the SMT solver.
We keep track of which terms are SMT variables $\cVAR[x,t]$ of
type $\sym{t}$ (generated by \rn{$c$-SMT-Var}) and which are plain SMT
formulas of type $\smt{t}$ (all other rules). We treat $\sym{t}$ as a
subtype of $\smt{t}$ (\rn{$\phi$-Promote}).

The $\qq{\phi}$ operator is an expression term that introduces
quoted SMT formulas represented as special, SMT constructors of the
form $\cSMT{...}$ (described below); the $\unq e$ operator is the corresponding
`unquote' operator that introduces an expression
(\rn{$\phi$-Unquote}).

While unquoting generally suffices for embedding the results of
expressions in formulas, we treat constructors specially so that we
can mix concrete and symbolic (i.e., SMT) arguments in a single
data type constructor (\rn{$c$-SMT-Ctor}): we assign them types that
are fully SMT-ized via the $\tosmt$ function, but unquoting allows for
easy mixing of values of SMT-types $t$ as if they were of type
$\smt{t}$.
The $\tosmt$ metafunction alters the type of $e$ to make sure it is
SMT representable; $\tosmt$ relies on an $\erase$ function to avoid
nesting $\smt{\dots}$ and $\sym{\dots}$ type constructors. One can
only run these functions on SMT types.
For example, we can write terms like the following in concrete syntax:
\[ \letin{H}{5}{\qq{\mathsf{cons}(H, \#l[\bv{32} ~ \mathsf{list}])}} \]
which desugars to the SMT constructors:
\[ \letin{H}{5}{\qq{\cCTOR[\mathsf{cons}](\unq{H}, \cVAR[l,\bv{32} ~ \mathsf{list}]())}} \]
Note that $H$ is an expression variable and $l$ is an SMT variable;
the type conversion in \rn{$\phi$-Unquote} lets us mix them in the
same list of 32-bit numbers.
The $\sym{t}$ type is used in \rn{$\phi$-Let} and \rn{$\phi$-Forall},
which construct SMT formuale that use binders.
The only way to get a value of type $\sym{t}$ is either
with \rn{$c$-SMT-Var}/\rn{$\phi$-Ctor} or with \rn{$\phi$-Unquote}, as in
$\letin{X}{\qq{\cVAR[x,\bv{32}]}}{\qq{\unq X}}$.

Uninterpreted functions must be applied to appropriate SMT types
(\rn{$\phi$-UFun}); recall that \rn{$\Phi$-UFun} ensures that each
uninterpreted function's types are SMT representable.

Finally, there are a suite of SMT constructors of the form
$\cSMT{\dots}$.
Each of these special $\cSMT{\dots}$ constructors is treated as an
ordinary constructor by the operational semantics, even though the
constructors don't appear in $\Delta$.
Rather than making SMT terms opaque, we model them with constructors
to allow for matching on generated formulae in \rn{$P$-EqSMT-BF}
and \rn{$P$-Eq-FB}. The types of the SMT constructors
$\cSMT{\dots}$ are given in Figure~\ref{fig:types-formulas}.
It is a crucial invariant that all of these types be SMT types: we
would not want to treat an SMT variable $\cVAR[x,\bool]$ as though it
were an \emph{actual} $\bool$!
Reusing the \rn{$c$-SMT-$\dots$} rules is convenient---we need only
state the types of these constructors once and we get precise types in
our premises.
Several SMT constructors take special arguments in square brackets: $\cVAR[x,t]$ is a 0-ary SMT constructor, while $\cVAR$ itself is a family of SMT constructors for given variable names $x$ and pre-types $t$; similarly, $\cCONST[k]$ is a 0-ary SMT constructor, while $\cCONST$ itself is a family of SMT constructors for given constants $k$. The embedding of data type constructors $\cCTOR[c]$ is similarly parameterized on a constructor name $c$, and the embedding of uninterpreted functions $\cUF[\uf]$ takes an uninterpereted function as a parameter.
Separating these parameters from the interesting, $\phi$-shaped
subparts of each SMT constructor lets us reuse
the \rn{$c$-SMT-$\dots$} rules when typing premises that might bind to
subparts of an SMT formula (\rn{$P$-EqSMT-$\dots$},
Figure~\ref{fig:types-rules}).

Our formal model elides some of the detail of our SMT encoding, such
as constructors for SMT operations like bit vector manipulation or
equality. These operations are all encoded as more SMT-specific
constructors, i.e.,
$\cCTOR[\mathsf{bv_{32}\_add}](\cVAR[x,\bv{32}], \cCONST[1]))$
represents result of adding the 32-bit vector $x$ and $1$.
There are some subtle issues around polymorphism and determinacy. We
give a monomorphic interpretation of SMT here, though our SMT
constructors can work with our polymorphic data types. Our
implementation treats equality and other polymorphic SMT operations
specially, where each use of a polymorphic operator must be fully
instantiated. In practice our implementation can usually infer the
instantiation; users must annotate in those places we cannot infer.

\section{Operational semantics}
\label{sec:opsem}

\begin{figure}[t]

  \[\begin{array}{lrcl}
    \multicolumn{4}{l}{\textbf{Namespaces}} \\
    \text{World} & \world &\in& \mathrm{PredVar} \rightarrow \mathcal{P}
                                (\mathrm{Val} \times \cdots \times \mathrm{Val}) \\
    \text{World or error} & \world_\bot &\in& \mathrm{World} + \mathrm{Error} \\
    \text{Substitution} & \theta     &\in& \mathrm{Var} \rightharpoonup \mathrm{Val} \\
    \text{Substitution or error} & \theta_\bot &\in& \mathrm{Substitution} + \mathrm{Error} \\
    \multicolumn{4}{l}{\textbf{Values}} \\
    \text{Results} & v_\bot &::=& v \BNFALT \bot \\
    \text{Values} & v \in \mathrm{Val} &::=& k \BNFALT c(\vec{v_i}) \\

  ~ \\[0.25em]

    \text{Unifiable term} & u \in \mathrm{UTerm} &::=& X \BNFALT k \BNFALT c(\vec{u_i}) \\
  \end{array} \]

  \hdr{Substitution and world well formedness}{
    \quad \fbox{$\implicit{\Delta; \Phi; {}} \Gamma \models \theta$}
    \quad \fbox{$\Delta; \Phi \models \world$}
  }

  \sidebyside[.45][.51][t]
  {\[
    \begin{array}{c}
    \Gamma \models \theta \\
    \Leftrightarrow \\
    \forall X \in \dom(\Gamma) ~ \left\{
      \begin{array}{l@{}}
        \text{(1) } X \in \dom(\theta) \\
        \text{(2) } \cdot \vdash \theta(X) : \Gamma(X) \\
      \end{array}
      \right.
    \end{array}
  \]}
  {\[ 
  \begin{array}{c}
    \Delta; \Phi \models \world \\
    \Leftrightarrow \\ 
    \forall p \subseteq \vec{\tau_i} \in \Phi ~ \left\{
      \begin{array}{l@{}}
          \text{(1) } p \in \dom(\world) \\
          \text{(2) } \vec{v_j} \in \world(p) \Rightarrow i = j \\
          \text{(3) } \forall \vec{v_i} \in \world(p), ~ \Delta; \Phi; \cdot \vdash v_i : \tau_i \\
      \end{array} 
      \right.
  \end{array}
  \]}

  \caption{Definitions for semantics}
  \label{fig:opsem-defs}
\end{figure}

\Name's operational semantics operates over \emph{worlds} $\world$ and substitutions $\theta$ (Figure~\ref{fig:opsem-defs}); the semantics is a mix of small-step rules modeling a single application of a Datalog rule (Figure~\ref{fig:opsem-clause}), which depend on a small-step rules explaining how premises unify (Figures~\ref{fig:opsem-premise} and~\ref{fig:opsem-unification}); the premise semantics in turn depends on a semantics of expressions (Figures~\ref{fig:opsem-expression} and~\ref{fig:opsem-expression-error}) and formulas (Figure~\ref{fig:opsem-formula}).

Our worlds $\world$ are (subsets of) Herbrand models.
Our small-step semantics iteratively builds up a world that is in fact
a Herbrand model of the original relations in the program. 
We could have modeled our semi-naive evaluation model for \Name in
more detail, showing that all programs generate a world $\world$ that
is a well typed Herbrand model of the user's program (possibly taking
infinite time to do so). Doing so wouldn't add anything materially
interesting to our formulation.

Throughout, the type system's goal is prevent a program yielding
$\bot$, the bottom ``wrong'' value.
Such a value denotes a serious, unrecoverable error, such as using a
relation with the wrong arity or conditioning on a non-boolean.
It is important to distinguish bad, $\bot$-yielding programs from
those that simply fail to step. The goal of Datalog evaluation is to
reach a fixed point, i.e., to be unable to step!
Finally, as is common, we assume that built-in operations do not yield
$\bot$, i.e., they are total. While we could in principle design a
type system for \Name that avoids, say, division by zero, we are more
interested in making the hard parts easy (generating well typed SMT
formulas) rather than making the easy parts foolproof (statically
protecting partial functions).

Rules of the form \rn{$\dots$-E$n$} denote $\bot$-yielding rules. Each
such rule characterizes a form of wrongness avoided by our static type
system.
We write $v_\bot$ to denote the disjoint sum of values $v$ and the
wrong value $\bot$.

\begin{figure}[t]
  \hdr{Clause semantics}{ \quad \fbox{$\vec{F}; \world \vdash
    H \stepsto \world_\bot$} }
  
  ~ \\[0.5em]

  \infrule[Clause]{
    \cdot    \vdash P_{0} \stepsto \theta_1 \quad \dots \quad
    \theta_i \vdash P_{i} \stepsto \theta_{i+1} \quad \dots \quad
    \theta_n \vdash P_{n} \stepsto \theta
  }{
    \vec{F}; \world \vdash p(\vec{X_j}) \horn \vec{P_i} \stepsto \world[p \mapsto \world(p) \cup \{\theta(\vec{X_j})\}]
  }

  \infrule[Clause-E1]{
    \cdot    \vdash  P_{0} \stepsto \theta_1 \quad \dots \quad
    \theta_i \vdash  P_{j} \stepsto \bot
  }{
    \vec{F}; \world \vdash p(\vec{X_j}) \horn \vec{P_i} \stepsto \bot
  }
  
  \infrule[Clause-E2]{
    \cdot    \vdash  P_{0} \stepsto \theta_1 \quad \dots \quad
    \theta_i \vdash  P_{i} \stepsto \theta_{i+1} \quad \dots \quad
    \theta_n \vdash  P_{n} \stepsto \theta \andalso
    \vec{X_j} \not \subseteq \dom(\theta) 
  }{
    \vec{F}; \world \vdash p(\vec{X_j}) \horn \vec{P_i} \stepsto \bot
  }%
  \caption{Clause semantics}
  \label{fig:opsem-clause}
\end{figure}

During correct execution, \rn{Clause} takes a Horn clause
$p(\vec{X_j}) \horn \vec{P_i}$, executes each premise $P_i$ from left
to right, yielding a final substitution for the variables $\vec{X_j}$
in the head of the rule.
There are two possible failing rules. \rn{Clause-E1} simply propagates
the first error from a premise; \rn{Clause-E2} fails because not every $X_j$ in
the head of the rule is bound by the end.
Since \Name enforces the range restriction
($H$-Clause), \rn{Clause-E2} can never apply in a well typed program.
Finally, the \rn{Clause*} operational rules and the \rn{$H$-Clause}
typing rule both use the fixed, given order of premises for
checking. Different orderings induce different binding orders, some of
which may succeed and some of which may not.

\begin{figure}[t]
  \hdr{Premise semantics}{
    \quad \fbox{$\implicit{\vec{F}; {}} \world; \theta \vdash P \stepsto \theta_\bot$}
  }
  
  ~ \\[0.5em]

  \sidebyside
  {
    \infrule[PosAtom]{
      \vec{v} \in \world(p) \andalso
      \theta \vdash \vec{X} \Unify \vec{v} : \theta'_\bot
    }{
      \world; \theta \vdash p(\vec{X}) \stepsto \theta'_\bot
    }
  }{
    \infrule[NegAtom]{
      \theta(\vec{X})=\vec{v} \andalso \vec{v} \not \in \world(p)
    }{
      \world; \theta \vdash \mathord{!}p(\vec{X}) \stepsto \theta
    }
  }

  ~ \\[.5em]

  \sidebyside
  {
    \infrule[EqCtor]{
      \theta \vdash Y \unify c(\vec{X}) : \theta'_\bot
    }{
      \world; \theta \vdash {Y = c(\vec{X})} \stepsto \theta'_\bot
    }
  }{
    \infrule[EqSMT]{
      \theta \vdash Y \unify \cSMT{c'}(\vec{X}) : \theta'_\bot
    }{
      \world; \theta \vdash {Y = \qq{\cSMT{c'}(\vec{\unq{X}})}} \stepsto \theta'_\bot
    }
  }

  {
    \infrule[EqExpr]{
      \text{$e$ is not a constructor} \andalso
      \world; \theta \vdash e \StepstoE v \andalso
      \theta \vdash Y \unify v : \theta'_\bot
    }{
      \world; \theta \vdash {Y = e} \stepsto \theta'_\bot
    }
  }

  ~ \\[.25em]

  \sidebyside[.4][.6]
  {
    \infrule[NegAtom-E]{
      \vec{X} \not \subseteq \dom(\theta)
    }{
      \world; \theta \vdash \mathord{!}p(\vec{X}) \stepsto \bot
    }
  }{
    \infrule[EqExpr-E]{
      \text{$e$ is not a constructor} \andalso
      \world; \theta \vdash e \StepstoE \bot
    }{
      \world; \theta \vdash {Y = e} \stepsto \bot
    }
  }

  ~ \\[.25em]

  \threesidebyside
  {
    \infrule[NegExpr]{
      \world; \theta \vdash e \StepstoE v \andalso
      \theta(Y) \ne v
    }{
      \world; \theta \vdash {!(Y = e)} \stepsto \theta
    }
  }{
    \infrule[NegExpr-E1]{
      \world; \theta \vdash e \StepstoE \bot
    }{
      \world; \theta \vdash {!(Y = e)} \stepsto \bot
    }
  }{
    \infrule[NegExpr-E2]{
      Y \not\in \dom(\theta)
    }{
      \world; \theta \vdash {!(Y = e)} \stepsto \bot
    }
  }
  
  \caption{Premise semantics}
  \label{fig:opsem-premise}
\end{figure}

The premise semantics (Figure~\ref{fig:opsem-premise}) uses unification (Figure~\ref{fig:opsem-unification}) to match and bind variables.
Positive atoms $p(\vec{X_i})$ try to unify their arguments with a
tuple for $p$ drawn from the world $\world$ (\rn{PosAtom}).
Negative atoms $p(\vec{X_i})$ require that all of their arguments
$X_i$ are already bound (\rn{NegAtom}); failing to find such bound
terms yields an error (\rn{NegAtom-E}).
Rules for equations also use unification, whether for a constructor
over variables (\rn{EqCtor}) or an expression (\rn{EqExpr}).
The latter can fail if evaluation fails (\rn{EqExpr-E}).
Before discussing term evaluation, we give rules for unification.

\begin{figure}[t]
  \hdr{Value unification}{
    \quad \fbox{$\theta \vdash u \unify v \rhd \theta$}
    \quad \fbox{$\theta \vdash \vec{u} \unify \vec{v} \rhd \theta$}
  }

  ~ \\[0.5em]

  \sidebyside
  {\infrule[$uv$-Eq-Var]
     {\theta(X) = v}
     {\theta \vdash X \unify v \rhd \theta}
  }
  {\infrule[$uv$-Bind-Var]
     {X \not\in \dom(\theta)}
     {\theta \vdash X \unify v \rhd \theta[X \mapsto v]}
  }

  ~ \\[0.5em]

  \sidebyside
  {\infrule[$uv$-Constant]
     {}
     {\theta \vdash k \unify k \rhd \theta}
  }
  {\infrule[$uv$-Ctor]
     {\theta \vdash \vec{u_i} \unify \vec{v_i} \rhd \theta'}
     {\theta \vdash c(\vec{u_i}) \unify c(\vec{v_i}) \rhd \theta'}
  }

  ~ \\[0.5em]

  {\infrule[$\vec{u}\vec{v}$-All]
     {\theta \vdash u_0 \unify v_0 \rhd \theta_1 \andalso \dots \andalso
      \theta_1 \vdash u_i \unify v_i \rhd \theta_i \andalso \dots \andalso
      \theta_n \vdash u_n \unify v_n \rhd \theta'}
     {\theta \vdash \vec{u_i} \unify \vec{v_i} \rhd \theta}
  }

  ~ \\[0.5em]

  \hdr{Unification}{
    \quad \fbox{$\theta \vdash u \unify u : \theta_\bot$}
    \quad \fbox{$\theta \vdash \vec{u} \unify \vec{u} : \theta_\bot$}
  }

  ~ \\[0.5em]

  \sidebyside[.46][.53]{
    \infrule[$uu$-BB]{
      \theta(u_1) = v_1 \andalso
      \theta(u_2) = v_2 \andalso
      v_1 = v_2
    }{
      \theta \vdash u_1 \unify u_2 : \theta
    }
  }{
    \infrule[$uu$-FB]{
      \nexists v_1, \theta(u_1) = v_1 \andalso
      \theta(u_2) = v_2 \andalso
      \theta \vdash u_1 \unify v_2 \rhd \theta'
    }{
      \theta \vdash u_1 \unify u_2 : \theta'
    }
  }

  ~ \\[0.5em]

  \sidebyside[.56][.43]
  {
    \infrule[$uu$-BF]{
      \theta(u_1) = v_1 \quad
      \nexists v_2, \theta(u_2) = v_2 \quad
      \theta \vdash u_2 \unify v_1 \rhd \theta'
    }{
      \theta \vdash u_1 \unify u_2 : \theta'
    }
  }{
    \infrule[$uu$-FF]{
      \nexists v_1, \theta(u_1) = v_1 \quad
      \nexists v_2, \theta(u_2) = v_2
    }{
      \theta \vdash u_1 \unify u_2 : \bot
    }
  }

  ~ \\[0.5em]

  \sidebyside
  {\infrule[$\vec{u}\vec{u}$-All]
     {\theta \vdash u_i \unify u_i' \rhd \theta_i}
     {\theta \vdash \vec{u_i} \unify \vec{u_i'} \rhd \theta\vec{\theta_i}}
  }
  {\infrule[$\vec{u}\vec{u}$-All-E]
     {\dots \andalso \theta \vdash u \unify u' \rhd \bot \andalso \dots}
     {\theta \vdash \vec{u_i} \unify \vec{u_i'} \rhd \bot}
  }

  \[ \theta(c(\vec{u_i})) = c(\overrightarrow{\theta(u_i)}) \]

  \caption{Unification}
  \label{fig:opsem-unification}
\end{figure}

Unification is split into two levels. \emph{Unification} proper takes a pair of unifiable terms---values with variables in them---and tries to yield a substitution. \emph{Value unification} takes a unifiable term and a value and tries to yield a substitution.
The unification rules are of the form \rn{$uu$-$\dots$}. These rules
analyze the two unifiable terms to find which side is completely
bound---i.e., applying $\theta$ can completely fill in the
variables---and so can be passed to value unification as a value.
The \rn{B} and \rn{F} in these rules stand for \rn{B}ound
and \rn{F}ree. The only error in unification is in \rn{$uu$-FF}, when
neither unifiable term is bound to a value.
We write a case for when both unifiable terms are bound (\rn{$uu$-BB})
and require that they are directly equal---but it would also work to
drop this rule and rely on value unification to identify the equality.

Value unification rules are of the form \rn{$uv$-$\dots$}. The rules
here lookup variables in the unifiable term and either check that the
binding conforms to the given value (\rn{$uv$-Eq-Var},
cf. \rn{$X\tau$-Check}) or binds the value (\rn{$uv$-Bind-Var},
cf. \rn{$X\tau$-Bind}).
The remaining value unification rules match the structure of the
unifiable term to the structure of the value
(\rn{$uv$-Constant}, \rn{$uv$-Ctor}) or fold value unification along a
vector (\rn{$\vec{u}\vec{v}$-All}).
Value unification never produces $\bot$. It isn't an error when two
values fail to unify, since one might have to search through many
tuples for a relation in $\world$ to find a one that matches, say, a
given constructor.

\begin{figure}[t]
  \hdr{Expression semantics}{
    \quad \fbox{$\implicit{\vec{F}; {}} \world; \theta \vdash e \StepstoE v_\bot$}
    \quad \fbox{$\implicit{\vec{F}; {}} \world; \theta \vdash \vec{e} \StepstoVecE \vec{v}_\bot$}
  }

  ~ \\[0.5em]

  \sidebyside
  {
    \infrule[$\StepstoVecE$-Empty]{
    }{
      \world; \theta \vdash \cdot \StepstoVecE \cdot
    }
  }
  {
    \infrule[$\StepstoVecE$-All]{
      \world; \theta \vdash e \StepstoE v \andalso \world; \theta \vdash \vec{e} \StepstoVecE \vec{v}
    }{
      \world; \theta \vdash e, \vec{e} \StepstoVecE v, \vec{v}
    }
  }

  ~ \\[0.5em]

  \sidebyside{
    \infrule[$\StepstoE$-Ctor]{
      \world; \theta \vdash \vec{e_i} \StepstoVecE \vec{v_i}
    }{
      \world; \theta \vdash c(\vec{e_i}) \StepstoVecE c(\vec{v_i})
    }
  }{
    \infrule[$\StepstoE$-Ctor-E]{
      \world; \theta \vdash \vec{e_i} \StepstoVecE \bot
    }{
      \world; \theta \vdash c(\vec{e_i}) \StepstoE \bot
    }
  }

  ~ \\[0.5em]

  \threesidebyside[.28][.28][.37]
  {
    \infrule[$\StepstoE$-Const]{
    }{
      \world; \theta \vdash k \StepstoE k
    }
  }
  {
    \infrule[$\StepstoE$-Var]{
      \theta(X) = v
    }{
      \world; \theta \vdash X \StepstoE v
    }
  }
  { 
    \infrule[$\StepstoE$-Quote]{
      \world; \theta \vdash \phi \StepstoPhi v_\bot
    }{
      \world; \theta \vdash \qq{\phi} \StepstoE v_\bot
    }
  }

  ~ \\[0.5em]

  {
    \infrule[$\StepstoE$-Op]{
      \world; \theta \vdash \vec{e} \StepstoVecE \vec{v}
      \andalso \denot{\otimes}(\vec{v}) = v
    }{
      \world; \theta \vdash \operation{\vec{e}} \StepstoE v
    }
  }
  ~ \\[0.5em]

  {
    \infrule[$\StepstoE$-Fun]{
      \fun{f}{\vec{X_i} : \vec{\tau_i}}{\tau}{e} \in \vec{F} \andalso
      \world; \theta \vdash \vec{e_i} \StepstoVecE \vec{v_i} \andalso
      \world; \theta[\vec{X_i} \mapsto \vec{v_i}] \vdash e \StepstoE v_\bot
    }{
      \world; \theta \vdash f(\vec{e_i}) \StepstoE v_\bot
    }
  }

  ~ \\[0.5em]

  \sidebyside
  {
    \infrule[$\StepstoE$-Rel-T]{
      \world; \theta \vdash \vec{e_i} \StepstoVecE \vec{v_i} \andalso
      \vec{v_i} \in \world(p)
    }{
      \world; \theta \vdash p(\vec{e_i}) \StepstoE \true
    }
  }
  {
    \infrule[$\StepstoE$-Rel-F]{
      \world; \theta \vdash \vec{e_i} \StepstoVecE \vec{v_i} \andalso
      \vec{v_i} \not\in \world(p)
    }{
      \world; \theta \vdash p(\vec{e_i}) \StepstoE \false
    }
  }

  {
    \infrule[$\StepstoE$-Let]{
      \world; \theta \vdash e_1 \StepstoE v_1 \andalso 
      \world; \theta[X \mapsto v_1] \vdash e_2 \StepstoE v_\bot
    }{
      \world; \theta \vdash \letin{X}{e_1}{e_2} \StepstoE v_\bot
    }
  }

  \infrule[$\StepstoE$-Match]{
    \world; \theta \vdash e \StepstoE c(\vec{v_i}) \andalso
    \world; \theta[\overrightarrow{X_i \mapsto v_i}] \vdash e \StepstoE v_\bot
  }{
    \world; \theta \vdash \match{e}{\dots c(\vec{X_i}) \rightarrow e \dots} \StepstoE v_\bot
  }

  ~ \\[0.5em]

  \sidebyside{
    \infrule[$\StepstoE$-IteT]{
      \world; \theta \vdash e_1 \StepstoE \true \andalso \world; \theta \vdash e_2 \StepstoE v_\bot
    }{
      \world; \theta \vdash \ite{e_1}{e_2}{e_3} \StepstoE v_\bot
    }
  }{ 
    \infrule[$\StepstoE$-IteF]{
      \world; \theta \vdash e_1 \StepstoE \false \andalso \world; \theta \vdash e_3 \StepstoE v_\bot
    }{
      \world; \theta \vdash \ite{e_1}{e_2}{e_3} \StepstoE v_\bot
    }
  }

  \caption{Expression semantics}
  \label{fig:opsem-expression}
\end{figure}

The expression semantics is an entirely conventional big-step
semantics using explicit substitutions.
The operational rules implicitly take the function definitions
$\vec{F}$ for use in applications (\rn{$\StepstoE$-Fun}).

\begin{figure}[t]
  \hdr{Expression semantics (continued)}{
    \quad \fbox{$\implicit{\vec{F}; {}} \world; \theta \vdash e \StepstoE v_\bot$}
    \quad \fbox{$\implicit{\vec{F}; {}} \world; \theta \vdash \vec{e} \StepstoVecE \vec{v}_\bot$}
  }

  ~ \\[0.5em]

  {
    \infrule[$\StepstoVecE$-All-E]{
      \world; \theta \vdash \vec{e_i} \StepstoVecE \vec{v_i} \andalso \world; \theta \vdash e \StepstoE \bot
    }{
      \world; \theta \vdash \vec{e_i},e,\vec{e_j} \StepstoVecE \bot
    }
  }

  ~ \\[0.5em]

  \sidebyside[.44][.55]
  {
    \infrule[$\StepstoE$-Var-E]{
      X \not \in \dom(\theta)
    }{
      \world; \theta \vdash X \StepstoE \bot
    }
  }
  {
    \infrule[$\StepstoE$-Let-E]{
      \world; \theta \vdash e_1 \StepstoE \bot
    }{
      \world; \theta \vdash \letin{X}{e_1}{e_2} \StepstoE \bot
    }
  }

  ~ \\[0.5em] 

  \sidebyside[.44][.55]
  {
    \infrule[$\StepstoE$-Op-E1]{
      \world; \theta \vdash \vec{e} \StepstoVecE \bot
    }{
      \world; \theta \vdash \operation{\vec{e}} \StepstoE \bot
    }
  }{
    \infrule[$\StepstoE$-Op-E2]{
      \world; \theta \vdash \vec{e} \StepstoVecE \vec{v} \andalso 
      \vec{v} \not\in \dom(\denot{\otimes}) 
    }{
      \world; \theta \vdash \operation{\vec{e}} \StepstoE \bot
    }
  }

  ~ \\[0.5em]

  \threesidebyside
  {
    \infrule[$\StepstoE$-Fun-E1]{
      \world; \theta \vdash \vec{e_i} \StepstoVecE \bot
    }{
      \world; \theta \vdash f(\vec{e_i}) \StepstoE \bot
    }
  }
  {
    \infrule[$\StepstoE$-Fun-E2]{
      \fun{f}{\vec{X_i} : \vec{\tau_i}}{\tau}{e} \in \vec{F} \andalso
      i \ne j
    }{
      \world; \theta \vdash f(\vec{e_j}) \StepstoE \bot
    }
  }
  {
    \infrule[$\StepstoE$-Fun-E3]{
      f \not\in \vec{F}
    }{
      \world; \theta \vdash f(\vec{e_j}) \StepstoE \bot
    }
  }

  ~ \\[0.5em]

  \threesidebyside[.29][.38][.29]
  {
    \infrule[$\StepstoE$-Rel-E1]{
      \world; \theta \vdash \vec{e_i} \StepstoVecE \bot
    }{
      \world; \theta \vdash p(\vec{e_i}) \StepstoE \bot
    }
  }
  {
    \infrule[$\StepstoE$-Rel-E2]{
      \world(p) \subseteq \mathcal{P}(\overrightarrow{\mathrm{Val}_i}) \andalso
      i \ne j
    }{
      \world; \theta \vdash p(\vec{e_j}) \StepstoE \bot
    }
  }
  {
    \infrule[$\StepstoE$-Rel-E3]{
      p \not\in \dom(\world)
    }{
      \world; \theta \vdash p(\vec{e_j}) \StepstoE \bot
    }
  }

  ~ \\[0.5em]

  \sidebyside[.5][.5]
  {
    \infrule[$\StepstoE$-Match-E1]{
      \world; \theta \vdash e \StepstoE \bot 
    }{
      \world; \theta \vdash \match{e}{\overrightarrow{c_i(\vec{X_j}) \rightarrow e_i}} \StepstoE \bot
    }
  }
  {
    \infrule[$\StepstoE$-Match-E2]{
      \world; \theta \vdash e \StepstoE v \andalso
      v \ne c(\vec{v'})
    }{
      \world; \theta \vdash \match{e}{\overrightarrow{c_i(\vec{X_j}) \rightarrow e_i}} \StepstoE \bot
    }
  }

  ~ \\[0.5em]

  \sidebyside[.5][.5]
  {
    \infrule[$\StepstoE$-Match-E3]{
      \world; \theta \vdash e \StepstoE c(\vec{v_k}) \andalso
      c \not\in \{ \vec{c_i} \}
    }{
      \world; \theta \vdash \match{e}{\overrightarrow{c_i(\vec{X_j}) \rightarrow e_i}} \StepstoE \bot
    }
  }
  {
    \infrule[$\StepstoE$-Match-E4]{
      \world; \theta \vdash e \StepstoE c(\vec{v_k}) \andalso
      j \ne k
    }{
      \world; \theta \vdash \match{e}{\overrightarrow{\dots c(\vec{X_j} \dots ) \rightarrow e_i}} \StepstoE \bot
    }
  }

  ~ \\[0.5em]

  \sidebyside[.45][.54]
  {
    \infrule[$\StepstoE$-Ite-E1]{
      \world; \theta \vdash e_1 \StepstoE \bot
    }{
      \world; \theta \vdash \ite{e_1}{e_2}{e_3} \StepstoE \bot
    }
  }
  {
    \infrule[$\StepstoE$-Ite-E2]{
      \world; \theta \vdash e_1 \StepstoE v \andalso
      v \not\in \{ \true, \false \}
    }{
      \world; \theta \vdash \ite{e_1}{e_2}{e_3} \StepstoE \bot
    }
  }

  \caption{Expression semantics (error rules)}
  \label{fig:opsem-expression-error}
\end{figure}

There are a variety of wrong behaviors prevented by our type system,
mostly concerning mismatches between values and elimination forms:
unbound variables (\rn{$\StepstoE$-Var-E});
mistyped arguments to built-in operations (\rn{$\StepstoE$-Op-E2});
function, relation and constructor arity errors
(\rn{$\StepstoE$-Fun-E2}, \rn{$\StepstoE$-Rel-E2}, \rn{$\StepstoE$-Match-E4});
non-existent functions and relations
(\rn{$\StepstoE$-Fun-E3}, \rn{$\StepstoE$-Rel-E3});
conditionals on inappropriate values
(\rn{$\StepstoE$-Match-E2}, \rn{$\StepstoE$-Ite-E2});
and ill formed constructor names (\rn{$\StepstoE$-Match-E3});
The remaining rules propagate errors
(\rn{$\StepstoE$-Let-E}, \rn{$\StepstoE$-Op-E1}, \rn{$\StepstoE$-Fun-E1}, \rn{$\StepstoE$-Rel-E1}, \rn{$\StepstoE$-Match-E1}, \rn{$\StepstoE$-Ite-E1}).
As mentioned in the early discussion of our semantics in this
section, \rn{$\StepstoE$-Op-E2} is \emph{not} about division by zero
(a form of going wrong our type system \emph{doesn't} prevent), but
about mis-application of built-in functions, e.g., taking the boolean
negation of a number.

\begin{figure}[pt]
  \hdr{Formula semantics}{
    \quad \fbox{$\implicit{\vec{F}; {}} \world; \theta \vdash \phi \StepstoPhi v_\bot$}
    \quad \fbox{$\implicit{\vec{F}; {}} \world; \theta \vdash \vec{\phi} \StepstoVecPhi \vec{v}_\bot$}
  }

  ~ \\[0.5em]

  \sidebyside[.4][.58]{
    \infrule[$\StepstoVecPhi$-Empty]{
    }{
      \world; \theta \vdash \cdot \StepstoVecPhi \cdot
    }
  }{
    \infrule[$\StepstoVecPhi$-All]{
      \world; \theta \vdash \phi \StepstoPhi v \andalso \world; \theta \vdash \vec{\phi_i} \StepstoVecPhi \vec{v_i}
    }{
      \world; \theta \vdash \phi, \vec{\phi_i} \StepstoVecPhi v, \vec{v_i}
    }
  }

  ~ \\[0.5em]

  {
    \infrule[$\StepstoVecPhi$-All-E]{
      \world; \theta \vdash \vec{\phi_i} \StepstoVecPhi \vec{v_i} \andalso 
      \world; \theta \vdash \phi \StepstoPhi \bot
    }{
      \world; \theta \vdash \vec{\phi_i},\phi,\vec{\phi_j} \StepstoVecPhi \bot
    }
  }

  ~ \\[0.5em]

  \sidebyside
  {
    \infrule[$\StepstoPhi$-Unquote]{
      \world; \theta \vdash e \StepstoE v
    }{
      \world; \theta \vdash \unq{e} \StepstoPhi \tosmt(v)
    }
  }
  {
    \infrule[$\StepstoPhi$-Unquote-E]{
      \world; \theta \vdash e \StepstoE \bot
    }{
      \world; \theta \vdash \unq{e} \StepstoPhi \bot
    }
  }
  ~ \\[0.5em]

  \sidebyside[.55][.44]
  {
    \infrule[$\StepstoPhi$-Ctor]{
      \world; \theta \vdash \vec{\phi_i} \StepstoVecPhi \vec{v_i} \andalso
    }{
      \world; \theta \vdash \cSMT{c}(\vec{\phi}) \StepstoPhi \cSMT{c}(\vec{v})
    }
  }{
    \infrule[$\StepstoPhi$-Ctor-E]{
      \world; \theta \vdash \vec{\phi_i} \StepstoVecPhi \bot
    }{
      \world; \theta \vdash \cSMT{c}(\vec{\phi_i}) \StepstoPhi \bot
    }
  }

  {
    \infrule[$\StepstoPhi$-SMT-Value]
      {}
      {\world; \theta \vdash \cSMT{c}(\vec{v_i}) \StepstoPhi \cSMT{c}(\vec{v_i})}
  }

  ~ \\[0.5em]

  \hdr{SMT conversion}{
    \quad \fbox{$\tosmt(v) = v$}
  }

  \sidebyside[.45][.47][t]
  {\[ 
   \begin{array}{@{}rcl@{}}
     \tosmt(k) &=& \cCONST[k]() \\
     \tosmt(c(\vec{v_i})) &=& \cCTOR[c](\overrightarrow{\tosmt(v_i)}) \\
     \tosmt(\cCONST[k]()) &=& \cCONST[k]() \\[0.35em]
     \tosmt(\cVAR[x,t]()) &=& \cVAR[x,t]() \\[0.35em]
    \end{array} 
  \]}
  {\[
   \begin{array}{@{}rcl@{}}
     \tosmt(\cCTOR[c](\vec{v_i})) &=& \cCTOR[c](\vec{v_i}) \\[0.35em]
     \tosmt(\cLET(v_1, v_2, v_3)) &=& \cLET(v_1, v_2, v_3) \\[0.35em]
     \tosmt(\cFORALL(v_1, v_2)) &=& \cFORALL(v_1, v_2) \\[0.35em]
     \tosmt(\cUF[\uf](\vec{v_i})) &=& \cUF[\uf](\vec{v_i}) \\[0.35em]
   \end{array}
  \]}

  \caption{Formula semantics}
  \label{fig:opsem-formula}
\end{figure}

The operational semantics on formulas is simple: the rules generate
ASTs for the SMT solver using the $\cSMT{\dots}$ constructors:
constants (\cCONST), SMT variables (\cVAR), SMT data types (\cCTOR),
let bindings (\cLET), quantification (\cFORALL), and uninterpreted
function application (\cUF).

When unquoting values resulting from evaluating expressions, we use
the $\tosmt$ function to translate expression values into the SMT's
AST.
The $\tosmt$ function is an identity on SMT ASTs, but it explicitly
tags the constants and \Name-defined constructors using $\cCONST$ and
$\cCTOR$.

\section{Metatheory}
\label{sec:metatheory}

We break the metatheory into two parts: lemmas characterizing the SMT
conversion (Section~\ref{sec:metatheory-smt}) and lemmas showing type
safety (Section~\ref{sec:metatheory-safety}).
The SMT lemmas culminate in two proofs:
first, regularity (Lemma~\ref{lem:regularity}) guarantees that (a) every type
or context generated by the operational semantics is well formed and
(b) that formula evaluation generates well typed SMT ASTs;
second, we show that SMT conversion of values agrees with SMT
conversion of types (Lemma~\ref{lem:tosmt-well-typed}).
Type safety culminates in theorems showing that premises don't yield
$\bot$ and generate well typed substitutions
(Lemma~\ref{lem:typesafe-premise}) and so Horn clauses (a) never yield
$\bot$ (Theorem~\ref{thm:typesafe-program-safe}) and (b) take well
typed worlds to well typed worlds
(Theorem~\ref{thm:typesafe-program-pres}).

\subsection{SMT Conversion}
\label{sec:metatheory-smt}

We show a variety of properties of the erasure and SMT conversion
functions:
$\mSMT$ is a sub-kind of $\mEXP$ (Lemma~\ref{lem:lift-msmt});
erasures and SMT conversion yield well formed types from SMT types
(Lemmas~\ref{lem:erase-wf},~\ref{lem:smt-wf-inversion},~\ref{lem:tosmt-wf},
and~\ref{lem:tosmt-defined});
weakening and strengthening of typing contexts (Lemmas~\ref{lem:weakening} and~\ref{lem:types-strengthen});
type variable substitution (Lemma~\ref{lem:substitution})---we have no
need of a value substitution lemma because our semantics uses
environments;
regularity (Lemma~\ref{lem:regularity});
and, finally, that SMT conversion of values agrees with SMT conversion
of types (Lemma~\ref{lem:tosmt-well-typed}).

\begin{lemma}[$\mSMT$ is a subkind of $\mEXP$]
  \label{lem:lift-msmt}
  If $\Gamma \vdash_\mSMT \tau$ then $\Gamma \vdash_\mEXP \tau$.
\begin{proof}
  By induction on $\tau$.
  \begin{proofcases}

  \item[($\tau=B$)] Immediate: \rn{$\tau$-Base} allows any $m$.
    
  \item[($\tau=\alpha$)] Contradictory---type variables are only well formed at $\mEXP$.
    
  \item[($\tau=D ~ \vec{\tau_i}$)] By the IH on each $\tau_i$ and \rn{$t$-ADT}.

  \item[($\tau=\smt{t}$)] Immediate: \rn{$\tau$-SMT} allows any $m$.

  \item[($\tau=\sym{t}$)] Immediate: \rn{$\tau$-SMT} allows any $m$.

  \item[($\tau=\MODEL$)] Contradictory---$\MODEL$ is only well formed at $\mEXP$.
    \qedhere

  \end{proofcases}
\end{proof}
\end{lemma}

\begin{lemma}[Erasure is well formed]
\label{lem:erase-wf}
  If $\Gamma \vdash_\mSMT \tau$ then $\Gamma \vdash_\mSMT \erase(\tau)$.
\begin{proof}
  By induction on the well formedness derivation.
\begin{proofcases}

  \item[(\rn{$t$-Base})] Immediate, since $\erase(B) = B$.

  \item[(\rn{$t$-TVar})] Contradictory---type variables aren't well typed at $\mSMT$.

  \item[(\rn{$t$-ADT})] By the IH on each constituent of $D
  ~ \vec{\tau_i}$, and then by \rn{$t$-ADT}.

  \item[(\rn{$\tau$-SMT})] Since $\erase(\smt{t}) = \erase(t)$, by the
  IH on $\Gamma \vdash_\mSMT t$.

  \item[(\rn{$\tau$-Sym})] Since $\erase(\sym{t}) = \erase(t)$, by the
  IH on $\Gamma \vdash_\mSMT t$.

  \item[(\rn{$\tau$-Model})] Contradictory---$\MODEL$ isn't well typed
    at $\mSMT$.
    \qedhere

\end{proofcases}
\end{proof}
\end{lemma}

\begin{lemma}[SMT types have only SMT parts]
  \label{lem:smt-wf-inversion}
  If $\Gamma \vdash_\mSMT \tau$, then all of $\tau$'s subparts are also well formed at $\mSMT$.
  \begin{proof}
    By induction on $\tau$.
    \begin{proofcases}
    \item[($\tau=B$)] Immediate.
      
    \item[($\tau=\alpha$)] Contradictory---type variables are only well formed at $\mEXP$.
      
    \item[($\tau=D ~ \vec{\tau_i}$)] By the IH on each $\tau_i$.
      
    \item[($\tau=\smt{t}$)] By the IH on $t$.
      
    \item[($\tau=\sym{t}$)] By the IH on $t$.
      
    \item[($\tau=\MODEL$)] Contradictory---$\MODEL$ is only well formed at $\mEXP$.
    \qedhere
    \end{proofcases}
  \end{proof}
\end{lemma}

\begin{lemma}[SMT conversion is well formed]
\label{lem:tosmt-wf}
  If $\Gamma \vdash_\mSMT \tau$ then $\tosmt(\tau) = \sym{t}$ or $\smt{t}$
  such that $\Gamma \vdash_\mSMT t$ (and so
  $\Gamma \vdash_\mSMT \tosmt(\tau)$).
\begin{proof}
  By induction on the well formedness derivation.
\begin{proofcases}

  \item[(\rn{$t$-Base})] $\tosmt(B) = \smt{B}$, which is well formed
  by \rn{$\tau$-SMT} and \rn{$t$-B}.

  \item[(\rn{$t$-TVar})] Contradictory---type variables are only well formed at $\mEXP$.

  \item[(\rn{$t$-ADT})] We know that $\erase(D ~ \vec{\tau_i})$ is
  still well formed by Lemma~\ref{lem:erase-wf}; then
  by \rn{$\tau$-SMT}.

  \item[(\rn{$\tau$-SMT})] Since it must be that $\Gamma \vdash_\mSMT
  t$, then $\erase(t)$ is also well formed by
  Lemma~\ref{lem:erase-wf}; then by \rn{$\tau$-SMT}.

  \item[(\rn{$\tau$-Sym})] Since it must be that $\Gamma \vdash_\mSMT
  t$, then $\erase(t)$ is also well formed by
  Lemma~\ref{lem:erase-wf}; then by \rn{$\tau$-Sym}.

  \item[(\rn{$\tau$-Model})] Contradictory---$\MODEL$ is only well formed at $\mSMT$
  \qedhere

\end{proofcases}
\end{proof}
\end{lemma}

\begin{lemma}[SMT conversion is only for SMT types]
  \label{lem:tosmt-defined}
  $\tosmt(\tau)$ is defined iff $\Gamma \vdash_\mSMT \tau$.
  \begin{proof}
    The right-to-left direction is proved by Lemma~\ref{lem:tosmt-wf}.
    For left-to-right, we go by induction on $\tau$.
    \begin{proofcases}
    \item[($\tau=B$)] By \rn{$t$-Base}.
      
    \item[($\tau=\alpha$)] Contradictory---type variables are undefined for $\erase$.
      
    \item[($\tau=D ~ \vec{\tau_i}$)] By the IH on each $\tau_i$ and \rn{$t$-ADT}.
      
    \item[($\tau=\smt{t}$)] By the IH on $t$ and \rn{$\tau$-SMT}.
      
    \item[($\tau=\sym{t}$)] By the IH on $t$ and \rn{$\tau$-Sym}.
      
    \item[($\tau=\MODEL$)] Contradictory---$\MODEL$ is undefined for $\erase$.
    \qedhere
    \end{proofcases}    
  \end{proof}  
\end{lemma}

We say a type $t$ is an ``SMT type'' when $\Gamma \vdash_\mSMT t$; a
type $\tau$ is an SMT type when it is equal to an SMT type $t$ or when
it is of the form $\smt{t}$ or $\sym{t}$. Note that $\tosmt$ always
produces an SMT type, but does not work on types that contain type
variables or the unrepresentable $\MODEL$ type.

\begin{lemma}[Weakening]
\label{lem:weakening}
  If $\vdash \Gamma$ and $\vdash \Gamma'$ and
  $\dom(\Gamma) \cap \dom(\Gamma') = \emptyset$ then:
\begin{enumerate}

\item \label{weaken:Gamma} $\vdash \Gamma,\Gamma'$

\item \label{weaken:tau} If $\Gamma \vdash_m \tau$ then $\Gamma,\Gamma' \vdash_m \tau$;

\item \label{weaken:e} If $\Gamma \vdash e : \tau$ then $\Gamma,\Gamma' \vdash e : \tau$; and

\item \label{weaken:phi} If $\Gamma \vdash \phi : \tau$ then $\Gamma,\Gamma' \vdash \phi : \tau$.

\end{enumerate}
\begin{proof}
  By mutual induction on the derivations.

\paragraph*{Contexts}
\begin{proofcases}

  \item[(\rn{$\Gamma$-Empty})] We have $\Gamma' = \cdot$; immediate by assumption.

  \item[(\rn{$\Gamma$-Var})] We have $\Gamma' = \Gamma'',X:\tau$. By
  the IH on $\Gamma''$ and \rn{$\Gamma$-Var}, finding
  $\Gamma,\Gamma'' \vdash_m \tau$ by part (\ref{weaken:tau}) of
  the IH.

  \item[(\rn{$\Gamma$-TVar})] We have $\Gamma' = \Gamma'',\alpha$. By
  the IH on $\Gamma''$ and \rn{$\Gamma$-TVar}.

\end{proofcases}

\paragraph*{Type well formedness}
\begin{proofcases}

  \item[(\rn{$t$-Base})] Immediate, by \rn{$t$-Base}.

  \item[(\rn{$t$-TVar})] Since $\Gamma$ and $\Gamma'$ have disjoint
  domains, we know $\alpha \in \Gamma$---by \rn{$t$-TVar}.

  \item[(\rn{$t$-ADT})] By the IH on each constituent of $D
  ~ \vec{\tau_i}$, followed by \rn{$t$-ADT}.

  \item[(\rn{$\tau$-SMT})] By the IH on $\Gamma \vdash_\mSMT t$ and
  then \rn{$\tau$-SMT}.

  \item[(\rn{$\tau$-Sym})] By the IH on $\Gamma \vdash_\mSMT t$ and
  then \rn{$\tau$-Sym}.

  \item[(\rn{$\tau$-Model})] Immediate, by \rn{$\tau$-Model}.
  \qedhere

\end{proofcases}

\paragraph*{Expressions}
\begin{proofcases}

  \item[(\rn{$e$-Var})] Since the domains are disjoint,
  $(\Gamma,\Gamma')(X) = \tau$ and we can still find \rn{$e$-Var}.

  \item[(\rn{$e$-Const})] Immediate, by \rn{$e$-Const}.

  \item[(\rn{$e$-Let})] By the \rn{$e$-Let} and the IH on $e_1$ and
  $e_2$, $\alpha$-renaming $X$ appropriately.

  \item[(\rn{$e$-Ctor})] By \rn{$e$-Ctor} and the IH, using part
  (\ref{weaken:tau}) on $\tau_j'$ and part (\ref{weaken:e}) on
  $e_i$.

  \item[(\rn{$e$-Quote})] By the part (\ref{weaken:phi}) of the IH.

  \item[(\rn{$e$-Rel})] By \rn{$e$-Rel} and the IH on each $e_i$.

  \item[(\rn{$e$-Fun})] By \rn{$e$-Fun} and the the IH, using part
  (\ref{weaken:tau}) on $\tau_j'$ and part (\ref{weaken:e}) on
  $e_i$.

  \item[(\rn{$e$-Op})] By \rn{$e$-Op} and the the IH, using part
  (\ref{weaken:tau}) on $\tau_j'$ and part (\ref{weaken:e}) on
  $e_i$.

  \item[(\rn{$e$-If})] By \rn{$e$-If} and the IH on each of the $e_i$.

  \item[(\rn{$e$-Match})] By \rn{$e$-Match} and the IH on $e$ and each
  of the $e_i$, $\alpha$-renaming each $X_k$ appropriately.

\end{proofcases}

\paragraph*{Formulas}

\begin{proofcases}
  \item[(\rn{$\phi$-Var})] By \rn{$\phi$-Var} and part
  (\ref{weaken:tau}) of the IH.

  \item[(\rn{$\phi$-Promote})] By \rn{$\phi$-Promote} and the IH.

  \item[(\rn{$\phi$-Unquote})] By \rn{$\phi$-Unquote} and part
  (\ref{weaken:e}) of the IH.

  \item[(\rn{$\phi$-Ctor})] By \rn{$\phi$-Ctor} and the IH, using part
    (\ref{weaken:tau}) on the $\tau_i$ and part (\ref{weaken:phi}) on
    $\phi_i$, observing that the actual types are unchanged, and so
    the $\tosmt$ conversions are the same.
  \qedhere

\end{proofcases}

\end{proof}
\end{lemma}

\begin{lemma}[Type well formedness strengthening]
\label{lem:types-strengthen}
  If $\Gamma,X:\tau,\Gamma' \vdash_m \tau'$ then $\Gamma,\Gamma' \vdash \tau'$.
\begin{proof}

\begin{proofcases}

  \item[(\rn{$t$-Base})] Immediate.

  \item[(\rn{$t$-TVar})] Immediate: removing the variable binding can't affect $\alpha$.

  \item[(\rn{$t$-ADT})] By the IH on each constituent of $D ~ \vec{\tau_i}$.

  \item[(\rn{$\tau$-SMT})] By the IH on $\Gamma \vdash_\mSMT t$.

  \item[(\rn{$\tau$-Sym})] By the IH on $\Gamma \vdash_\mSMT t$.

  \item[(\rn{$\tau$-Model})] Immediate.
  \qedhere

\end{proofcases}
\end{proof}
\end{lemma}

\begin{lemma}[Type variable substitution]
\label{lem:substitution}
  If $\vdash \Gamma,\alpha,\Gamma'$ and $\Gamma \vdash_m \tau'$, then:
\begin{enumerate}
\item \label{subst:ctx} $\vdash \Gamma,\Gamma'[\tau/\alpha]$;

\item \label{subst:type-wf} If $\Gamma,\alpha,\Gamma' \vdash_m \tau$ then $\Gamma,\Gamma'[\tau/\alpha] \vdash_m \tau'[\tau/\alpha]$;

\item \label{subst:binding} If $\Gamma,\alpha,\Gamma' \vdash X, \tau' \rhd \Gamma''$ then
$\Gamma,\Gamma'[\tau/\alpha] \vdash X, \tau'[\tau/\alpha] \rhd \Gamma''[\tau/\alpha]$; and

\item \label{subst:binding-vec} If $\Gamma,\alpha,\Gamma' \vdash \vec{X_i}, \vec{\tau'_i} \rhd \Gamma''$ then
$\Gamma,\Gamma'[\tau/\alpha] \vdash \vec{X_i}, \vec{\tau'_i}[\tau/\alpha] \rhd \Gamma''[\tau/\alpha]$.

\end{enumerate}
\begin{proof}

  For parts (\ref{subst:ctx}) and (\ref{subst:type-wf}), by mutual
  induction on the derivations. Note that if $\alpha$ actually occurs
  in the type, we could only have found well formedness at $\mEXP$.
\begin{proofcases}

\item[(\rn{$\Gamma$-Empty})] Contradictory: $\cdot \neq \Gamma,\alpha,\Gamma'$.

\item[(\rn{$\Gamma$-Var})] We have $\Gamma' = \Gamma'',X:\tau'$ where $\Gamma,\Gamma'' \vdash_\mEXP \tau'$. By the IH on $\Gamma''$, we know that $\vdash \Gamma,\Gamma''[\tau/\alpha]$; by part (\ref{subst:type-wf}), we have $\Gamma,\Gamma''[\tau/\alpha] \vdash_\mEXP \tau'[\tau/\alpha]$; and so we have $\vdash \Gamma,\Gamma'[\tau/\alpha]$ by \rn{$\Gamma$-Var}.

\item[(\rn{$\Gamma$-TVar})] We have $\Gamma' = \Gamma'',\beta$; by the IH on $\Gamma''$, we have $\vdash \Gamma,\Gamma''[\tau/\alpha]$; since $\beta[\tau/\alpha] = \beta$, we can apply \rn{$\Gamma$-TVar} to find $\vdash \Gamma,\Gamma'[\tau/\alpha]$ as desired.

\item[(\rn{$t$-B})] Immediate by \rn{$t$-B}, since $B[\tau/\alpha] = B$.

\item[(\rn{$t$-TVar})] We have $\tau' = \beta$. If $\alpha = \beta$, then we have $\Gamma \vdash_m \alpha[\tau/\alpha]$ by assumption. If $\alpha \ne \beta$, then it must be that $\beta \in \Gamma$ or $\Gamma'$---either way, $\beta$ is unaffected by the substitution and we have $\beta \in \Gamma,\Gamma'[\tau/\alpha]$ and so $\Gamma,\Gamma'[\tau/\alpha] \vdash_m \beta$ by \rn{$t$-TVar}.

\item[(\rn{$t$-ADT})] By the IH on each premise, followed by \rn{$t$-ADT}.

\item[(\rn{$\tau$-SMT})] By the IH on $\Gamma,\alpha,\Gamma' \vdash_\mSMT t$ and then by \rn{$\tau$-SMT}.

\item[(\rn{$\tau$-Sym})] By the IH on $\Gamma,\alpha,\Gamma' \vdash_\mSMT t$ and then by \rn{$\tau$-Sym}.

\item[(\rn{$\tau$-Model})] Immediate by \rn{$\tau$-Model}.

\end{proofcases}

  For parts (\ref{subst:binding}) and (\ref{subst:binding-vec}), by
  mutual induction on the derivations.
\begin{proofcases}

\item[(\rn{$X\tau$-Bind})] We have $X \not\in \dom(\Gamma,\alpha,\Gamma')$, so it must also be the case that $X \not\in \dom(\Gamma,\Gamma'[\tau/\alpha])$. We therefore find $\Gamma,\Gamma'[\tau/\alpha] \vdash X, \tau'[\tau/\alpha] \rhd \Gamma,\Gamma'[\tau/\alpha],\tau'[\tau/\alpha]$ by \rn{$X\tau$-Bind}.

\item[(\rn{$X\tau$-Check})] We have $(\Gamma,\alpha,\Gamma')(X) = \tau'$. Is $X:\tau'$ in $\Gamma$ or $\Gamma'$? Either way we will find $\Gamma,\Gamma'[\tau/\alpha] \vdash X,\tau'[\tau/\alpha] \rhd \Gamma,\Gamma'[\tau/\alpha]$ by \rn{$X\tau$-Check}.

If $\tau' \in \dom(\Gamma)$, then $\Gamma \vdash \tau'$ and so
$\tau'[\tau/\alpha] = \tau'$ $(\Gamma,\Gamma'[\tau/\alpha])(X)
= \tau'$ and we have $\Gamma,\Gamma'[\tau/\alpha] \vdash
X, \tau' \rhd \Gamma,\Gamma'[\tau/\alpha]$.

If, on the other hand, $\tau' \in \dom(\Gamma')$, then
$(\Gamma,\Gamma'[\tau/\alpha])(X) = \tau'[\tau/\alpha]$. We therefore
have $\Gamma,\Gamma'[\tau/\alpha] \vdash
X, \tau'[\tau/\alpha] \rhd \Gamma,\Gamma'[\tau/\alpha]$.

\item[(\rn{$\vec{X}\vec{\tau}$-All})] By part (\ref{subst:binding}) of the IH on each premise.
\qedhere

\end{proofcases}
\end{proof}
\end{lemma}

\begin{lemma}[Regularity; formulas have SMT types]
\label{lem:regularity}
\begin{enumerate}

  \item \label{regularity:gamma} If $\vdash \Gamma$ and $\Gamma(X)
  = \tau$ then $\Gamma \vdash_\mEXP \tau$.

  \item \label{regularity:Phi} If $\vdash \Phi$ then (a) if $f
  : \forall \vec{\alpha_j}, \vec{\tau_i} \rightarrow \tau \in \Phi$
  then $\vec{\alpha_j} \vdash_\mEXP \tau$, and (b) if $\uf
  : \vec{t_i'} \rightarrow t \in \Phi$ then $\cdot \vdash_\mSMT t$.

  \item \label{regularity:e} If $\Delta; \Phi; \Gamma \vdash e
  : \tau$ then $\Gamma \vdash_\mEXP \tau$.

  \item \label{regularity:c} If
  $\Delta; \Phi; \Gamma \vdash \cSMT{c}(\vec{\phi_i})
  : \vec{\tau_i} \rightarrow \tau$ then $\tau_i = \smt{t_i}$ or
  $\tau_i = \sym{t_i}$ and $\tau = \smt{t}$ or $\tau = \sym{t}$ and
  $\Gamma \vdash_\mSMT \tau_i$ and
  $\Gamma \vdash_\mSMT \tau$.

  \item \label{regularity:phi} If $\Delta; \Phi; \Gamma \vdash \phi
  : \tau$ then $\tau = \smt{t}$ or $\tau = \sym{t}$ and
  $\Gamma \vdash_\mSMT \tau$.

  \item \label{regularity:evec} If $\Delta; \Phi; \Gamma \vdash \vec{e_i}
  : \vec{\tau_i}$ then $\Gamma \vdash_\mEXP \tau_i$.

  \item \label{regularity:phivec} If
  $\Delta; \Phi; \Gamma \vdash \vec{\phi_i} : \vec{\tau_i}$ then
  $\tau_i = \smt{t_i}$ or $\tau = \sym{t_i}$ and $\Gamma \vdash_\mSMT
  t_i$ (and so $\Gamma \vdash_\mEXP \tau_i$).
\end{enumerate}

\begin{proof}
  By induction on the typing derivation.
\paragraph*{Contexts}
\begin{proofcases}

  \item[(\rn{$\Gamma$-Empty})] Contradictory---there's no way $\cdot$
  has a binding for $X$.

  \item[(\rn{$\Gamma$-Var})] $\Gamma = \Gamma', Y:\tau$. If $X=Y$,
  then we know $\Gamma \vdash_\mEXP \tau$ by assumption; otherwise, by
  the IH on $\Gamma'$.

  \item[(\rn{$\Gamma$-TVar})] $\Gamma = \Gamma', \alpha$. By the IH on
  $\Gamma$.

\end{proofcases}

\paragraph*{Program signatures}
\begin{proofcases}

  \item[(\rn{$\Phi$-Empty})] Contradictory---there are no function definitions in $\cdot$.

  \item[(\rn{$\Phi$-Fun})] $\Phi = \Phi', g : \dots$. For case (a)
  when $f = g$, then by assumption. Otherwise, by the IH on $\Phi'$.

  \item[(\rn{$\Phi$-Rel})] $\Phi = \Phi', p \subseteq \vec{\tau_i}$. By the IH on $\Phi$.

  \item[(\rn{$\Phi$-UFun})] $\Phi = \Phi', \uf'
  : \vec{t_i'} \rightarrow t$. For case (b) when $\uf = \uf'$, then by
  assumption. Otherwise, by the IH on $\Phi'$.

\end{proofcases}

\paragraph*{Expressions}
\begin{proofcases}

  \item[(\rn{$e$-Var})] By the part (\ref{regularity:gamma}) on $\vdash \Gamma$.

  \item[(\rn{$e$-Const})] By assumption, we know that
  $\Gamma \vdash_\mSMT \typeof(k)$; by Lemma~\ref{lem:lift-msmt}
  we can find $\Gamma \vdash_\mEXP \typeof(k)$.

  \item[(\rn{$e$-Let})] By the IH on $\Gamma, X:\tau_1 \vdash e_2
  : \tau_2$, using strengthening (Lemma~\ref{lem:types-strengthen}) to
  find that if $\Gamma, X:\tau_1 \vdash_\mEXP \tau_2$ then
  $\Gamma \vdash_\mEXP \tau_2$.

  \item[(\rn{$e$-Ctor})] Since $\Gamma \vdash_\mEXP \tau_j'$, we know
  by \rn{$t$-ADT} that $\Gamma \vdash_\mEXP D ~ \vec{\tau_j'}$.

  \item[(\rn{$e$-Quote})] By the IH on part (\ref{regularity:phi}), we
    know that $\Gamma \vdash_\mSMT \tau$ (and, less relevantly, that
    $\tau = \smt{t}$ or $\sym{t}$). We can find the same well
    formedness at $\mEXP$ by Lemma~\ref{lem:lift-msmt}.

  \item[(\rn{$e$-Rel})] Immediate by \rn{$t$-B}.

  \item[(\rn{$e$-Fun})] Since $f
  : \forall \vec{\alpha_j}, \vec{\tau_i} \rightarrow \tau \in \Phi$
  and $\vdash \Phi$, we know by part (\ref{regularity:Phi}) of the IH
  know that $\vec{\alpha_j} \vdash_\mEXP \tau$. By weakening
  (Lemma~\ref{lem:weakening}) we can lift that well formedness
  judgment to $\Gamma$. Since each $\Gamma \vdash_\mEXP \tau_j'$, we
  can find that $\Gamma \vdash \tau[\tau_j'/\alpha_j]$ by substitution
  (Lemma~\ref{lem:substitution}).

  \item[(\rn{$e$-Fun})] We have by assumption that $\typeof(\otimes)$
  yields a well-formed type, i.e., $\vec{\alpha_j} \vdash_\mEXP \tau$
  (and also for each $\tau_i$). By weakening
  (Lemma~\ref{lem:weakening}) we can lift that well formedness
  judgment to $\Gamma$. Since each $\Gamma \vdash_\mEXP \tau_j'$, we
  can find that $\Gamma \vdash \tau[\tau_j'/\alpha_j]$ by substitution
  (Lemma~\ref{lem:substitution}).

  \item[(\rn{$e$-If})] By the IH on $\Gamma \vdash e_2 : t$.

  \item[(\rn{$e$-Match})] By the IH on
  $\Gamma, \overrightarrow{X_1:\tau_1[\tau_j/\alpha_j]} \vdash e_1
  : \tau$ we have
  $\Gamma, \overrightarrow{X_1:\tau_1[\tau_j/\alpha_j]} \vdash \tau$;
  we can use strengthening (Lemma~\ref{lem:types-strengthen}) to find
  $\Gamma \vdash_\mEXP \tau$.

\end{proofcases}

\paragraph*{SMT constructors}
\begin{proofcases}
  \item[(\rn{$c$-SMT-Var})] Immediate, with $\Gamma \vdash_\mSMT t$
  coming from the rule itself.

  \item[(\rn{$c$-SMT-Const})] Immediate, since we have by assumption
  that $\cdot \vdash_\mSMT \typeof(k)$.

  \item[(\rn{$c$-SMT-Let})] Immediate, with the necessary well
  formedness assumptions coming from the rule itself.

  \item[(\rn{$c$-SMT-Ctor})] We know that $D ~ \vec{t_j'}$ is well
    formed by \rn{$t$-ADT}. We can find the translation of the
    argument types well formed by Lemma~\ref{lem:tosmt-wf} on each of
    the $\Gamma \vdash_\mSMT \tau_i[t_j'/\alpha_j]$ derivations.

  \item[(\rn{$c$-SMT-Forall})] Immediate, with the necessary
  $\Gamma \vdash_\mSMT t_1$ coming from the rule itself.

  \item[(\rn{$c$-SMT-UFun})] Since $\vdash \Phi$ and $\uf
  : \vec{t_i} \rightarrow t \in \Phi$, we know that
  $\cdot \vdash_\mSMT t$ by part (\ref{regularity:Phi}) of the IH and
  so $\cdot \vdash_\mSMT \smt{t}$, which we can lift to $\Gamma$ by
  weakening (Lemma~\ref{lem:weakening}).

\end{proofcases}

\paragraph*{Formulas}
\begin{proofcases}

  \item[(\rn{$\phi$-Promote})] Since $\Delta; \Phi; \Gamma \vdash \phi
  : \sym{t}$, we know that $\Gamma \vdash_\mSMT t$ and so we are
  correct in yielding $\tau = \smt{t}$.

  \item[(\rn{$\phi$-Unquote})] We have $\Delta; \Phi; \Gamma \vdash e
    : \tau$ such that $\Gamma \vdash_\mSMT \tau$. By
    Lemma~\ref{lem:tosmt-wf} we know that $\tosmt(\tau)$ is a well
    formed SMT type.

  \item[(\rn{$\phi$-Ctor})] By the IH part (\ref{regularity:c}) on $\Gamma \vdash \cSMT{c} : \vec{\tau_i} \rightarrow \tau$.
\end{proofcases}

\paragraph*{Vectored expressions and formulas}
By the IH for parts (\ref{regularity:e}) and (\ref{regularity:phi}),
respectively.

\end{proof}
\end{lemma}

\begin{lemma}[SMT value conversion is type correct]
\label{lem:tosmt-well-typed}
  If $\Delta; \Phi; \Gamma \vdash v : \tau$ and $\Gamma \vdash_\mSMT \tau$ then
  $\Delta; \Phi; \Gamma \vdash \tosmt(v) : \tosmt(\tau)$.
  \begin{proof}
  First, observe that $\tau$ and all of its parts must be well formed
  at $\mSMT$, by Lemmas~\ref{lem:tosmt-defined}
  and~\ref{lem:smt-wf-inversion}.
  By induction on the typing derivation. In expression mode, the
  applicable rules are \rn{$e$-Const} and \rn{$e$-Ctor}; only a few
  typing rules could even have applied to a value in formula mode:
  the \rn{$\phi$-Ctor} and \rn{$\phi$-Promote}.
\begin{proofcases}

  \item[(\rn{$e$-Const})] We have $\Gamma \vdash k : \typeof(k)$;
  since $\tosmt(k) = \cCONST[k]()$ and $\tosmt(\typeof(k)) = \smt{k}$
  (since $\Gamma \vdash_\mSMT \typeof(k)$ by assumption), we must show
  that $\Gamma \vdash \cCONST[k]() : \smt{\typeof(k)}$, which we have
  by \rn{$\phi$-SMT-Const}.

  \item[(\rn{$e$-Ctor})] We have $v = c(\vec{v_i})$ and:
  \[ \Delta(D) = \forall \vec{\alpha_j}, \{ \dots, c : \vec{\tau_i}, \dots \} \andalso
      \Gamma \vdash_\mEXP \tau_j' \andalso
      \Gamma \vdash v_i : \tau_i[\tau_j'/\alpha_j]
  \]
  Further, we know that $\Gamma \vdash_\mSMT D ~ \vec{\tau_j'}$ (and
  so each of the subderivations must also be $\mSMT$) and that
  $\tosmt(c(\vec{v_i})) = \cCTOR[c](\overrightarrow{\tosmt(v_i)})$.
  By the IH on each of these $v_i$, we know that we find appropriate
  values at appropriately converted types, i.e.,
  $\Gamma \vdash \tosmt(v_i) : \tosmt(\tau_i[\tau_j'/\alpha_j])$. 
  By Lemma~\ref{lem:tosmt-wf}, we know
  $\tosmt(\tau_i[\tau_j'/\alpha_j])$ is some well formed SMT type. We
  are \emph{almost} able to apply \rn{$\phi$-SMT-Ctor}, but we must
  pick appropriate $t_j'$. We know that $\tosmt(\tau_j')$ is a well
  formed SMT type of the form $\smt{t_j'}$ or $\sym{t_j'}$
  (Lemma~\ref{lem:tosmt-wf}). Whether it's symbolic or not, let
  the inner $t_j'$ there be our $t_j'$. We can now apply \rn{$\phi$-SMT-Ctor} to
  find that $\Gamma \vdash \tosmt(c(\vec{v_i})) : \tosmt(D ~
  \vec{\tau_j'})$.

  \item[(\rn{$\phi$-Promote})] By the IH on $\Gamma \vdash v
  : \sym{t}$, we know that $\Gamma \vdash \tosmt(v)
  : \tosmt(\sym{t})$, i.e., $\Gamma \vdash \tosmt(v) : \sym{t}$. By
  reapplying \rn{$\phi$-Promote} we can find that
  $\Gamma \vdash \tosmt(v) : \smt{v}$.

  \item[(\rn{$\phi$-Ctor})] Immediate: $\tosmt$ does nothing to the
  $\cSMT{\dots}$ constructed value nor to the SMT-type $\tau$ assigned
  to it (which is $\smt{t}$ in all cases except for $\cVAR$).

\end{proofcases}
\end{proof}
\end{lemma}

\section{Type safety}
\label{sec:metatheory-safety}

To prove type safety, we prove two properties for every mode of evaluation:
first, it is \emph{safe}, i.e, never yields $\bot$;
and second, it is \emph{type preserving}, i.e., well typed inputs yield
well typed outputs.

The proofs are fairly conventional. For all but the last step, we
prove safety and type preservation simultaneously.
We start with expressions and formulas
(Lemma~\ref{lem:typesafe-term-formula}), which requires a modest
notion of canonical forms (Lemma~\ref{lem:canonical-forms-sym}).
Next, we prove that value unification
(Lemma~\ref{lem:typesafe-value-unification}) is type preserving,
reasoning about unification in general within the lemma showing safety
and type preservation for premises (Lemma~\ref{lem:typesafe-premise}).
After a brief lemma about bindings (Lemma~\ref{lem:bind-contain}), we
can prove that program evaluation is safe
(Theorem~\ref{thm:typesafe-program-safe}) and type preserving
(Theorem~\ref{thm:typesafe-program-pres}).

\begin{lemma}[Canonical forms for $\sym{t}$]
\label{lem:canonical-forms-sym}
If $\Delta; \Phi; \Gamma \vdash v : \sym{t}$ then $v = \cVAR[x, t]()$.
\begin{proof}
  The only typing rule that could have applied is \rn{$\phi$-SMT-Var}.
\end{proof}
\end{lemma}

\begin{lemma}[Term and formula type safety]
\label{lem:typesafe-term-formula}
  If $\Delta; \Phi \models \world$ and $\Delta; \Phi \vdash \vec{F}$
  and $\Gamma \models \theta$, when either:
  \begin{enumerate}

  \item\label{typesafety:e} $\Delta; \Phi; \Gamma \vdash e : \tau$ and
  $\world; \theta \vdash e \StepstoE v_\bot$; or

  \item\label{typesafety:phi} $\Delta; \Phi; \Gamma \vdash \phi
  : \tau$ and $\world; \theta \vdash \phi \StepstoPhi v_\bot$

  \item\label{typesafety:F} $\Delta; \Phi \vdash \fun{f}{\vec{X_i}
  : \vec{\tau_i}}{\tau}{e}$ and $\vec{\alpha_j}, \overrightarrow{X_i
  : \tau_i} \models \theta'$ and $\world; \theta' \vdash e \StepstoE
  v_\bot$

  \end{enumerate}
then $v_\bot = v$ (i.e., $v_\bot \ne \bot$) and
$\Delta; \Phi; \Gamma \vdash v : \tau$.

Similarly, when either:

\begin{enumerate}
  \item\label{typesafety:vece} if $\Delta; \Phi; \Gamma \vdash e_i
  : \tau_i$ and $\world; \theta \vdash \vec{e_i} \StepstoE \vec{v_i}$
  then $\Delta; \Phi; \Gamma \vdash \vec{v_i}_\bot : \vec{\tau_i}$; and

  \item\label{typesafety:vecphi} if
  $\Delta; \Phi; \Gamma \vdash \phi_i : \tau_i$ and
  $\world; \theta \vdash \vec{\phi_i} \StepstoVecPhi \vec{v_i}$ then
  $\Delta; \Phi; \Gamma \vdash \vec{v_i}_\bot : \tau$
\end{enumerate}
then $\vec{v_i}_\bot = \vec{v_i}$ (i.e., it is not $\bot$) and
$\Delta; \Phi; \Gamma \vdash v_i : \tau_i$.
\begin{proof}
  By mutual induction on derivations and the length of the vectored
  expressions/formulas, leaving $\theta$ general (for,
  e.g., \rn{$e$-Let} and \rn{$e$-Match}).

  \paragraph*{Expressions}

\begin{proofcases}

  \item[(\rn{$e$-Var})] We have $\Gamma(X) = \tau$; since
  $\Gamma \models \theta$, we have $\theta(X) = v$ (and
  so \rn{$\StepstoE$-Var-E} didn't apply). So it must be the case
  that \rn{$\StepstoE$-Var} applied. We can see further that
  $\Delta; \Phi; \cdot \vdash v : \tau$, and we are done by weakening
  (Lemma~\ref{lem:weakening}).

  \item[(\rn{$e$-Const})] It must be that \rn{$\StepstoE$-Const}
  applied, and we immediately see that $v_\bot \ne \bot$ and $k$ is
  well typed in any well formed context by assumption
  and \rn{$e$-Const}.

  \item[(\rn{$e$-Let})] We know that $\Gamma \vdash e_1 : \tau_1$ and
  $\Gamma, X:\tau_1 \vdash e_2 : \tau_2$. By the IH on $e_1$, we know
  that $\theta; \world \vdash e_1 \StepstoE v_1$, so it can't be the
  case that \rn{$\StepstoE$-Let-E} applied---it must hae
  been \rn{$\StepstoE$-Let}. By the IH on $e_2$, we know that the
  final result is also not $\bot$ and is well typed.

  \item[(\rn{$e$-Ctor})] We have $\Delta(D)
  = \forall \vec{\alpha_j}, \{ \dots, c : \vec{\tau_i}, \dots \}$ and
  $\Gamma \vdash e_i : \tau_i[\tau_j'/\alpha_j]$. By the IH, we know
  that each of the $\vec{e_i}$ must have reduced to non-$\bot$ values,
  and so \rn{$\StepstoE$-Ctor-E} could not have applied. We can
  therefore see that each $e_i$ reduces to an appropriately typed
  $v_i$, and our resulting value is well typed by \rn{$e$-Ctor}.

  \item[(\rn{$e$-Quote})] Only \rn{$\StepstoE$-Quote} could have
  applied. By the IH, we know that $\phi$ reduces to a non-$\bot$
  value $v$ well typed at $\tau$.

  \item[(\rn{$e$-Rel})] We know that
  $p \subseteq \vec{\tau_i} \in \Phi$ and $\Gamma \vdash e_i
  : \tau_i$. The IH on $\vec{e_i}$ rules out \rn{$StepstoE$-Rel-E1};
  the typing rule rules out the arity mismatch
  in \rn{$\StepstoE$-Rel-E2} and the missing relation
  in \rn{$\StepstoE$-Rel-E3}. So it must be the case
  that \rn{$\StepstoE$-Rel-True} or \rn{$\StepstoE$-Rel-False}
  applied; either way, we yield a $\bool$, which is appropriately
  typed by \rn{$e$-Const}.

  \item[(\rn{$e$-Fun})] We know that $f
  : \forall \vec{\alpha_j}, \vec{\tau_i} \rightarrow \tau \in \Phi$
  and $\Gamma \vdash e_i : \tau_i[\tau_j'/\alpha_j]$. The IH on
  $\vec{e_i}$ rules out \rn{$StepstoE$-Fun-E1}; the typing rule rules
  out the arity mismatch in \rn{$\StepstoE$-Fun-E2} and the missing
  function in \rn{$\StepstoE$-Fun-E3}. So it must be the case
  that \rn{$\StepstoE$-Fun} applied. Since $\Delta; \Phi \vdash F$, we
  know by the IH on part (\ref{typesafety:F}) that the resulting value
  is non-$\bot$ and well typed at $\tau[\tau_j'/\alpha_j]$.

  \item[(\rn{$e$-Op})] We know that $\typeof(\otimes)
  = \forall \vec{\alpha_j}, \vec{\tau_i} \rightarrow \tau$ and
  $\Gamma \vdash e_i : \tau_i[\tau_j'/\alpha_j]$. The IH on
  $\vec{e_i}$ rules out \rn{$StepstoE$-Op-E1}; the typing rule rules
  out the arity/domain mismatch in \rn{$\StepstoE$-Op-E2}. So it must
  be the case that \rn{$\StepstoE$-Op} applied. We know that the
  result is well typed by our assumption that $\typeof(\otimes)$ and
  $\denot{\otimes}$ agree.

  \item[(\rn{$e$-If})] We have $\Gamma \vdash e_1 : \bool$ and
      $\Gamma \vdash e_2 : \tau$ and $\Gamma \vdash e_3 : \tau$.
  By the IH on $e_1$, we know that $e_1$ reduces to $\true$ or
  $\false$ (since those are the only values of type $\bool$). So we
  can rule out \rn{$\StepstoE$-Ite-E1}
  and \rn{$\StepstoE$-Ite-E2}---we must have stepped by
  either \rn{$\StepstoE$-Ite-T} or \rn{$\StepstoE$-Ite-F}.
  The IH on $e_2$ or $e_3$ (respectively) guarantees we step to a
  non-$\bot$, well typed value.

  \item[(\rn{$e$-Match})] We have $\Gamma \vdash e : D ~ \vec{\tau_j}$
  and $\Delta(D) = \forall \vec{\alpha_j}, \{ \dots, c_i
  : \vec{\tau_k}, \dots \}$ and
  $ \Gamma,\overrightarrow{X_k:\tau_k[\tau_j/\alpha_j]} \vdash e_i
  : \tau$. 
  The IH on $e$ guarantees that we get a non-$\bot$ value at type $D
  ~ \vec{\tau_j}$, which rules out the error
  case \rn{$\StepstoE$-Match-E1}, the non-constructor value
  of \rn{$\StepstoE$-Match-E2}, the mis-named constructor
  of \rn{$\StepstoE$-Match-E3}, and the arity error
  of \rn{$\StepstoE$-Match-E4}.
  So it must be the case that we applied \rn{$\StepstoE$-Match}; by
  the IH, the matching pattern reduces to a well typed non-$\bot$
  value.

\end{proofcases}

  \paragraph*{Formulas}

\begin{proofcases}

  \item[(\rn{$\phi$-Promote})] We have
  $\Delta; \Phi; \Gamma \vdash \phi : \sym{t}$; by the IH, we know
  that $\phi$ steps to a non-$\bot$ value $v$ well typed at $\sym{t}$;
  by \rn{$\phi$-Promote} we can see that $v$ is also well typed at
  $\smt{t}$.

  \item[(\rn{$\phi$-Unquote})] We have $\unq e$; since
  $\Delta; \Phi; \Gamma \vdash e : \tau$, we know by the IH that $e$
  reduces to a non-$\bot$ value $v$ that is also well typed at $\tau$.
  We can therefore rule out \rn{$\StepstoPhi$-Unquote-E}, so we must
  have stepped by \rn{$\StepstoPhi$-Unquote}.

  Since $\Delta; \Phi; \Gamma \vdash v : \tau$ and $\Gamma \vdash_\mSMT \tau$, we have
  $\Delta; \Phi; \Gamma \vdash \tosmt(v) : \tosmt(\tau)$ by
  Lemma~\ref{lem:tosmt-well-typed}, as desired.

  \item[(\rn{$\phi$-Ctor})] We have $\cSMT{c}(\vec{\phi_i})$ such
  that $\Gamma \vdash \cSMT{c} : \vec{\tau_i} \rightarrow \tau$ and
  $\Gamma \vdash \phi_i : \tau_i$. We know by the IH that each
  $\phi_i$ is well typed at $\tau_i$ and so none of them step to
  $\bot$, and so \rn{$\StepstoPhi$-Ctor-E} cannot apply.

  Therefore either \rn{$\StepstoPhi$-Ctor} or \rn{$\StepstoPhi$-Value}
  applied; the resulting value is well typed by the IH or remains well
  typed, respectively.

\end{proofcases}

  \paragraph*{Functions}

By part (\ref{typesafety:e}) on
$\vec{\alpha_j}, \overrightarrow{X_i:\tau_i} \vdash e : \tau$, using
weakening (Lemma~\ref{lem:weakening}) to recover typing in $\Gamma$.

  \paragraph*{Vectored expressions and formulas}

By induction on the vector length, using parts (\ref{typesafety:e})
and (\ref{typesafety:phi}) in each case. \qedhere

\end{proof}
\end{lemma}

\begin{lemma}[Value unification preservation]
\label{lem:typesafe-value-unification}
  If $\Gamma \models \theta$ when either:
  \begin{enumerate}

  \item $\Gamma \vdash \vec{X}, \vec{\tau} \rhd \Gamma'$ and
  $\Gamma \vdash \vec{v} : \vec{\tau}$ and
  $\theta \vdash \vec{X} \sim \vec{v} : \theta'$; or

  \item $\Gamma \vdash X, \tau \rhd \Gamma'$ and $\Gamma \vdash v
  : \tau$ and $\theta \vdash X \sim v \rhd \theta'$;

  \end{enumerate}
  then $\Gamma' \models \theta'$.

\begin{proof}
  By induction on the derivation of well typing.
\begin{proofcases}

  \item[(\rn{$X\tau$-Bind})] Only \rn{$uv$-Bind-Var} could have
  applied, so we have $\Gamma \vdash v : \tau$ and
  $\Gamma \models \theta$ and must show that
  $\Gamma,X:\tau \models \theta[X \mapsto v]$, which we have
  immediately.
  
  \item[(\rn{$X\tau$-Check})] Here $X \in \Gamma$, so it must be that
  $\theta(X)$ is defined. One of three rules could have applied:

  \begin{proofcases}

    \item[(\rn{$uv$-Eq-Var})] We have $\Gamma' = \Gamma$ and $\theta'
    = \theta$, so $\Gamma' \models \theta'$ by assumption.

    \item[(\rn{$uv$-Ctor})] By the IH, we know that $\Gamma' models \theta'$.

    \item[(\rn{$uv$-Constant})] As for \rn{$uv$-Eq-Var}, we have
    $\Gamma' = \Gamma$ and $\theta' = \theta$, so
    $\Gamma' \models \theta'$ by assumption.

  \end{proofcases}

  \item[(\rn{$X\tau$-All})] It must be that \rn{$\vec{u}\vec{v}$-All}
  applied; by the IH on each sub-derivation, we can find that
  $\Gamma_i \models \theta_i$, and so $\Gamma' \models \theta'$ in
  particular.

\end{proofcases}
\end{proof}
\end{lemma}

User code will never \emph{directly} trigger a use
of \rn{$uv$-Eq-Var} directly, because the unification rules won't call
value unification with a defined LHS (we'd just use \rn{$uu$-BB}
instead). But a use of \rn{$\vec{u}\vec{v}$-All} could lead to a
variable being unified early on and then used again in the same
unification process.

\begin{lemma}[Premise preservation and safety]
\label{lem:typesafe-premise}
  If $\Delta; \Phi; \Gamma \vdash P \rhd \Gamma'$ and
  $\Delta; \Phi \models \vec{F}$ and $\Delta; \Phi \models \world$ and
  $\Gamma \models \theta$ then
  if $\vec{F}; \world; \theta \vdash P \stepsto \theta'_\bot$ then:
  \begin{enumerate}

  \item\label{typesafety:premise-safe} $\theta'_\bot = \theta'$ (i.e.,
  it is not $\bot$); and

  \item\label{typesafety:premise-pres}
  $\Gamma' \models \theta'$.  
  \end{enumerate}
\begin{proof}
  By induction on the premise typing derivation, followed by cases on the step taken.
\begin{proofcases}

  \item[(\rn{$P$-PosAtom})] We have:
  \[ p \subseteq \vec{\tau_i} \in \Phi \andalso
     \Gamma \vdash \vec{X_i}, \vec{\tau_i} \rhd \Gamma'
  \]
  The only rule that could have applied is \rn{PosAtom}, i.e.,
  $\vec{v} \in \world(p)$ and
  $\theta \vdash \vec{X_i} \unify \vec{v_i} : \theta'_\bot$. We must
  show that $\theta'_\bot = \theta'$ and $\Gamma' \models \theta'$.

  Since $\Delta; \Phi \models \world$, we know that
  $\cdot \vdash \vec{v_i} : \vec{\tau_i}$; by weakening we have
  $\Gamma \vdash \vec{v_i} : \vec{\tau_i}$
  (Lemma~\ref{lem:weakening}).

  Syntactically, we know that $\vec{X_i}$ are all variables and that
  $\vec{v_i}$ are all values. For each one, therefore only two
  unification rules could possibly apply: \rn{$uu$-BB} ($X_i$ is
  bound) and \rn{$uu$-FB} ($X_i$ is free). In
  particular, \rn{$uu$-FF} cannot apply, and so we cannot
  produce $\bot$, so $\theta'_\bot = \theta' = \theta\vec{\theta_i}$.
  By Lemma~\ref{lem:typesafe-value-unification}, we know that
  $\Gamma' \models \theta\theta_i$ for each $i$, and so
  $\Gamma' \models \theta\vec{\theta_i}$.

  \item[(\rn{$P$-NegAtom})] We have:
  \[ \begin{array}{c}
      p \subseteq \vec{\tau_i} \in \Phi \andalso
      \Gamma \vdash \vec{X_i}, \vec{\tau_i} \rhd \Gamma
  \end{array} \]
  Two rules are possible: \rn{NegAtom} and \rn{NegAtom-E}. We must
  show that the latter cannot apply and that the former preserves
  typing.

  Since $\Gamma \models \theta$, it must be that case that each
  $\vec{X_i} \in \dom(\theta)$, and so \rn{NegAtom-E} cannot have
  applied.
  It remains to be seen that $\Gamma \models \theta'$---but
  in \rn{NegAtom} we have $\theta = \theta'$, and so we are done.

  \item[(\rn{$P$-EqCtor-BF})] We have:
  \[ \begin{array}{c}
    \Delta(D) = \forall \vec{\alpha_j}, \{ \dots, c : \vec{\tau_i}, \dots \} \\
    \Gamma \vdash Y, \tau[\vec{\tau_j'}/\vec{\alpha_j}] \rhd \Gamma \andalso
    \vec{X_i} \nsubseteq \Gamma \andalso
    \Gamma \vdash \vec{X_i}, \vec{\tau_i}[\vec{\tau_j'}/\vec{\alpha_j}] \rhd \Gamma'
  \end{array} \]
  The only rule that could have applied is \rn{EqCtor}, where
  $\theta \vdash Y ~ c(\vec{X_i}) : \theta'_\bot$. We must show that
  $\theta'_\bot = \theta'$ (i.e., it is not $\bot$) and that
  $\Gamma' \models \theta'$.

  Since $\Gamma \vdash
  Y, \tau[\vec{\tau_j'}/\vec{\alpha_j}] \rhd \Gamma$, it must be the
  case that $Y \in \dom(\Gamma)$ and so $\theta(Y) = v$ (and so
  $\cdot \vdash v : \tau[\vec{\tau_j'}/\vec{\alpha_j}]$, which also
  holds under $\Gamma$ thanks to weakening
  (Lemma~\ref{lem:weakening})).

  Only two rules could have applied to show $\theta \vdash Y ~
  c(\vec{X_i}) : \theta'_\bot$: \rn{$uu$-BF} (when some of $\vec{X_i}$
  are unbound) or \rn{$uu$-BB} (when all of the $\vec{X_i}$ are
  bound). In either case, \rn{$uu$-FF} can't have a applied, and so
  $\theta'_\bot = \theta'$.

  One of two rules could have applied: \rn{$uv$-Eq-Var}
  or \rn{$uv$-Ctor}. 

  In the former case, we applied \rn{$uu$-BB}, because
  $\theta(c(\vec{X_i})) = c(\vec{v_i})$. We have $\theta'
  = \theta[X \mapsto c(\vec{v_i})$ and $\Gamma' \models \theta'$ by
  substitution on
  $\Gamma \vdash \vec{X_i}, \vec{\tau_i}[\vec{\tau_j'}/\vec{\alpha_j}] \rhd \Gamma'$
  (Lemma~\ref{lem:substitution}).

  In the latter case, we can find that $\Gamma' \models \theta'$ by
  Lemma~\ref{lem:typesafe-value-unification} on the assumption that
  $\Gamma \vdash \vec{X_i}, \vec{\tau_i}[\vec{\tau_j'}/\vec{\alpha_j}] \rhd \Gamma'$,
  and the fact $\theta(Y) = v$ is well typed in $\Gamma$.

  \item[(\rn{$P$-EqSMT-BF})] We have:
  \[ \begin{array}{c}
    \Gamma \vdash \cSMT{c'} : \vec{\tau_i} \rightarrow \tau \andalso
    \Gamma \vdash Y, \tau \rhd \Gamma \andalso
    \vec{X_i} \nsubseteq \Gamma \andalso
    \Gamma \vdash \vec{X_i}, \vec{\tau_i} \rhd \Gamma'
  \end{array} \]
  The only rule that could have applied is \rn{EqSMT}, where
  $\theta \vdash Y \unify \cSMT{c'}(\vec{X_i}) : \theta'_\bot$. We must
  show that $\theta'_\bot = \theta'$ (i.e., it is not $\bot$ and that
  $\Gamma \models \theta'$.

  Since $\Gamma \vdash Y, \tau \rhd \Gamma$, it must be the case that
  $Y \in \dom(\Gamma)$ and so $\theta(Y) = v$. We can conclude that
  $\cdot \vdash v : \tau$ and so $\Gamma \vdash v : \tau$
  (Lemma~\ref{lem:weakening}).

  Only two rules could have applied to show $\theta \vdash Y
  ~ \cSMT{c'}(\vec{X_i}) : \theta'_\bot$, noting the removal of the
  unquote, since unification doesn't care: \rn{$uu$-BF} (when some of
  $\vec{X_i}$ are unbound) or \rn{$uu$-BB} (when all of the
  $\vec{X_i}$ are bound). In either case, \rn{$uu$-FF} can't have a
  applied, and so $\theta'_\bot = \theta'$.

  It remains to show that $\Gamma \models \theta'$. If the outer
  unification rule was \rn{$uu$-BB}, we have $\theta = \theta'$ and so
  $\Gamma \models \theta'$ by assumption.  If outer unification rule
  was \rn{$uu$-BF}, one of two rules could have applied to find value
  unification: either \rn{$uv$-Eq-Var} or \rn{$uv$-Ctor}.

  In the former case, we apply \rn{$uu$-BB} inside, because
  $\theta(Y) = \cSMT{c'}(\vec{v_i})$. We have
  $\theta' = \theta[X \mapsto \cSMT{c'}(\vec{v_i})$ and
  $\Gamma' \models \theta'$ by substitution on
  $\Gamma \vdash \vec{X_i}, \vec{\tau_i} \rhd \Gamma'$
  (Lemma~\ref{lem:substitution}).

  In the latter case, we can find that $\Gamma' \models \theta'$ by
  Lemma~\ref{lem:typesafe-value-unification} on the assumption that
  $\Gamma \vdash \vec{X_i}, \vec{\tau_i} \rhd \Gamma'$,
  and the fact $\theta(Y) = v$ is well typed in $\Gamma$.

  \item[(\rn{$P$-Eq-FB})] We have:
  \[ \begin{array}{c}
    \Gamma \vdash e : \tau \andalso
    \Gamma \vdash Y, \tau \rhd \Gamma'
  \end{array} \]
  The two possible rules are \rn{EqExpr} and \rn{EqExpr-E}. We must
  show that the latter could not have applied (and so $\theta'_\bot
  = \theta'$) and that $\Gamma' \models \theta'$.
  By Lemma~\ref{lem:typesafe-term-formula}, we know that \rn{EqExpr-E}
  cannot apply and that $\world; \theta \vdash e \StepstoE v$ (and so
  $\Gamma \vdash v : \tau$).

  Since $v$ is a value, either \rn{$uu$-FB} or \rn{$uu$-BB} applied,
  depending on whether or not $Y$ is bound. Either way, \rn{$uu$-FF}
  couldn't have applied, and so $\theta'_\bot = \theta'$.

  We can find that $\Gamma' \models \theta'$ by
  Lemma~\ref{lem:typesafe-value-unification} on $\Gamma \vdash
  Y, \tau \rhd \Gamma'$ (along with the well typing of $v$).

  \item[(\rn{$P$-NegEq})] We have $\Gamma \vdash e : \tau$ and $\Gamma
    \vdash Y, \tau \rhd \Gamma$.
    By Lemma~\ref{lem:typesafe-term-formula}, we know that
    \rn{NegEqExpr-E1} cannot apply and that $\world; \theta \vdash e
    \StepstoE v$ (and so $\Gamma \vdash v : \tau$).
  Since $\Gamma \models \theta$, it must be that case that $Y \in
  \dom(\theta)$, and so \rn{NegAtom-E2} cannot have applied.
  It remains to be seen that $\Gamma \models \theta'$---but
  in \rn{NegExpr} we have $\theta = \theta'$, and so we are done.
  \qedhere

\end{proofcases}
\end{proof}
\end{lemma}

\begin{lemma}[Identical bindings implies containment]
\label{lem:bind-contain}
  If $\Gamma \vdash X,\tau \rhd \Gamma$, then $X \in \dom(\Gamma)$.

  Similarly, if $\Gamma \vdash \vec{X_i},\vec{\tau_i} \rhd \Gamma$,
  then $\vec{X_i} \subseteq \dom(\Gamma)$.
\begin{proof}
  By induction on the derivation.
\begin{proofcases}

  \item[(\rn{$X\tau$-Bind})] Contradictory: this rule could not have
  applied, since $\Gamma \ne \Gamma,X:\tau$.
       
  \item[(\rn{$X\tau$-Check})] We have $X \in \dom(\Gamma)$ by
  assumption.
  
  \item[(\rn{$\vec{X}\vec{\tau}$-All})] By the IH on each of our
  premises. 
  \qedhere
       
\end{proofcases}
\end{proof}
\end{lemma}

\begin{theorem}[Program safety]
\label{thm:typesafe-program-safe}
  If $\Delta; \Phi \vdash \vec{F_i} ~ \vec{H_j}$ and $\Delta; \Phi \models \world$ then
  for all $H \in \vec{H_j}, ~ \neg (\vec{F_i}; \world \vdash H \stepsto \bot)$.
\begin{proof}
  The program $\prog = \vec{F_i} ~ \vec{H_j}$ must have been well
  typed according to \rn{$\prog$-WF}, and so we have $\vdash \Delta$
  and $\vdash \Phi$ along with derivations for each $F$ and $H$:
  \[ \begin{array}{c} 
     \Delta; \Phi \vdash F_0 \andalso \dots \andalso
     \Delta; \Phi \vdash F_i \andalso \dots \andalso
     \Delta; \Phi \vdash F_n \\
     \Delta; \Phi \vdash H_0 \andalso \dots \andalso
     \Delta; \Phi \vdash H_j \andalso \dots \andalso
     \Delta; \Phi \vdash H_m \\
   \end{array} \]
  Let an $H = p(X_k) \horn \vec{P_\ell} \in \vec{H_j}$ be given.
  We know that $\Delta; \Phi \vdash H$ by \rn{$H$-Clause}, i.e.:
  \[ \begin{array}{c}
     \cdot \vdash P_0 \rhd \Gamma_1 \andalso \dots \andalso
     \Gamma_{\ell} \vdash P_{\ell} \rhd \Gamma_{\ell+1} \andalso \dots \andalso
     \Gamma_p \vdash P_p \rhd \Gamma' \\
     p \subseteq \vec{\tau_k} \in \Phi \andalso
     \Gamma' \vdash \vec{X_k}, \vec{\tau_k} \rhd \Gamma'
  \end{array} \]
  Let $\world$ be given such that $\Delta; \Phi \models \world$.
  We must show that it is not the case that $\vec{F_i}; \world \vdash
  H \stepsto \bot$, i.e., \rn{Clause-E1} and \rn{Clause-E2} cannot
  apply.
  We can rule out \rn{Clause-E1} by
  Lemma~\ref{lem:typesafe-premise}(\ref{typesafety:premise-safe}: it
  is not the case that a typesafe premise steps to $\bot$.
  To rule out \rn{Clause-E2}, we need to know that if we can build a
  final substitution, i.e.:
  \[ \begin{array}{c}
    \cdot    \vdash  P_{0} \stepsto \theta_1 \quad \dots \quad
    \theta_\ell \vdash  P_{\ell} \stepsto \theta_{\ell+1} \quad \dots \quad
    \theta_p \vdash  P_{p} \stepsto \theta
  \end{array} \]
  then $\vec{X_k} \in \dom(\theta)$. We know that
  $\vec{X_k} \subseteq \dom(\Gamma')$ by Lemma~\ref{lem:bind-contain}
  on $\Gamma' \vdash \vec{X_k}, \vec{\tau_k} \rhd \Gamma'$; since
  $\Gamma_p \vdash P_{p} \rhd \Gamma'$, we know by
  Lemma~\ref{lem:typesafe-premise}(\ref{typesafety:premise-pres}) that
  $\Gamma' \models \theta$. We can therefore conclude that $\forall
  X \in \dom(\Gamma'), ~ X \in \dom(\theta)$, and so
  $\vec{X_k} \in \dom(\theta)$... and \rn{Clause-E2} cannot apply.
\end{proof}
\end{theorem}

\begin{theorem}[Program preservation]
\label{thm:typesafe-program-pres}
  If $\Delta; \Phi \vdash \vec{F_i} ~ \vec{H_j}$ and $\Delta; \Phi \models \world$ and
  $\vec{F_i}; \world \vdash H \stepsto \world'$ for some $H \in \vec{H_j}$ then
  $\Delta; \Phi \models \world'$.
\begin{proof}
 The program $\prog = \vec{F_i} ~ \vec{H_j}$ must have been well
  typed according to \rn{$\prog$-WF}, and so we have $\vdash \Delta$
  and $\vdash \Phi$ along with derivations for each $F$ and $H$:
  \[ \begin{array}{c} 
     \Delta; \Phi \vdash F_0 \andalso \dots \andalso
     \Delta; \Phi \vdash F_i \andalso \dots \andalso
     \Delta; \Phi \vdash F_n \\
     \Delta; \Phi \vdash H_0 \andalso \dots \andalso
     \Delta; \Phi \vdash H_j \andalso \dots \andalso
     \Delta; \Phi \vdash H_m \\
   \end{array} \]
  Let an $H = p(X_k) \horn \vec{P_\ell} \in \vec{H_j}$ be given.
  We know that $\Delta; \Phi \vdash H$ by \rn{$H$-Clause}, i.e.:
  \[ \begin{array}{c}
     \cdot \vdash P_0 \rhd \Gamma_1 \andalso \dots \andalso
     \Gamma_{\ell} \vdash P_{\ell} \rhd \Gamma_{\ell+1} \andalso \dots \andalso
     \Gamma_p \vdash P_p \rhd \Gamma' \\
     p \subseteq \vec{\tau_k} \in \Phi \andalso
     \Gamma' \vdash \vec{X_k}, \vec{\tau_k} \rhd \Gamma'
  \end{array} \]
  Let $\world$ be given such that $\Delta; \Phi \models \world$.
  It must have been the case that we stepped by \rn{Clause}, and so:
  \[ \begin{array}{c}
     \cdot    \vdash P_{0} \stepsto \theta_1 \quad \dots \quad
     \theta_i \vdash P_{i} \stepsto \theta_{i+1} \quad \dots \quad
     \theta_n \vdash P_{n} \stepsto \theta \\
     \world' = \world[p \mapsto \world(p) \cup \theta(\vec{X_j})] \\
  \end{array} \]
  By Lemma~\ref{lem:typesafe-premise}(\ref{typesafety:premise-pres}),
  we know that $\Gamma_i \models \theta_i$ and
  $\Gamma' \models \theta$. We have
  $\vec{X_k} \subseteq \dom(\Gamma')$ by Lemma~\ref{lem:bind-contain}
  on $\Gamma' \vdash \vec{X_k}, \vec{\tau_k} \rhd \Gamma'$, we can
  conclude that $\vec{X_k} \subseteq \dom(\theta)$ and that
  $\Delta; \Phi; \cdot \vdash \theta(X_k) : \tau_k$ by
  Lemma~\ref{lem:typesafe-value-unification} on
  $\Gamma' \vdash \vec{X_k}, \vec{\tau_k} \rhd \Gamma'$.

  To see that $\Delta; \Phi \models \world$, we need to see that
  adding $\theta(\vec{X_k})$ to $\world(p)$ is safe. We already knew
  that $p \subseteq \vec{\tau_k} \in \Phi$ and $p \in \dom(\world)$; we
  have $k = k$ immediately, and we have seen that each $\theta(X_k)$
  is well typed at $\tau_k$.
\end{proof}
\end{theorem}

\section{Model-theoretic semantics}\label{sec:modeltheory}

We have focused on the operational semantics of \Name, as it helps us to
reason about type safety.
However, since we have kept \Name close to Datalog, it is also possible to give
a model-theoretic semantics to a \Name program.
First, all ML functions and expressions are desugared into Datalog rules; this
translation is relatively straightforward, with the trickiest part being the
translation of non-mutually exclusive patterns occurring in match expressions.
For each primitive operator, we introduce a (possibly infinite) EDB relation
that defines that operator; for example, the addition operator $~\ic{+}$ is
represented through a ternary relation \ic{add($x$, $y$, $z$)}, which states
that $z$ is the sum of $x$ and $y$.
Terms of the form $p(w^*)$ (i.e., invocations of predicates as functions) are
translated to aggregate predicates.
\Name requires the use of these terms, as well as negation, to be stratified;
thus, the program resulting from the translation can be given a perfect model
semantics in line with stratified
negation~\citep{apt1988towards,przymusinski1988declarative,vangelder1989negation}
and stratified aggregation~\citep{mumick1990magic}.

For a small example, consider this \Name program:
\begin{lstlisting}
fun length(Xs: 'a list) : bv[32] =
  match Xs with
  | [] => 0
  | _ :: T => 1 + length(T)
  end
ok :- length([1, 2, 3]) = 3.
\end{lstlisting}
This would be translated into a program like this:
\begin{lstlisting}
length([], 0).
length(_ :: T, Z) :-
  length(T, L),
  add(1, L, Z).
ok :- length([1, 2, 3], 3).
\end{lstlisting}
Note that the rules defining the \ic{length} predicate violate the range
restriction, and in fact define an infinite relation.
This does not pose a fundamental problem to the model theory.
To make this program evaluable, we could rewrite the \ic{length} predicate's
definition (and its uses) via the magic set transformation; the resulting
relations would meet the range restriction.
While the magic set transformation can turn a stratified program into a
non-stratified program, there are techniques to either restore
stratification~\citep{meskes1993generalized} or correctly evaluate the
non-stratified program~\citep{mumick1990magic,balbin}.
\fi % \iffull

\end{document}